\documentclass{aastex631}
\usepackage{rotating}
\usepackage{amsmath}

\shorttitle{91T-like Supernovae}
\shortauthors{Yang et al.}

\graphicspath{{figures/}}
\def\edithere#1{\textcolor{black}{#1}}  
\def \Si {Si II $\lambda\lambda$6355 }
\def \Simax {$pEW_{max}$(Si II $\lambda\lambda$6355)}

\begin{document}

\title{Using 1991T/1999aa-like Type Ia Supernovae as Standardizable Candles}

\author[0000-0002-1376-0987]{Jiawen Yang}
\affiliation{George P. and Cynthia Woods Mitchell Institute for Fundamental Physics and Astronomy, Department of Physics and Astronomy, Texas A\&M University, College Station, TX 77843, USA}

\author[0000-0001-7092-9374]{Lifan Wang}
\affiliation{George P. and Cynthia Woods Mitchell Institute for Fundamental Physics and Astronomy, Department of Physics and Astronomy, Texas A\&M University, College Station, TX 77843, USA}

\author[0000-0002-8102-181X]{Nicholas Suntzeff}
\affiliation{George P. and Cynthia Woods Mitchell Institute for Fundamental Physics and Astronomy, Department of Physics and Astronomy, Texas A\&M University, College Station, TX 77843, USA}

\author[0000-0001-7201-1938]{Lei Hu}
\affiliation{Purple Mountain Observatory, Nanjing 210023, People's Republic of China}

\author[0000-0003-0183-451X]{Lauren Aldoroty}
\affiliation{George P. and Cynthia Woods Mitchell Institute for Fundamental Physics and Astronomy, Department of Physics and Astronomy, Texas A\&M University, College Station, TX 77843, USA}

\author[0000-0001-6272-5507]{Peter J. Brown}
\affiliation{George P. and Cynthia Woods Mitchell Institute for Fundamental Physics and Astronomy, Department of Physics and Astronomy, Texas A\&M University, College Station, TX 77843, USA}

\author[0000-0002-6650-694X]{Kevin Krisciunas}
\affiliation{George P. and Cynthia Woods Mitchell Institute for Fundamental Physics and Astronomy, Department of Physics and Astronomy, Texas A\&M University, College Station, TX 77843, USA}

\author[0000-0001-7090-4898]{Iair Arcavi}
\affiliation{The School of Physics and Astronomy, Tel Aviv University, Tel Aviv 69978, Israel}

\author[0000-0003-0035-6659]{Jamison Burke}
\affiliation{Las Cumbres Observatory, 6740 Cortona Drive Suite 102, Goleta, CA, 93117-5575 USA}
\affiliation{Department of Physics, University of California, Santa Barbara, CA 93106-9530, USA}

\author[0000-0002-1296-6887]{Lluís Galbany}
\affiliation{Institute of Space Sciences (ICE, CSIC), Campus UAB, Carrer de Can Magrans, s/n, E-08193 Barcelona, Spain}
\affiliation{Institut d’Estudis Espacials de Catalunya (IEEC), E-08034 Barcelona, Spain}

\author[0000-0002-1125-9187]{Daichi Hiramatsu}
\affiliation{Center for Astrophysics | Harvard \& Smithsonian, 60 Garden Street, Cambridge, MA 02138-1516, USA}
\affiliation{The NSF AI Institute for Artificial Intelligence and Fundamental Interactions}

\author[0000-0002-0832-2974]{Griffin Hosseinzadeh}
\affiliation{Steward Observatory, University of Arizona, 933 North Cherry Avenue, Tucson, AZ 85721-0065, USA}

\author[0000-0003-4253-656X]{D. Andrew Howell}
\affiliation{Las Cumbres Observatory, 6740 Cortona Drive Suite 102, Goleta, CA, 93117-5575 USA}
\affiliation{Department of Physics, University of California, Santa Barbara, CA 93106-9530, USA}

\author[0000-0001-5807-7893]{Curtis McCully}
\affiliation{Las Cumbres Observatory, 6740 Cortona Drive Suite 102, Goleta, CA, 93117-5575 USA}

\author[0000-0002-7472-1279]{Craig Pellegrino}
\affiliation{Las Cumbres Observatory, 6740 Cortona Drive Suite 102, Goleta, CA, 93117-5575 USA}
\affiliation{Department of Physics, University of California, Santa Barbara, CA 93106-9530, USA}

\author[0000-0001-8818-0795]{Stefano Valenti}
\affiliation{Department of Physics and Astronomy, University of California, Davis, 1 Shields Avenue, Davis, CA 95616-5270, USA}

\parskip = 5 mm

\begin{abstract}
We present the photometry of 16 91T/99aa-like Type Ia Supernovae (SNe~Ia) observed by the Las Cumbres Observatory. We also use an additional set of 21 91T/99aa-like SNe~Ia and 87 normal SNe~Ia from the literature for an analysis of the standardizability of the luminosity of 91T/99aa-like SNe. We find that 91T/99aa-like SNe are 0.2 mag brighter than normal SNe~Ia, even when fully corrected by the light curve shapes and colors. The weighted root-mean-square of 91T/99aa-like SNe (with $z_{CMB}>0.01$) Hubble residuals is $0.25\pm0.03$ mag, suggesting that 91T/99aa-like SNe are also excellent relative distance indicators to $\pm$12\%. We compare the Hubble residuals with the pseudo-equivalent width (pEW) of \Si around the date of maximum brightness. We find that there is a broken linear correlation in between those two measurements for our sample including both 91T/99aa-like and normal SNe~Ia. As the \Simax\ increasing, the Hubble residual increases when \Simax$<55.6$ \AA. However, the Hubble residual stays constant beyond this. Given that 91T/99aa-like SNe possess shallower Si II lines than normal SNe~Ia, the linear correlation at \Simax$<55.6$ \AA\ can account for the overall discrepancy of Hubble residuals derived from the two subgroups. Such a systematic effect needs to be taken into account when using SNe~Ia to measure luminosity distances.

\end{abstract}


\section{Introduction} \label{sec:intro}

Type Ia supernovae (SNe~Ia) have played an important role in modern cosmology as cosmological distance indicators which led to the discovery of the accelerating expansion of the Universe and ever-improving values of the Hubble constant \citep{Hamuy_1996_hubble,Riess_etal_1998,Perlmutter_etal_1999,Suntzeff_1999,Freedman_etal_2001,Riess_2021}. Although they are not perfect standard candles, their luminosities can be calibrated using light curve shapes and colors of SNe~Ia (e.g. see \citealp{Phillips_1993,Riess_etal_1996_mlcs,Hamuy_etal_1996abs,Phillips_etal_1999}), by means of which the precision of distance determination can be improved to $5\sim 10\%$.

It is generally agreed that SNe~Ia result from the thermonuclear runaway of a carbon-oxygen (CO) white dwarf (WD) accreting mass from a companion star in a binary system while the nature of the companion star remains unclear. Potential progenitor systems can be broadly categorized into two groups: single degenerate (SD) where the companion star is a main-sequence, subgiant, red giant, or helium star, and double degenerate (DD) where the companion star is another WD. Within this scheme, three major triggering mechanisms have been proposed. The explosion can be triggered by compressional heating near the WD center as a deflagration front as the WD accretes matter from a degenerate or nondegenerate companion to a mass close to the Chandrasekhar limit \citep[e.g.][]{Whelan_Iben_1973}. The explosion can also be triggered by the heat of the dynamical merging process of two CO WDs in a binary system with angular momentum loss via gravitational radiation \citep[e.g.][]{Iben_Tutukov_1984,Webbink_1984}. Another triggering mechanism is through the detonation of a layer of helium on the surface of the WD, which causes a detonation front propagating inwards the WD \citep[e.g.][]{Taam_1980,Woosley_Weaver_1994}. 

Despite the homogeneity assumed for corrected peak luminosities in these supernovae, there remains a persistent intrinsic scatter amongst SNe~Ia beyond just photometric error and the uncertainties in the reddening law, both photometrically and spectroscopically. Mounting evidences are revealed by the spectral diversity among SNe~Ia. It is clear that they are not identical standard candles and are most likely to originate from more than one explosion mechanism as studied in the literature. Overall, SNe~Ia have been classified into four major sub-groups based on their maximum light spectra: core-normal, shallow-silicon, cool, and broad line \citep{Branch_etal_2006,Burrow_2020_csp}. They are also grouped into different categories based on the evolution of the \Si velocities \citep{Benetti_etal_2005,Wang_etal_2009_HV}. There are other frequently used sub-types based on the observed spectral features. The representative examples include: 1) the luminous 91T-like SNe, which show weak features of intermediate-mass elements (IMEs) and prominent Fe II/Fe III lines \citep{Phillips_etal_1992,Filippenko_etal_1992_91T}; 2) the sub-luminous 91bg-like \citep{Filippenko_etal_1992_91bg,Leibundgut_etal_1993} characterized by strong features of IMEs and prominent Ti II lines; 3) the 02cx-like \citep{Li_etal_2003} showing weak Si II lines before maximum light similar to that of  91T-like but with 91bg-like luminosity and significantly lower ejecta velocities than 91T-like SNe; 4) the over-luminous 03fg-like \citep{Scalzo_2012,Howell_etal_2006} with strong absorption lines due to unburned carbon in their spectra which could be identified with explosions of super-Chandrasekhar mass WDs; and 5) the so-called SNe~Ia-CSM 02ic-like \citep{Hamuy_2003} which show strong, broad emission lines of hydrogen presumably caused by strong interaction with circumstellar matter.

The 99aa-like SNe are similar to 91T-like SNe with the subtle difference of having weak Ca II H\&K and Si II absorption lines before maximum light. Their pre-maximum spectral evolution is intermediate of 91T-like SNe~Ia and normal SNe~Ia \citep{Krisciunas_etal_2000,Garavini_etal_2004}. Whether 91T-like SNe are extreme events of SNe~Ia or together with 99aa-like SNe they form a separate class arising from systematically different progenitors is still unknown. In this paper we consider them together as 91T/99aa-like SNe~Ia. So far, cosmological studies using SNe~Ia usually exclude spectroscopically peculiar SNe~Ia including 91T/99aa-like SNe Ia. However, the intrinsic rate of peculiar SNe~Ia can be as high as 36\% with $\sim20\%$ being 91T/99aa-like SNe~Ia \citep{Li_etal_2001}. Since they are among the most luminous SNe~Ia and are located in regions with comparatively young stars \citep{Hamuy_etal_1996abs}, future flux-limited surveys such as LSST/Rubin, WFIRST/Roman, and JWST will inevitably be biased towards finding this class. A potential Malmquist bias then becomes one of the critical issues for these future surveys if the luminosities are not properly taken into account. Not only 91T/99aa-like objects are more luminous, previous studies have also shown that Hubble residuals of 91T/99aa-like objects are in average brighter than the residuals of normal SNe~Ia even after light curve shape and color corrections \citep{Reindl_etal_2005,2021ApJ...912...71B}. Understanding the intrinsic properties of these SNe Ia is crucial to mitigate inevitable systematic errors in supernova cosmology. 

In this paper, we present the photometry of 16 91T/99aa-like SNe observed extensively in $Bg'Vr'i'$ as part of the Las Cumbres Observatory \citep{Brown_etal_2013_lcogt} Supernova Key Project and Global Supernova Project. Together we gather 21 91T/99aa-like objects (as well as additional data of our Las Cumbres 91T/99aa-like object SN2016hvl) and 87 normal SNe~Ia from the literature. With this sample, we analyse the standardizability of 91T/99aa-like objects as relative distance indicators and search for potential systematic effects that may bias the cosmological measurement.
The remainder of this paper is organized as follows.
In \autoref{sec:photometry}, we present the photometric observations, data reduction, calibration, and light curve fitting techniques. The spectral and line strength calculations of \Si are presented in \autoref{sec:spectra}. In \autoref{sec:standardization} we introduce the method to standardize the peak luminosity of SNe~Ia. In \autoref{sec:results} we present the residual Hubble diagram made with our sample, and the use of \Si for distance measurements. The paper concludes with a summary and discussion in \autoref{sec:discussion}. Throughout this paper, we assume a Hubble constant of $H_0$ = 72 km/s/Mpc and a standard cosmology of $\Omega_{M}=0.28, \Omega_{\Lambda}=0.72$.

\section{Photometry}
\label{sec:photometry}
 
\subsection{Data}

\subsubsection{Las Cumbres Data}
\begin{deluxetable}{ccccccc}
\centering
\tabletypesize{\scriptsize}
\tablecaption{General Properties of \edithere{Las Cumbres} 91T/99aa-like Supernovae.\label{tab:generalprop_LCO}}
\tablehead{\colhead{SN Name} & \colhead{$\alpha$(2000)\tablenotemark{a}} & \colhead{$\delta$(2000)\tablenotemark{a}} & \colhead{Host Galaxy\tablenotemark{a}} & \colhead{$z_{helio}$\tablenotemark{b}} & \colhead{Discovery Reference} & \colhead{Spectroscopic Reference}}
\startdata
SN2014dl & 16:29:46.09 & +08:38:30.6 & UGC 10414 & 0.03297 & CBET 3995 & CBET 3995\\
SN2014eg & 02:45:09.27 & -55:44:16.9 & ESO 154-G10 & 0.01863 & CBET 4062 & CBET 4062\\
PS15sv & 16:13:11.74 & +01:35:31.1 & ... & 0.033 & ATEL 7280 & ATEL 7308 \\
LSQ15aae & 16:30:15.70 & +05:55:58.7 & 2MASX J16301506+0555514\tablenotemark{c} & 0.0516\tablenotemark{c} & LSQ & ATEL 7325\\
SN2016gcl & 23:37:56.62 & +27:16:37.7 & AGC 331536 & 0.028 & TNSTR-2016-644 & TNSCR-2016-655 \\
SN2016hvl & 06:44:02.16 & +12:23:47.8 & UGC 3524 & 0.01308 & TNSTR-2016-884 & TNSCR-2016-892\\
SN2017awz & 11:07:35.50 & +22:51:04.7 & SDSS J110735.46+225104.2 & 0.02222 & TNSTR-2017-200 & TNSCR-2017-210 \\
SN2017dfb & 15:39:05.04 & +05:34:17.0 & ARK 481 & 0.02593 & TNSTR-2017-449 & TNSCR-2017-476 \\
SN2017glx & 19:43:40.30 & +56:06:36.4 & NGC 6824 & 0.01183 & TNSTR-2017-963 & TNSCR-2017-970 \\
SN2017hng & 04:21:40.58 & -03:32:26.5 & 2MASX J04214029-0332267 & 0.039 & TNSTR-2017-115 & TNSCR-2017-117 \\
SN2018apo & 12:45:05.30 & -44:00:23.0 & ESO 268- G 037 & 0.01625 & TNSTR-2018-440 & TNSCR-2018-468 \\
SN2018bie & 12:35:44.33 & -00:13:16.0 & ... & 0.022 & TNSTR-2018-609 & TNSCR-2018-626 \\
SN2018cnw & 16:59:05.06 & +47:14:11.0 & ... & 0.02416 & TNSTR-2018-832 & TNSCR-2018-833 \\
SN2019dks & 11:44:05.59 & -04:40:25.3 & ... & 0.057 & TNSTR-2019-566 & TNSCR-2019-601 \\
SN2019gwa & 15:58:41.18 & +11:14:25.1 & ... & 0.054988 & TNSTR-2019-930 & TNSCR-2019-965\\
SN2019vrq & 03:04:21.46 & -16:01:26.4 & GALEXASC J030421.67-160124.7 & 0.01308 & TNSTR-2019-247 & TNSCR-2019-248 
\enddata

\tablenotetext{a} {Basic information for each SN, including its J2000 right ascension and declination, and its host galaxy, were sourced from TNS or the discovery references.} 
\tablenotetext{b} {Host-galaxy heliocentric redshifts are from the NASA/IPAC Extragalactic (NED) or TNS Database unless otherwise indicated.}
\tablenotetext{c}{\citet{Phillips_2019}}
\label{tab:LCOGT_generalProp}
\end{deluxetable}

\begin{deluxetable}{ccccc}
\tablecaption{General Properties of 91T/99aa-like SNe Ia in the Literature.\label{tab:generalprop_91T_Literature}}
\tablehead{\colhead{SN Name} & \colhead{Host Galaxy} & \colhead{$z_{helio}$\tablenotemark{a}} & \colhead{Photometry Reference} & \colhead{Classification Reference}}
\startdata
SN1991T & NGC 4527 & 0.00579 & 1 & IAUC 5239 \\
SN1995bd & UGC 3151 & 0.0146 & 2 & IAUC 6278 \\
SN1998ab & NGC 4704 & 0.02715 & 4 & IAUC 7054 \\
SN1998es & NGC 632 & 0.01057 & 4, 5 & IAUC 7054 \\
SN1999aa & NGC 2595 & 0.01446 & 4, 5 & IAUC 7108 \\
SN1999ac & NGC 6063 & 0.0095 & 4, 5 & IAUC 7122 \\
SN1999aw & Anon 1101-06 & 0.0379\tablenotemark{6} & 8 & 7 \\
SN1999dq & NGC 976 & 0.01433 & 4, 5 & IAUC 7250 \\
SN1999gp & UGC 1993 & 0.02675 & 4, 5 & 7 \\
SN2001V & NGC 3987 & 0.01502 & 5, 6 & 7 \\
SN2001eh & UGC 1162 & 0.03704 & 5, 6 & 7 \\
SN2002hu & MCG +06-6-12 & 0.0374\tablenotemark{9} & 6 & 7 \\
SN2003fa & ARK 527 & 0.039\tablenotemark{5} & 5, 6 & 7 \\
SN2004br & NGC 4493 & 0.02308 & 5 & 10 \\
SN2005M & NGC 2930 & 0.022\tablenotemark{5} & 5, 6, 11 & IAUC 8474 \\
SN2005eq & MCG -01-9-6 & 0.02891 & 5, 6, 11 & 7 \\
SN2005hj & SDSS J012648.45-011417.3 & 0.05738 & 6, 11 & 7 \\
SN2007S & UGC 5378 & 0.01385 & 6, 11, 12 & CBET 839 \\
SN2007cq & 2MASX J22144070+0504435 & 0.02604 & 5, 6, 12 & 7 \\
SN2011hr & NGC 2691 & 0.01339 & 13 & CBET 2901 \\
SN2016hvl & UGC 3524 & 0.01308 & 14, 15, 16 & ATEL 9720 \\
iPTF14bdn & UGC 8503 & 0.01558 & 12, 17 & 17 \\
\enddata

\tablenotetext{a} {Heliocentric redshift are taken from the NASA/IPAC Extragalactic Database (NED) unless indicated otherwise.} 
\tablecomments {(1) \citet{Lira_1998}; (2) \citet{Riess_1999}; (3) \citet{Altavilla_etal_2004}; (4) \citet{Jha_2006}; (5) \citet{Ganeshalingam_2010}; (6) \citet{Hicken_2009}; (7) \citet{Blondin_etal_2012}; (8) \citet{Strolger_2002}; (9) \cite{Blondin_etal_2012}; (10) \citet{Silverman_etal_2012}; (11) \citet{Krisciunas_2017}; (12) \citet{Brown_etal_2014_SOUSA}; (13) \citet{Zhang2016}; (14) \citet{Stahl_2019}; (15) \citet{Foley_etal_2018_foundation}; (16) \citet{Baltay_2021}; (17) \citet{Smitka_etal_2015}.}
\end{deluxetable}

Here we adopt a loose classification scheme of 91T/99aa-like SNe, where an SN~Ia is classified as 91T/99aa-like as long as it is reported as a 91T/99aa-like object from at least one source (listed in \autoref{tab:generalprop_LCO} and \autoref{tab:generalprop_91T_Literature}) in the literature. As will be shown later in this paper, the definition of 91T/99aa-like SNe can be replaced by a more quantitative scheme and the results of our analysis will not be affected by this initial classification. 

The general properties of our 16 Las Cumbres Observatory 91T/99aa-like SNe are given in \autoref{tab:generalprop_LCO}. The Las Cumbres Observatory operates a fleet of 1-m class telescopes distributed around the globe. Las Cumbres Observatory images were pre-processed by the \texttt{BANZAI} pipeline \citep{McCully_2018} (after June 2016) and \texttt{ORAC} pipeline (before June 2016), during which bad pixel masking, bias subtraction, dark subtraction, and flat-field correction are done.
Most SNe in the sample are located in the vicinity of their host galaxy nuclei. Without template subtraction, the galaxy background with nonuniform brightness can significantly degrade the photometric accuracy. To remove the galaxy background contamination, we subtract a reference image in the same passband for each exposure of the SN. Since the pre-explosion observations are generally unavailable for Las Cumbres Observatory telescopes, we adopt the observations taken after the SNe have sufficiently faded (typically more than one year after explosion) to create the reference image. More specifically, the late-time observations are coadded and subsequently aligned to the science frame using \texttt{SWarp} \citep{Bertin_2002}. We use the Saccadic Fast Fourier Transform algorithm \citep[\texttt{SFFT} hereafter,][]{Hu_2022} to perform the galaxy subtractions. The SFFT method is a novel approach that presents the question of image subtraction in the Fourier domain, which brings a significant computational speed-up by leveraging GPU acceleration. Unlike the widely used \texttt{HOTPANTS} image subtraction program, SFFT uses a delta function basis which allows for kernel flexibility and requires minimal user-adjustable parameters. 
In our work, \texttt{SFFT} successfully carries out the galaxy subtraction for all images with the default software configuration (see SFFT GitHub\footnote{\url{https://github.com/thomasvrussell/sfft}}). By contrast, \citet{Baltay_2021} found that \texttt{HOTPANTS} produced satisfactory subtractions only for two thirds supernovae in their sample observed by Las Cumbres Observatory in their work.

For precision photometry, a small stamp centered at the SN position (typically 200 pixels $\times$ 200 pixels) is cut out from the difference image and pasted back to the original science image. The cutout stamp is added with a constant background to match the background level of the science image. Point spread function (PSF) photometry is then done on the new set of images with pasted stamps using \texttt{DAOPHOT} \citep{Stetson_1987}. We first run \texttt{ALLSTAR} on each frame separately to generate the entire star list for each frame. Then the instrumental magnitudes for images taken with the same filter for the same object are solved simultaneously using \texttt{ALLFRAME}. To do so, the coordinate transformations between the images are found by \texttt{DAOMATCH} and refined by \texttt{DAOMASTER}. Then \texttt{ALLFRAME} performs simultaneous profile fits for all stars in given images while maintaining a consistent star list and their positions for all images. Rather than solving for an independent position for each star in each image like in \texttt{ALLSTAR}, \texttt{ALLFRAME} solves for a single position for each star, and transforms that to the coordinate system of other images using the transformations found by \texttt{DAOMASTER}. This way the star's centroid, especially the centroid of the SN of interest when the SN becomes dimmer, can be better determined and results in higher photometric precision. Without a common position, the fainter photometry is biased brighter as the centrioding algorithm will tend toward peaks due to low signal to noise in the PSF location.
\begin{deluxetable}{cccccc}
\caption{\edithere{Las Cumbres} 91T/99aa-like SNe photometry in natural systems.\label{tab:natural_photometry_head}}
\tablehead{\colhead{SN} & \colhead{Filter} & \colhead{MJD} & \colhead{Magnitude} & \colhead{Magnitude Error} & \colhead{Site, Telescope and Instrument}}
\startdata
SN2014dl & B & 56929.3814 & 16.473 & 0.053 & coj1m003-kb71 \\
SN2014dl & B & 56929.3839 & 16.473 & 0.058 & coj1m003-kb71 \\
SN2014dl & B & 56932.3744 & 16.303 & 0.057 & coj1m011-kb05 \\
SN2014dl & B & 56932.3769 & 16.292 & 0.056 & coj1m011-kb05 \\
SN2014dl & B & 56935.0704 & 16.272 & 0.044 & elp1m008-kb74 \\
SN2014dl & B & 56937.0705 & 16.299 & 0.054 & elp1m008-kb74 \\
SN2014dl & B & 56937.073 & 16.309 & 0.053 & elp1m008-kb74 \\
SN2014dl & B & 56940.0746 & 16.432 & 0.045 & elp1m008-kb74 \\
SN2014dl & B & 56940.0771 & 16.436 & 0.043 & elp1m008-kb74 \\
SN2014dl & B & 56942.0803 & 16.544 & 0.05 & elp1m008-kb74
\enddata

\end{deluxetable}

\begin{deluxetable}{ccccc}
\caption{\edithere{Las Cumbres} 91T/99aa-like SNe photometry in standard systems.\label{tab:standard_phometry_head}}
\tablehead{\colhead{SN} & \colhead{Filter} & \colhead{MJD} & \colhead{Magnitude} & \colhead{Magnitude Error}}
\startdata
SN2014dl & B & 56929.3814 & 16.458 & 0.057 \\
SN2014dl & B & 56929.3839 & 16.458 & 0.061 \\
SN2014dl & B & 56932.3744 & 16.287 & 0.06 \\
SN2014dl & B & 56932.3769 & 16.276 & 0.059 \\
SN2014dl & B & 56935.0704 & 16.255 & 0.048 \\
SN2014dl & B & 56937.0705 & 16.281 & 0.058 \\
SN2014dl & B & 56937.073 & 16.291 & 0.057 \\
SN2014dl & B & 56940.0746 & 16.413 & 0.049 \\
SN2014dl & B & 56940.0771 & 16.418 & 0.047 \\
SN2014dl & B & 56942.0803 & 16.527 & 0.055
\enddata
\end{deluxetable}

\begin{deluxetable}{cccccc}
\caption{\edithere{Las Cumbres} 91T/99aa-like SNe photometry in $UR$ bands.\label{tab:LCOGT_UR_photometry}}
\tablehead{\colhead{SN} & \colhead{Filter} & \colhead{MJD} & \colhead{Magnitude} & \colhead{Magnitude Error} & \colhead{Site, Telescope and Instrument}}
\startdata
SN2018cnw & U & 58287.365 & 17.436 & 0.074 & elp1m008-fl05 \\
SN2018cnw & U & 58287.3712 & 17.447 & 0.061 & elp1m008-fl05 \\
SN2018cnw & U & 58289.3094 & 16.48 & 0.08 & elp1m008-fl05 \\
SN2018cnw & U & 58289.3134 & 16.479 & 0.058 & elp1m008-fl05 \\
SN2018cnw & U & 58290.3227 & 16.137 & 0.085 & elp1m008-fl05 \\
SN2018cnw & U & 58290.3267 & 16.145 & 0.086 & elp1m008-fl05 \\
SN2018cnw & U & 58292.3054 & 15.707 & 0.073 & elp1m008-fl05 \\
SN2018cnw & U & 58292.3094 & 15.706 & 0.083 & elp1m008-fl05 \\
SN2018cnw & U & 58293.157 & 15.588 & 0.111 & elp1m008-fl05 \\
SN2018cnw & U & 58293.161 & 15.642 & 0.059 & elp1m008-fl05
\enddata

\end{deluxetable}

Photometric measurements are calibrated by means of field stars in the APASS catalog DR9 \citep{Henden_2016} with catalog magnitudes in the range 10 mag $<m<$18 mag and uncertainties $\Delta m \le 0.1$ mag in each calibration band. The APASS catalog calibration was done using Landolt and SDSS standard stars. Therefore, the Las Cumbres Observatory magnitudes presented here are based on the Vega standard system for $B$ and $V$, and the Sloan BD17 system for $g'r'i'$ commonly called AB magnitudes. Photometric coefficients appropriate to the Las Cumbres Observatory 1-m telescopes \citep[see Table B1 in][]{Valenti_etal_2016} were used to convert APASS catalog magnitudes to values in the Las Cumbres Observatory natural system. We use the color terms to convert the APASS magnitude to Las Cumbres Observatory natural system magnitude as follows:

\begin{equation}
m_{nat} \; = \; m_{APASS} \; + \; C\times color \; ,
\end{equation}
\parindent = 0 mm

where $m_{nat}$ is the natural system magnitude, $m_{APASS}$ is the APASS magnitude, and C and ``color'' are the appropriate color term and color for each filter\footnote{APASS DR9 does not contain any $U$ or $R$ band magnitudes. We have $U$ and $R$ data for three objects. For completeness, here we use the transformation given by \cite{Jordi_2006_transform} to derive $U$ and $R$ magnitudes of reference stars from filters with known magnitudes. Also, note that \cite{Valenti_etal_2016} did not provide color terms for $U$ and $R$. Thus we did not convert the magnitudes of reference stars to the natural system.}. Zero points for each exposure were then determined by the difference between instrumental magnitudes of field stars and the corresponding APASS magnitudes converted to the Las Cumbres Observatory natural system. After that, the supernovae natural system photometry is brought onto standard systems using S-corrections (see \autoref{s-correction}). Final photometry data (both in natural system and in standard system) are given in \autoref{tab:natural_photometry_head} and \autoref{tab:standard_phometry_head}. Light curves are shown in \autoref{fig: LCOGT_lc} together with light curve fits (see \autoref{method: FPCA}). Photometry in U and R filters without color term corrections and S-corrections are also provided in \autoref{tab:LCOGT_UR_photometry} for completeness.

\parindent = 9 mm

\newcommand{\widthlc}{0.246}
\begin{figure}
\gridline{\fig{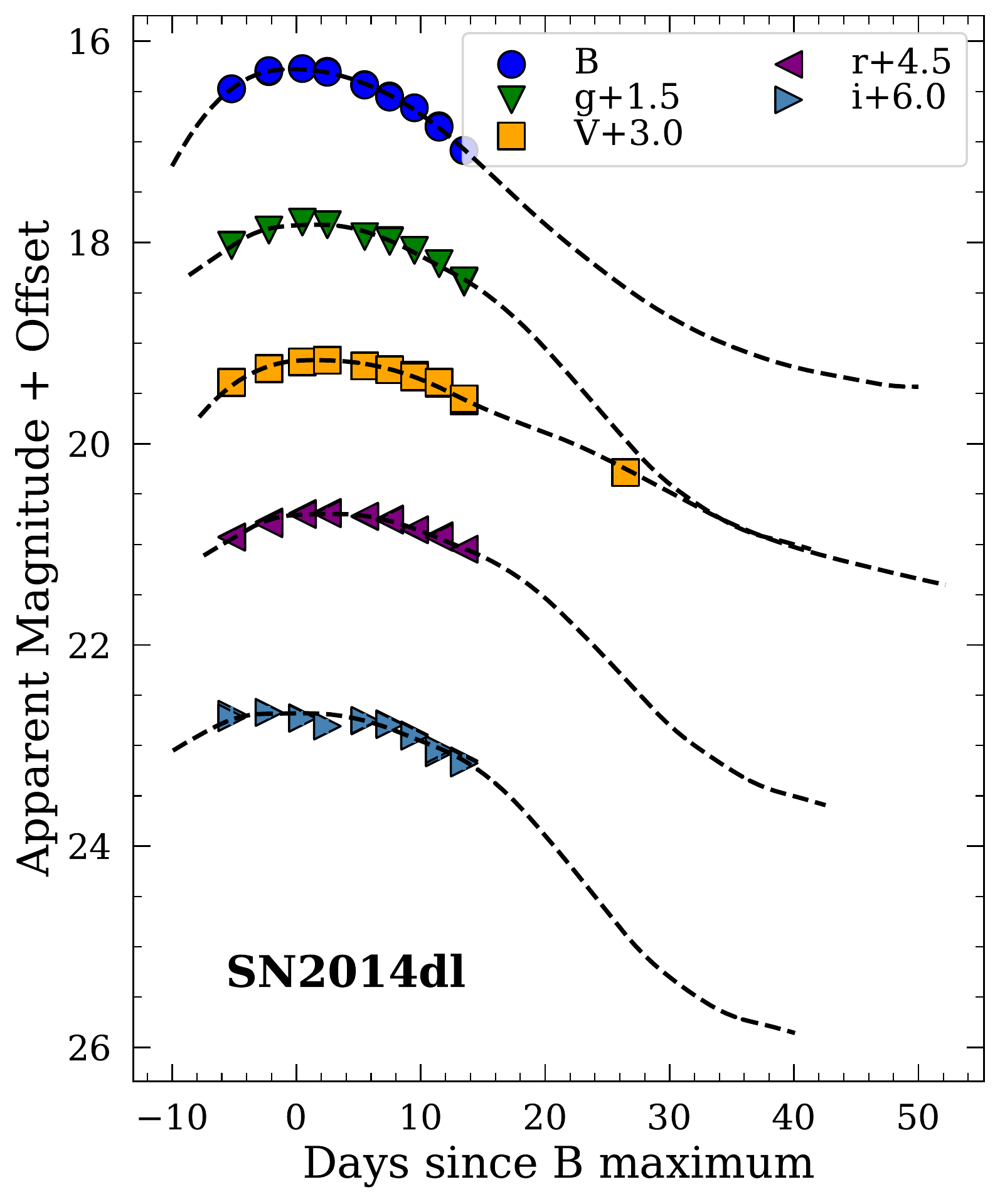}{\widthlc\textwidth}{}
          \fig{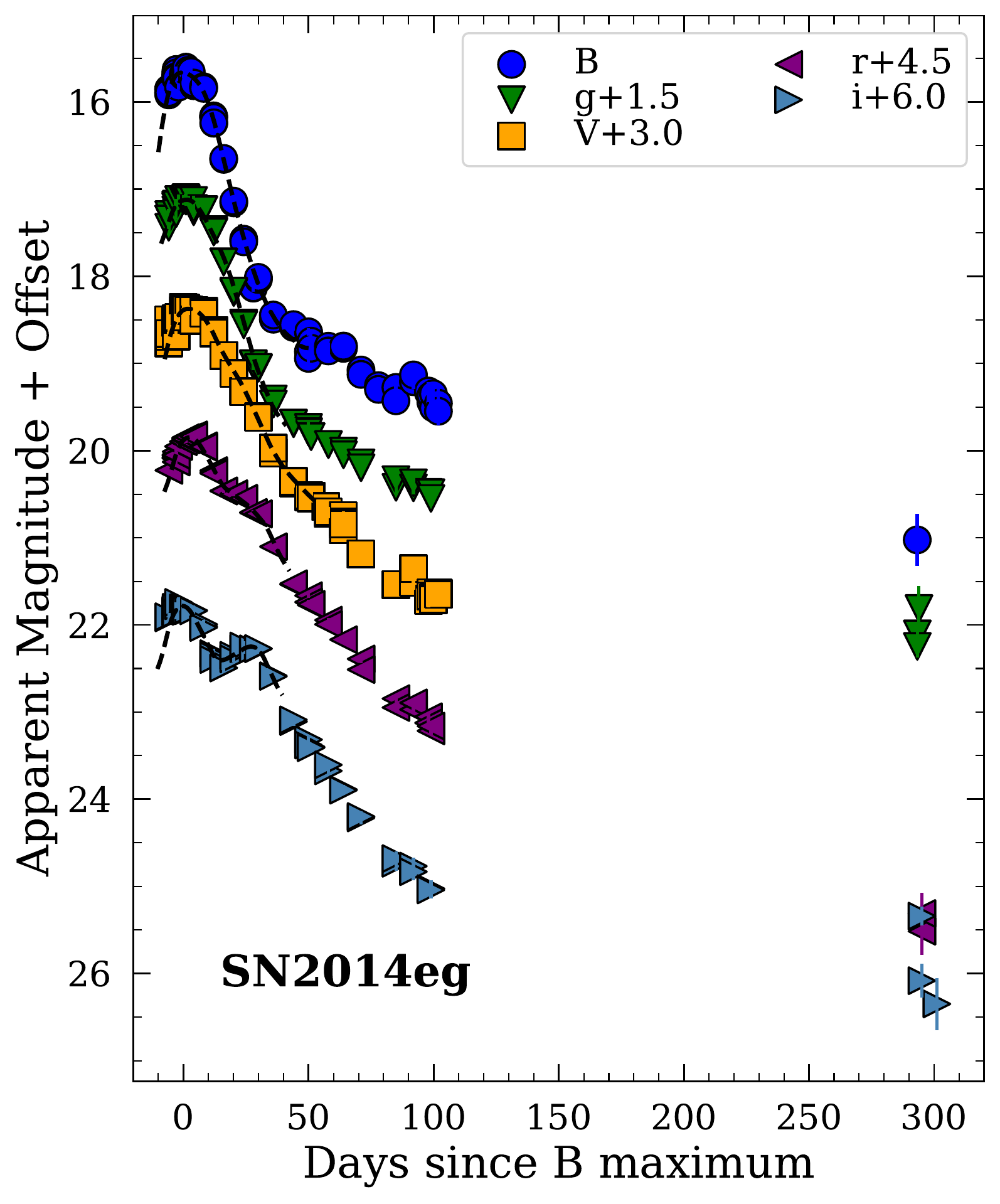}{\widthlc\textwidth}{}
          \fig{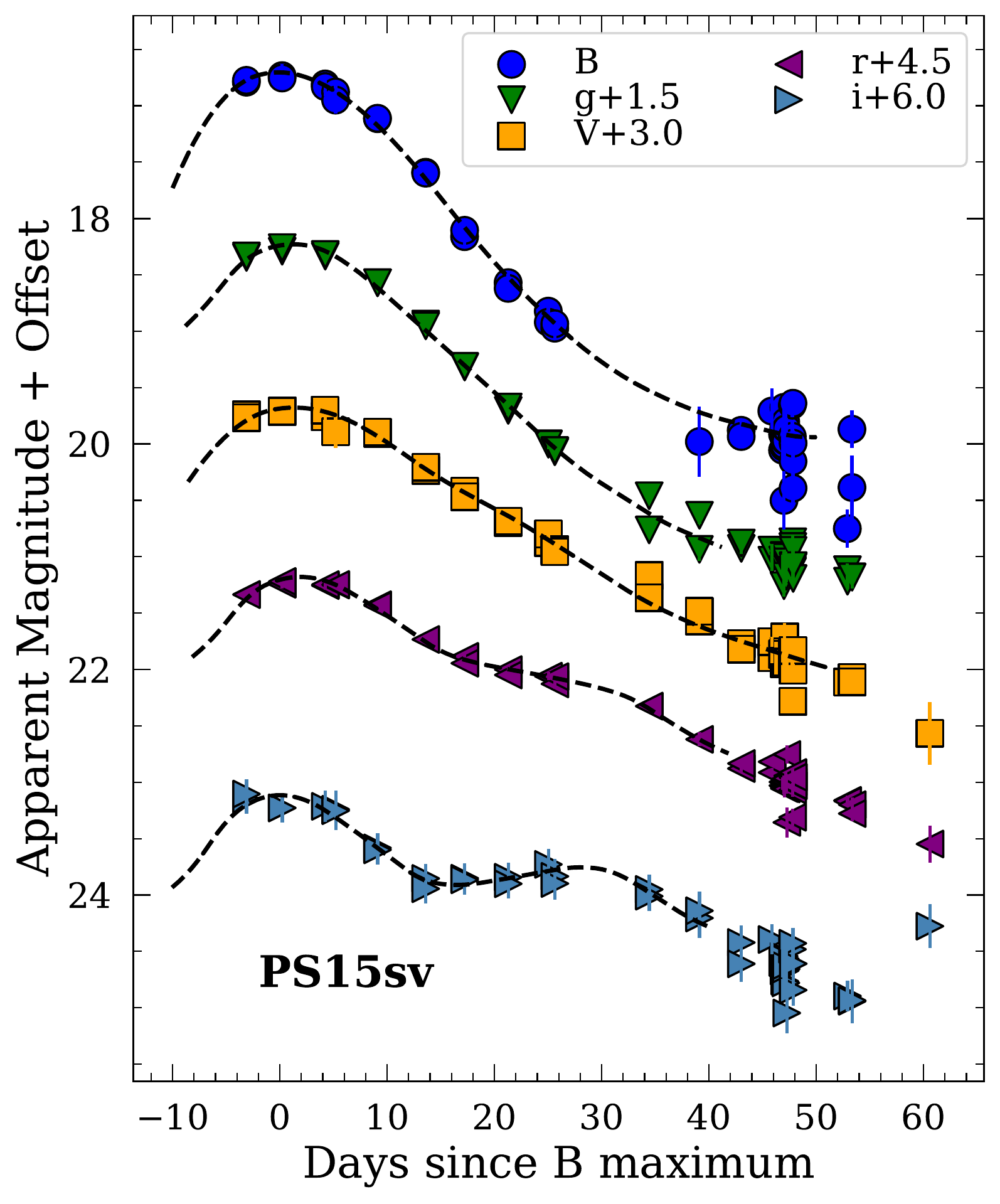}{\widthlc\textwidth}{}
          \fig{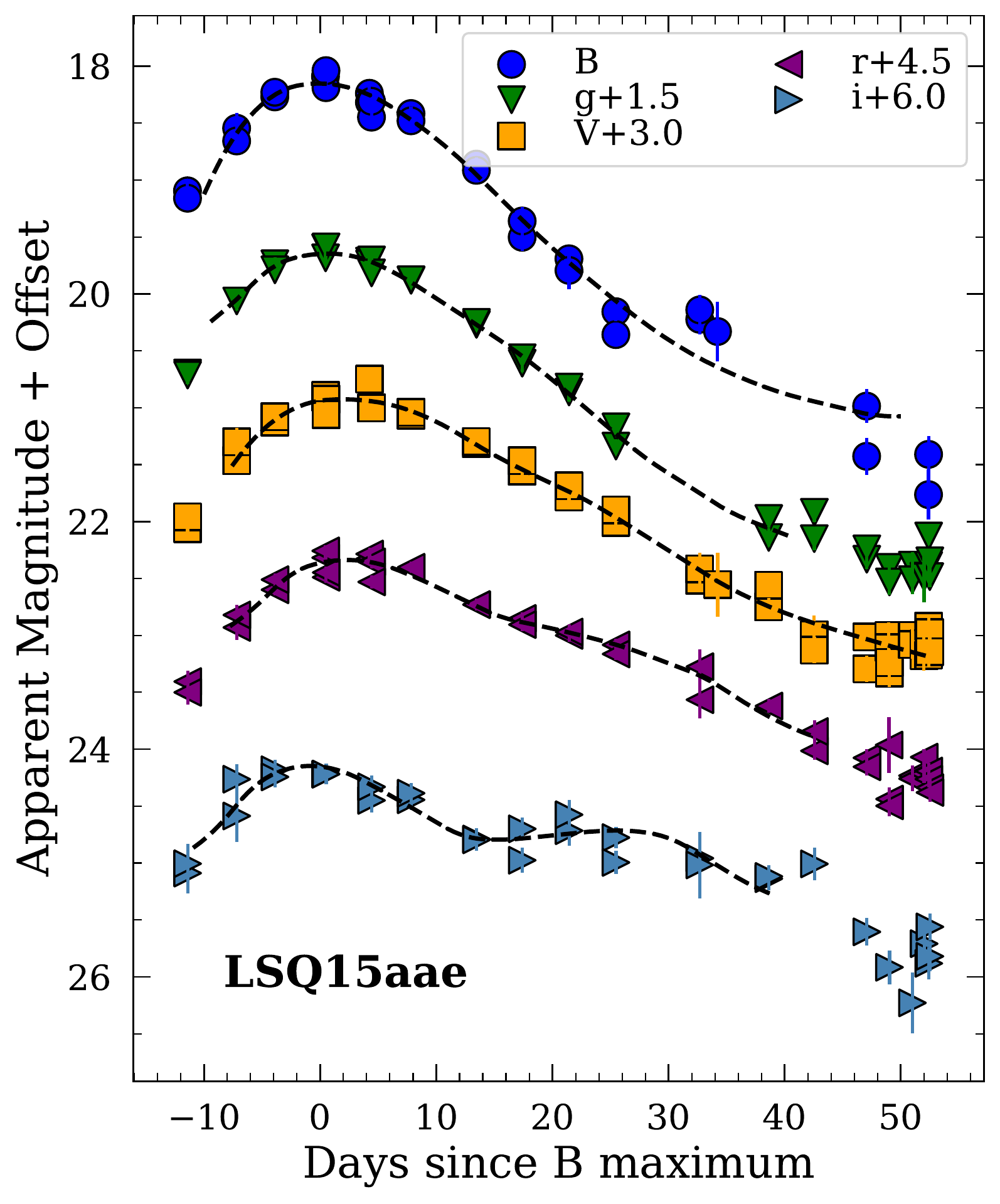}{\widthlc\textwidth}{}
          }
          \vspace{-0.8cm}
\gridline{
          \fig{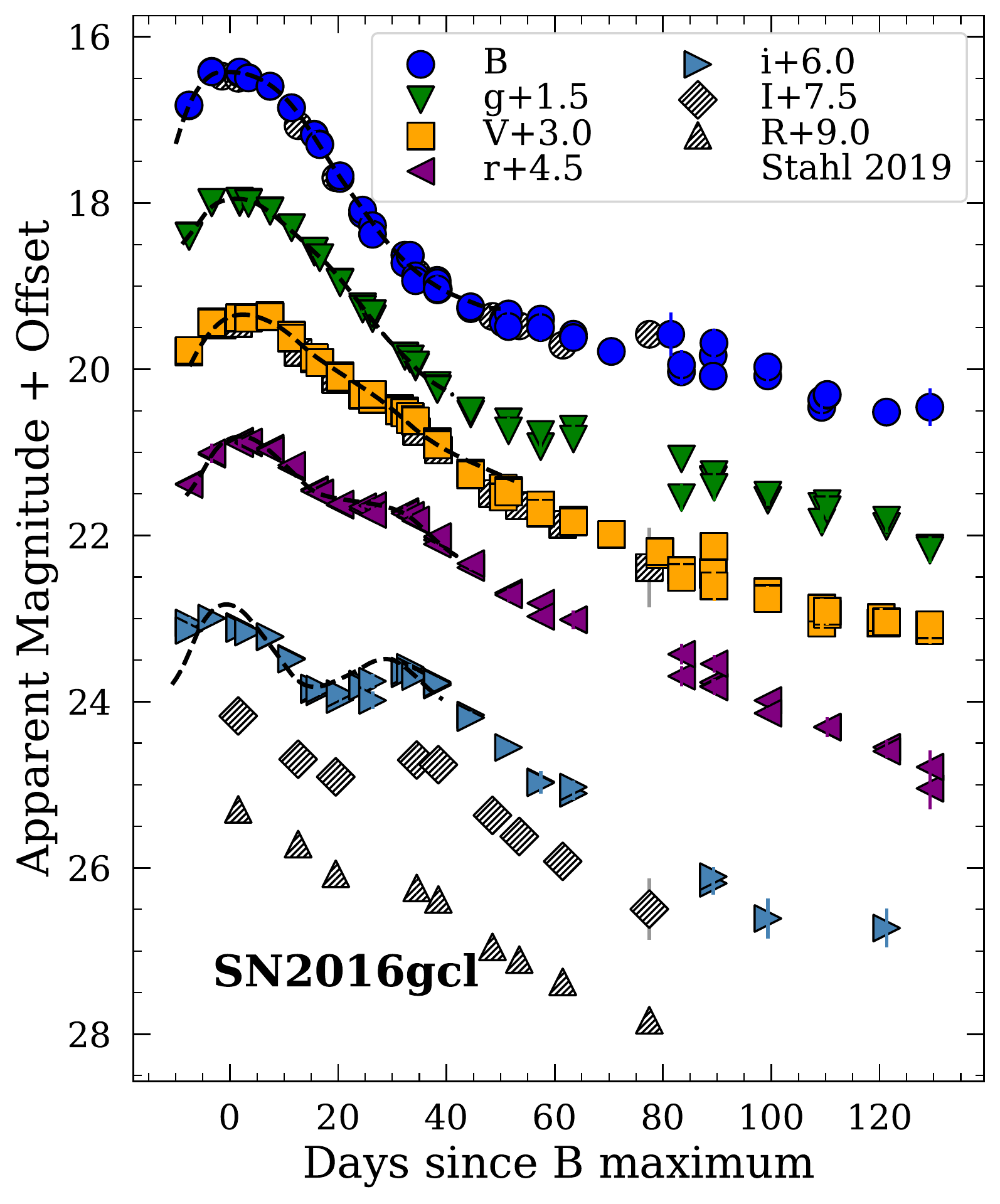}{\widthlc\textwidth}{}
          \fig{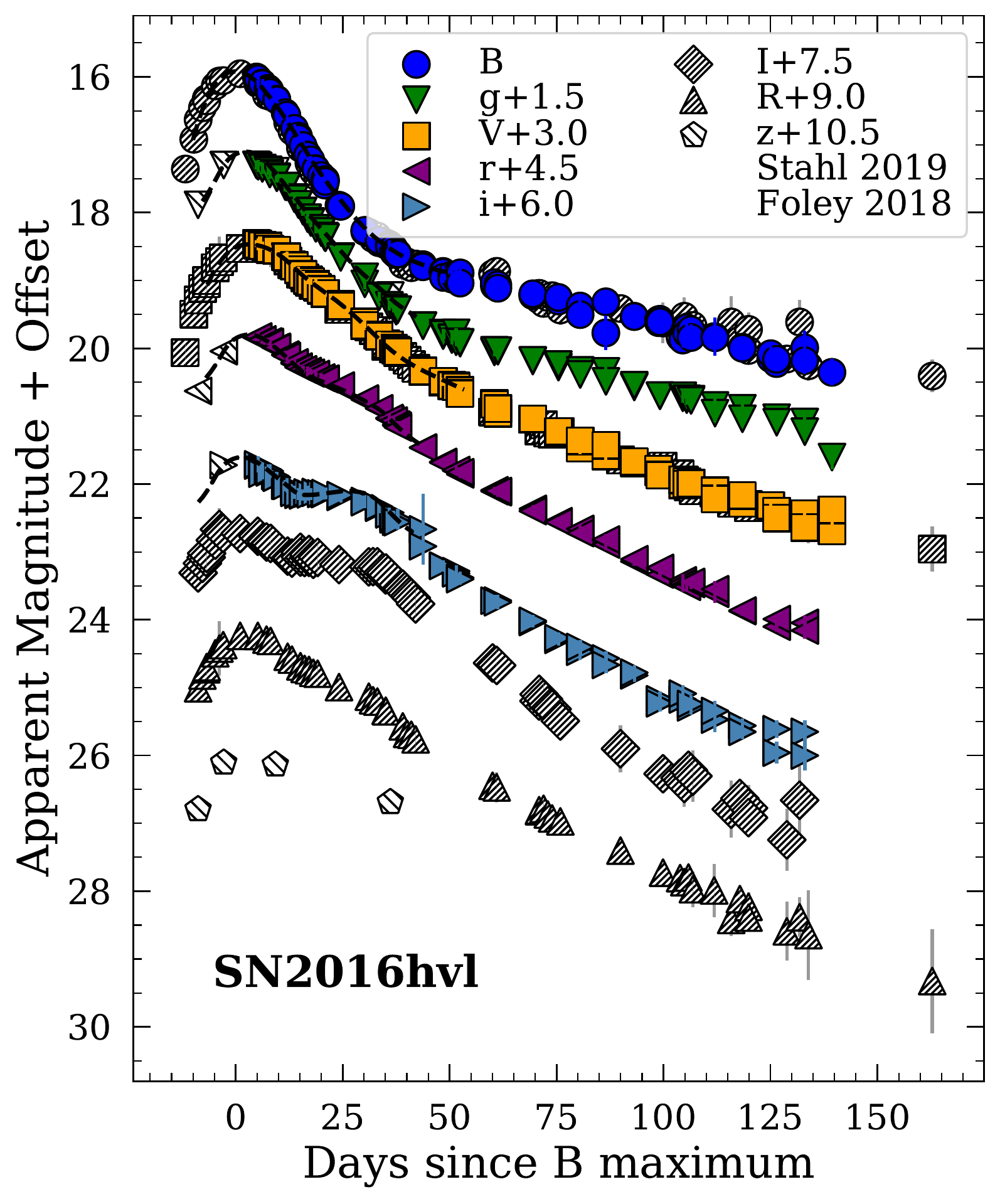}{\widthlc\textwidth}{}
          \fig{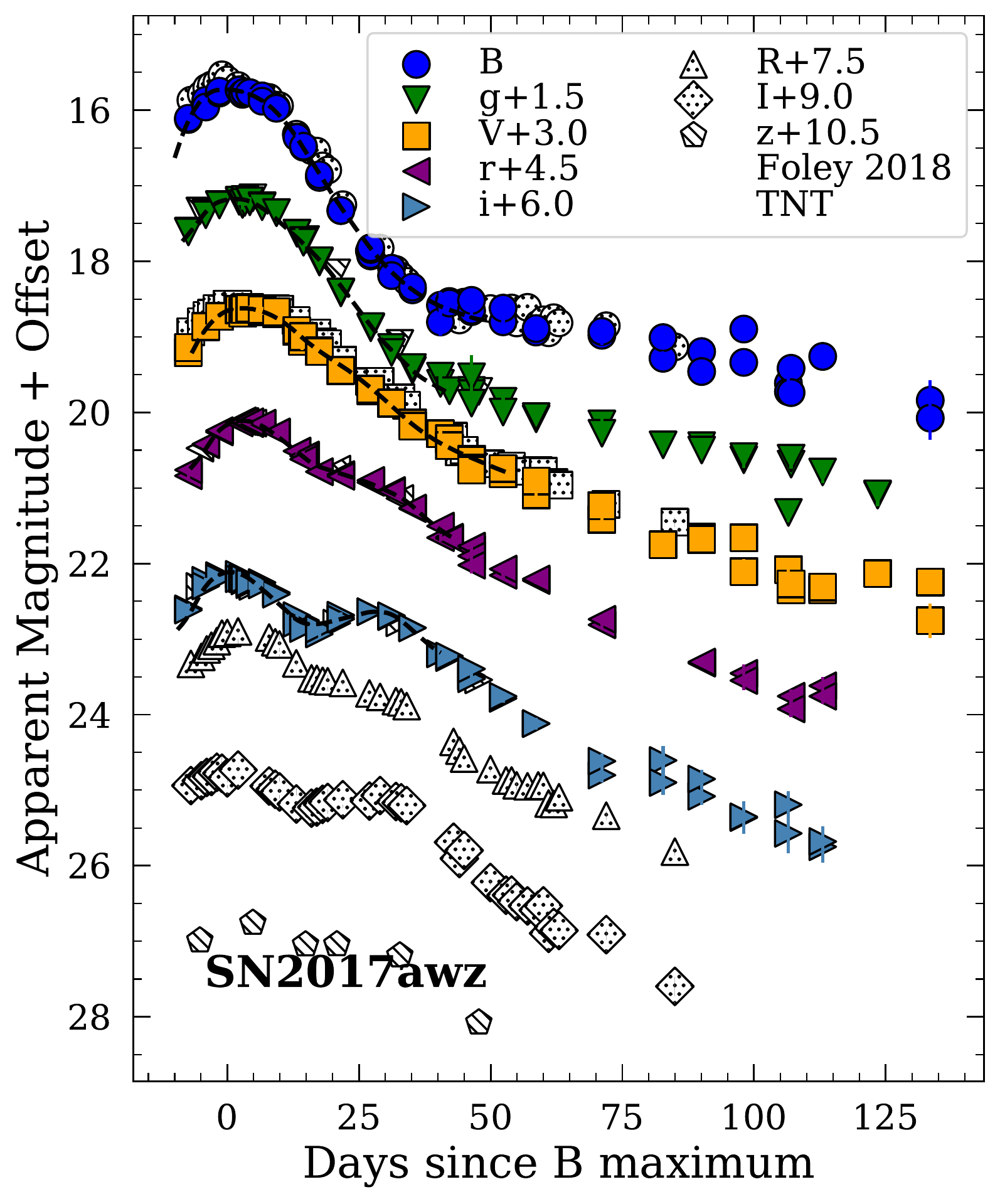}{\widthlc\textwidth}{}
          \fig{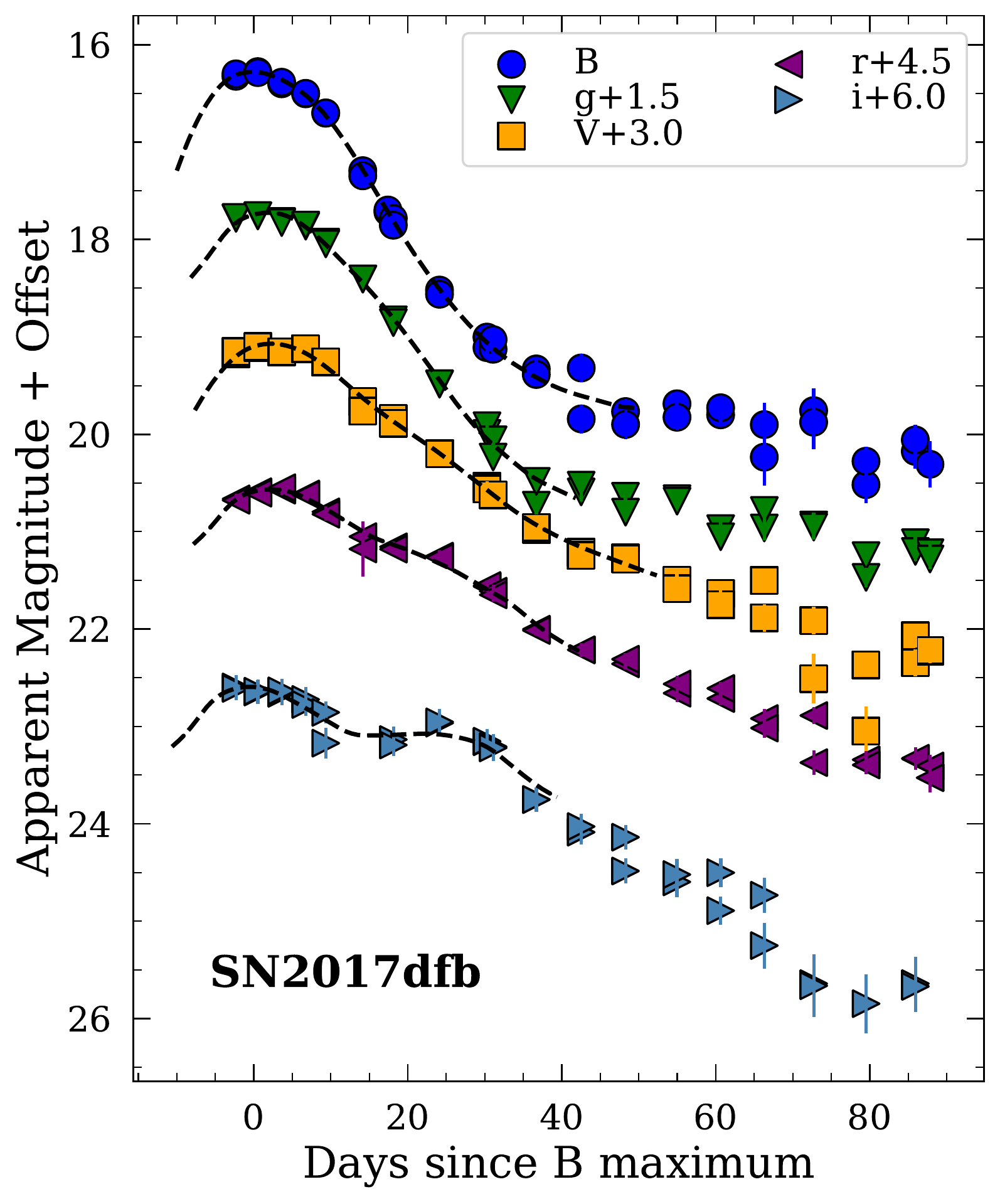}{\widthlc\textwidth}{}
          }
          \vspace{-0.8cm}
\gridline{
          \fig{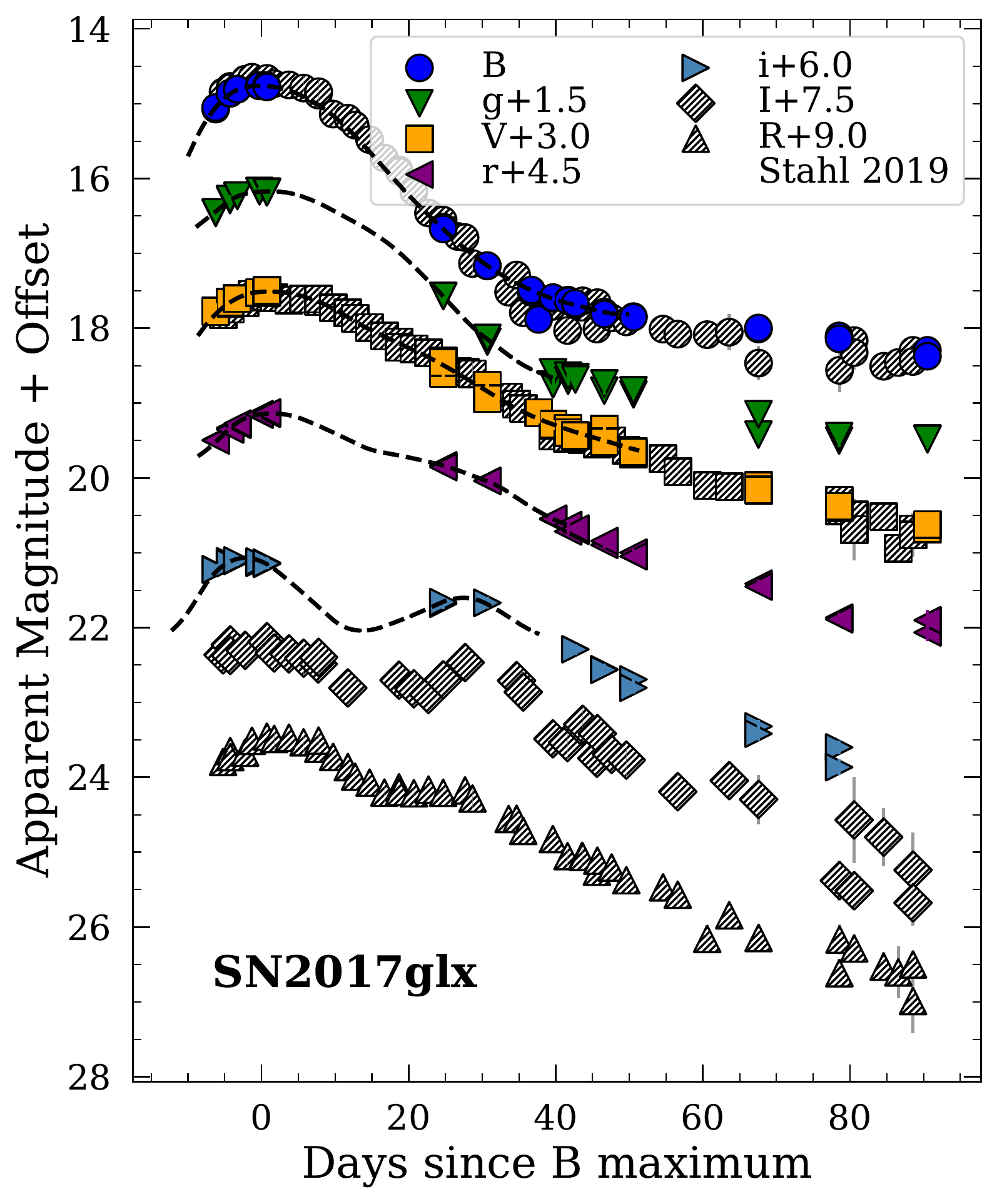}{\widthlc\textwidth}{}
          \fig{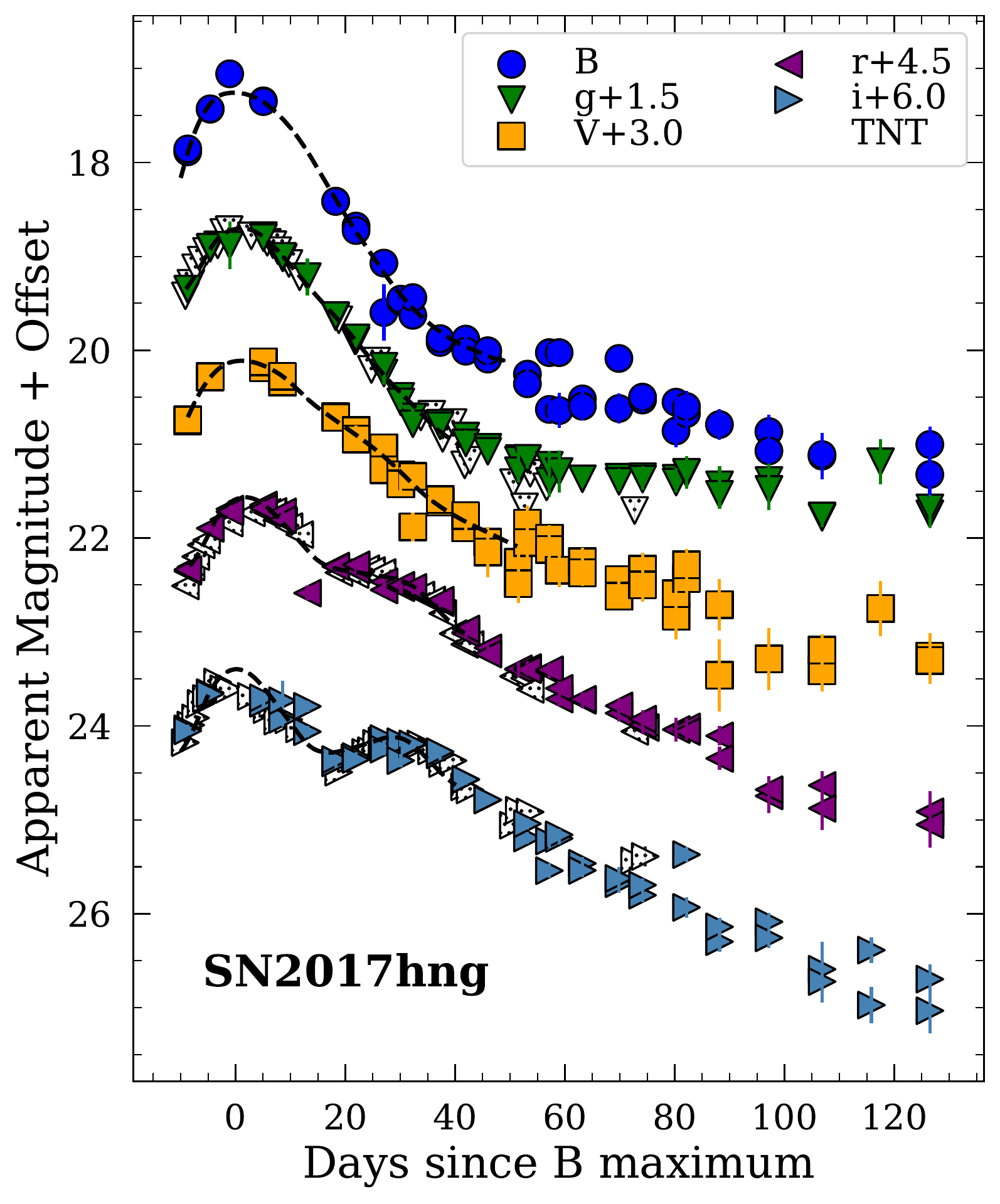}{\widthlc\textwidth}{}
          \fig{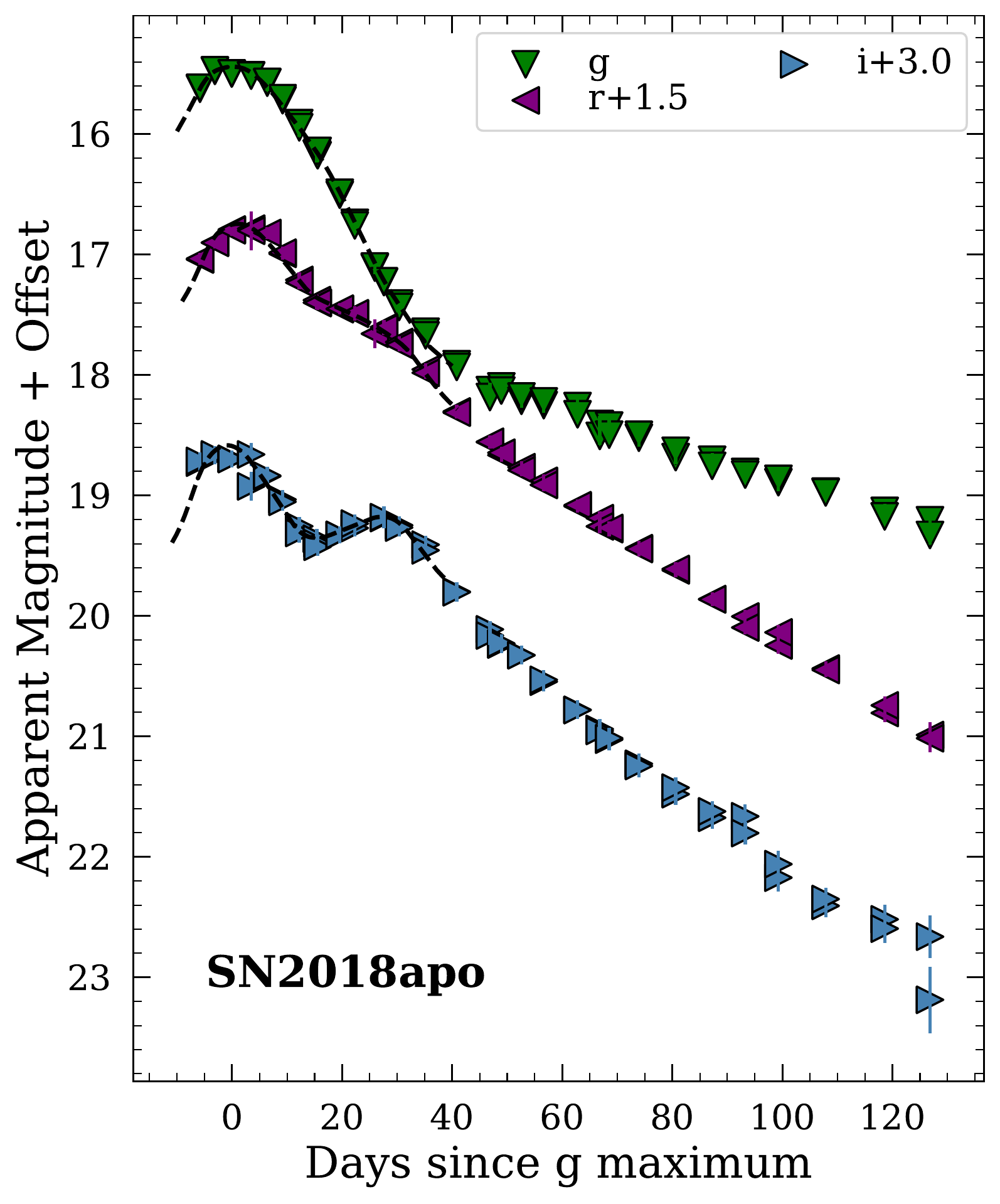}{\widthlc\textwidth}{}
          \fig{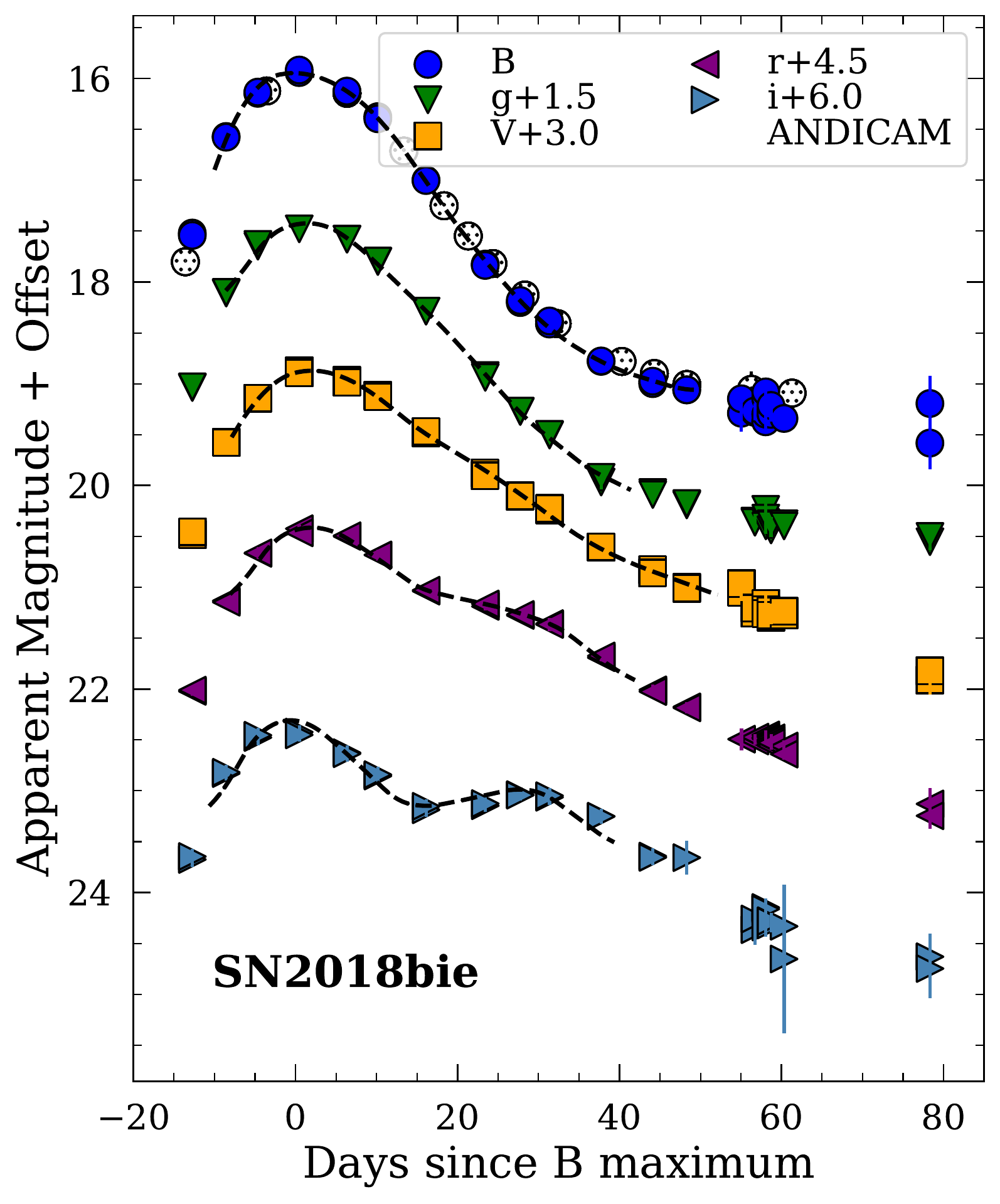}{\widthlc\textwidth}{}
          }
          \vspace{-0.8cm}
\gridline{
          \fig{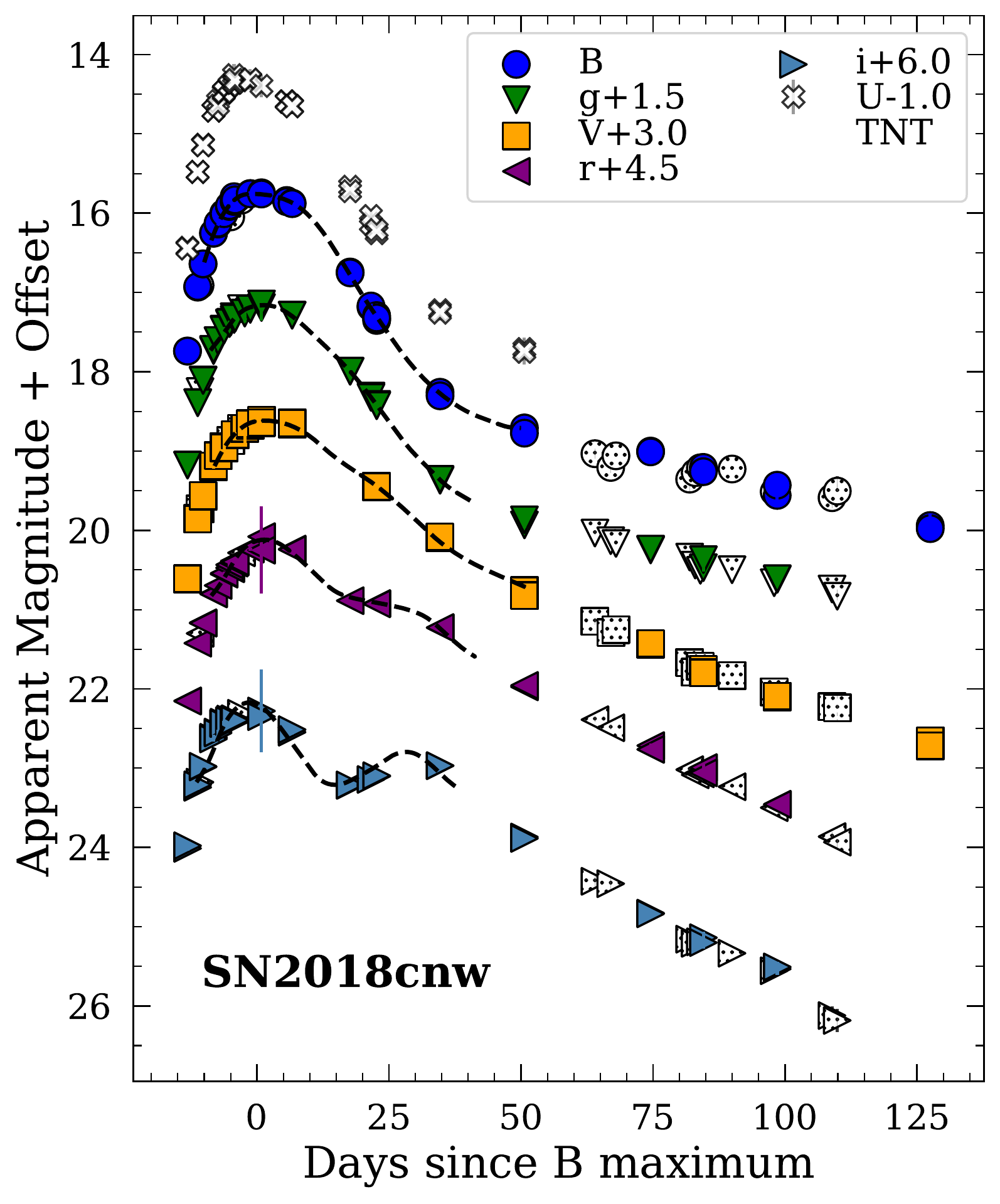}{\widthlc\textwidth}{}
          \fig{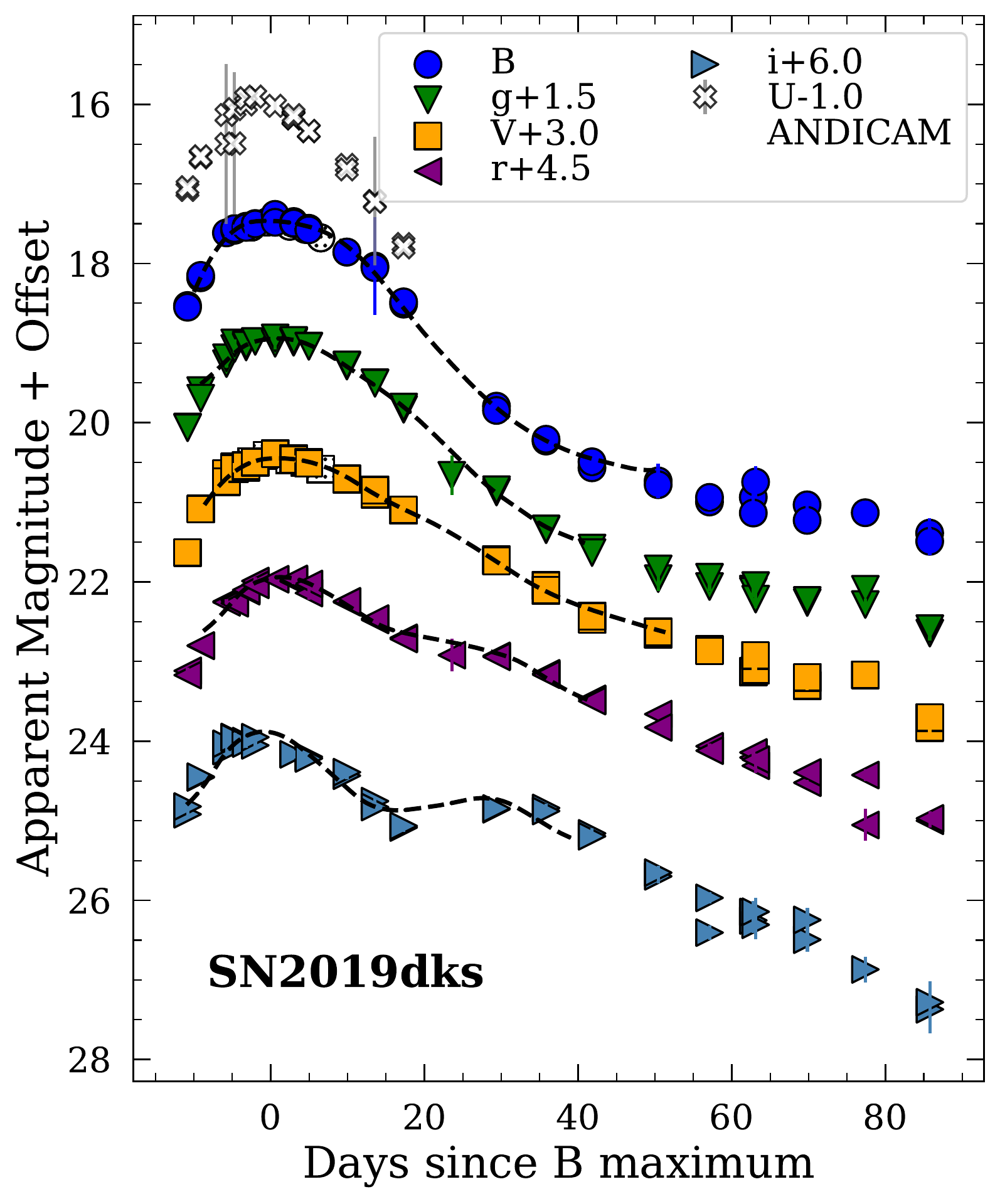}{\widthlc\textwidth}{}
          \fig{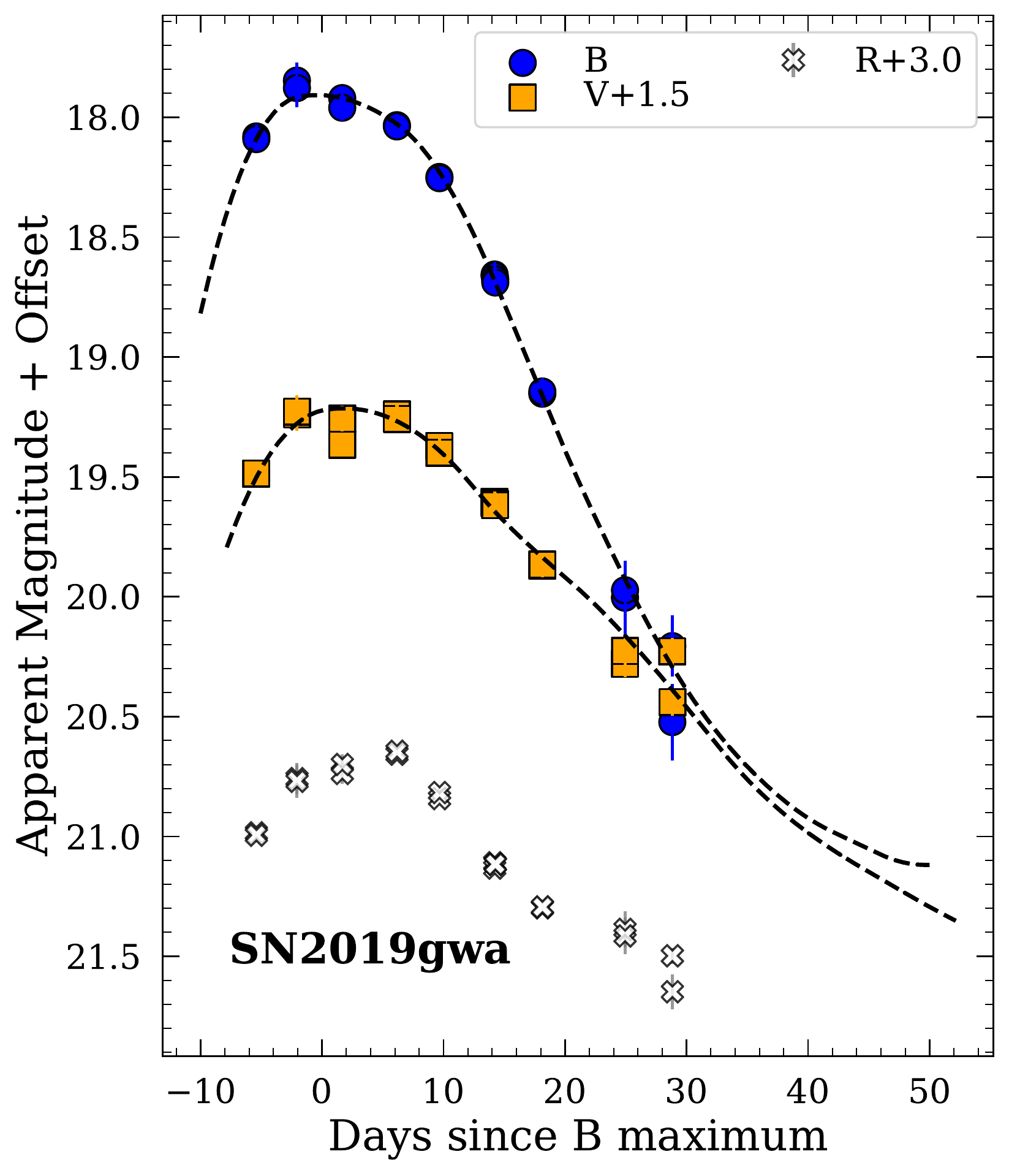}{\widthlc\textwidth}{}
          \fig{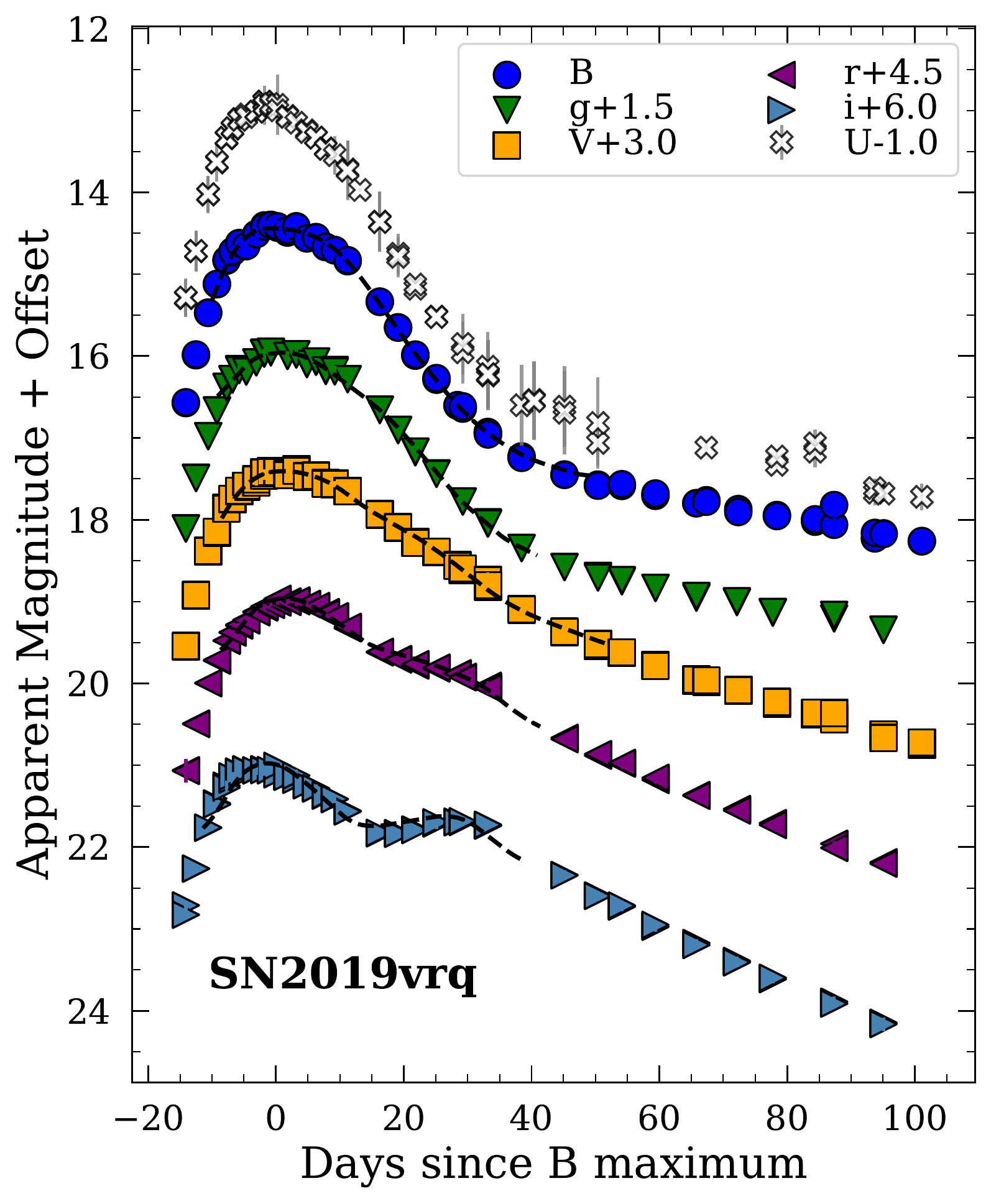}{\widthlc\textwidth}{}
          }
\caption{Light curves of Las Cumbres Observatory 91T/99aa-like supernovae in natural systems. Light curve fits for $BVg'r'i'$ bands using FPCA \citep{He_2018} are shown as dashed lines overlayed on top of their corresponding data points. For completeness, we also plot $U$ and $R$ photometry when available, though for those two filters, we did not correct for color terms. For SNe with data from non-Las Cumbres observing programs, they are included in the figure for comparison, with the sources of the data shown in the insert legends at the upper-right corner of each panel. The relevant references are: Stahl 2019: \citet{Stahl_2019}, Foley 2018: \citet{Foley_etal_2018_foundation}, and TNT: Tsinghua-NAOC Telescope.}

\label{fig: LCOGT_lc}
\end{figure}

\subsubsection{ANDICAM Data}
In \autoref{tab:Andicam_photometry} we present complementary $BV$ photometry for two objects, SN2019dks and SN2018bie, obtained under programmes NOAO-18A-0047, NOAO-18B-0016, and NOAO-19A-0081 with the {\it A Novel Double Imaging Camera} \citep[ANDICAM;][]{2003SPIE.4841..827D} instrument mounted on the 1.3m telescope and operated by the SMARTS Consortium at the Cerro Tololo Inter-American Observatory (CTIO). Images were reduced by the SMARTS consortium using standard routines, and were retrieved from their ftp server. Photometry on the reduced images were performed following standard procedures. Aperture photometry was performed on the SN and field stars, and then the instrumental magnitudes of the stars were calibrated to the APASS catalogue; once the magnitude conversion was found, we applied ANDICAM color terms and applied an airmass correction to find the SN magnitude in each image.
\begin{deluxetable}{ccccc}
\caption{ANDICAM SNe photometry in standard systems.\label{tab:Andicam_photometry}}
\tablehead{\colhead{SN} & \colhead{MJD} & \colhead{Filter} & \colhead{Magnitude} & \colhead{Magnitude Error}}
\startdata
SN2018bie & 58251.0668 & B & 17.802 & 0.067 \\
SN2018bie & 58261.1015 & B & 16.123 & 0.026 \\
SN2018bie & 58271.0232 & B & 16.146 & 0.062 \\
SN2018bie & 58278.0392 & B & 16.713 & 0.029 \\
SN2018bie & 58283.004 & B & 17.251 & 0.053 \\
SN2018bie & 58285.9925 & B & 17.548 & 0.066 \\
SN2018bie & 58289.0281 & B & 17.819 & 0.034 \\
SN2018bie & 58292.985 & B & 18.129 & 0.08 \\
SN2018bie & 58296.9623 & B & 18.408 & 0.035 \\
SN2018bie & 58304.9651 & B & 18.781 & 0.052 \\
SN2018bie & 58308.9624 & B & 18.9 & 0.1 \\
SN2018bie & 58312.9721 & B & 19.007 & 0.082 \\
SN2018bie & 58320.954 & B & 19.055 & 0.175 \\
SN2018bie & 58325.9612 & B & 19.092 & 0.118 \\
SN2019dks & 58595.1002 & B & 17.543 & 0.046 \\
SN2019dks & 58597.0843 & B & 17.48 & 0.054 \\
SN2019dks & 58600.05 & B & 17.532 & 0.056 \\
SN2019dks & 58602.0576 & B & 17.57 & 0.048 \\
SN2019dks & 58604.1016 & B & 17.679 & 0.053 \\
SN2019dks & 58595.1097 & V & 17.49 & 0.049 \\
SN2019dks & 58597.0938 & V & 17.411 & 0.034 \\
SN2019dks & 58600.0595 & V & 17.466 & 0.042 \\
SN2019dks & 58602.0681 & V & 17.498 & 0.053 \\
SN2019dks & 58604.1111 & V & 17.583 & 0.05
\enddata
\end{deluxetable}

\subsubsection{Data from the Literature}
To supplement our 91T/99aa-like SNe sample we gathered all SNe~Ia that are classified as 91T/99aa-like\footnote{SN2013dh is excluded, as it is a possible Iax supernova (see ATEL 5143). SN1997br is also excluded due to bad light curve fits.} from at least one source and have at least one $B$-band data point in the Open Supernova Catalogue \citep{Guillochon_etal_2017}. Then we make the quality cut following these criteria: 1) $B$ and $V$ data must be present; 2) at least one band of $i$, $R$, and $I$ must have data (in order to calculate K-corrections in the $V$-band (see \autoref{k_corrections})); 3) for the above three bands, there must be at least one data point after 15 days post maximum (to ensure a good light curve fit); and 4) for the above three bands, there must be at least six data points observed in each band. After making the cuts, we are left with 21 91T/99aa-like objects. Their general properties are listed in \autoref{tab:generalprop_91T_Literature}.

To compare 91T/99aa-like SNe with normal SNe~Ia, we include normal SNe~Ia from \citet{Jha_2006, Hicken_2009, Hicken_2012, Ganeshalingam_2010} and \citet{Stahl_2019}. All above papers provide photometry data in standard system, so no further S-correction is needed (see \autoref{s-correction}). Data are further selected according to the same criteria in the above paragraph, with extra criteria: for each of the three bands mentioned above, there must be at least two data points before maximum, and we tighten the constraint on the minimum total number of data points in each band from six to eight. After applying these requirements, we are left with 87 normal SNe~Ia.

\subsection{S-corrections}
\label{s-correction}
Photometric systems of different telescope and instrument combinations never match exactly. Since we try to compare photometry of supernovae taken by different instruments, each one of which defines its own photometric system, it is necessary to transform all of the data to a common standard system. Although the color equations are meant to transform between natural system magnitudes and standard system magnitudes, they lose accuracy when transforming natural system magnitudes of supernovae onto the standard system, as color terms are derived from stellar objects while the spectral energy distributions (SEDs) of supernovae are quite different from stellar spectra. This problem is mitigated through the use of S-corrections \citep{Stritzinger_etal_2002, Krisciunas_2003_Scorr}. 

S-corrections specifically account for differences in filter band passes for non-stellar objects. The magnitude correction is calculated by finding the difference between synthetic photometry generated by integrating the spectra or spectral templates in the old system and new systems:

\begin{equation}
S_{s, corr} \; = \; m_{s, standard} \; - \; m_{s, natural} \; 
\end{equation}
\parindent = 0 mm

where $S_{s, corr}$ is the photometric S-correction in magnitudes in filter s, $m_{s, standard}$ is the synthetic photometry of the object's SED using the standard filter band pass, and $m_{s, natural}$ is the synthetic magnitude calculated using the natural filter band pass. Thus, passbands $S(\lambda)$ of the two systems and the SED at each photometric epoch are both needed to calculate S-corrections. For our Las Cumbres Observatory instrumental system, we use the filter transmission curves and quantum efficiencies given by \citet{Baltay_2021}. For the standard band passes, we adopt filter functions given by \citet{Bessell_1990} for $BVRI$ and \citet{Fukugita_etal_1996} for $g'r'i'$. Since we do not have spectra at all photometry epochs, we use spectral templates from \citet{Hsiao_etal_2007} and color match them with observed photometry using the same method described in \autoref{k_corrections}. We use those as our SED approximations for our sample. These spectral templates only cover -19 $\sim$ 70 days with respect to B-band maximum. For data outside this range, we simply use the earliest or latest SED as the SED approximations for data points earlier/later than the range.

\parindent = 9 mm

\subsection{Light curve fitting technique}
\label{method: FPCA}
Direct measurements of the peak magnitudes and the 15 day decline rates ($\Delta m_{15}(B)$) are uncertain for most SNe Ia light curve because the lack of high cadence photometric observations. Thus, different models have been used to fit the discrete light curves (e.g., the light curve stretch method \citep{Goldhaber_etal_2001}, MLCS \citep{Riess_etal_1996_mlcs}, and SNooPy \citep{Burns_etal_2011}). Here we use the functional principal component analysis (FPCA) template fitting technique introduced by \citet{He_2018}. The light curves in each band are fitted as follows:

\begin{equation}
m(t-t_0) \; = \; m_0 \; + \; \phi_0(t-t_0) \; + \; \sum_{i=1}^K\beta_i\phi_i(t-t_0) \; \; \; ,
\end{equation}

\parindent = 0 mm

where $m(t-t_0)$ is the magnitude at time t in a specific band, $t_0$ is the time of maximum light, and $m_0$ is the maximum magnitude. The functions $\phi_i$ for $i=0$ to $K$ are templates constructed from a collection of well-observed SNe~Ia light curves, where $\phi_0$ is the mean template that captures the general light curve shape, and $\phi_i$ for $i=1$ to $K$ are $K$ fixed principal component templates that capture any additional variations of the observed light curves from the mean function $\phi_0$. Lower order functions of $\phi_i$ capture more variability \citep{He_2018}. 

\parindent = 9 mm

Two types of templates were introduced in \citet{He_2018}. One is ``filter-specific templates'', where the templates for each filter were trained with data only in that filter. The other is ``filter-vague templates'', where the templates were trained with data from all filters ($BVRI$) together. Since \citet{He_2018} only introduced the filter-specific templates for $BVRI$ filters, in this work, filter-specific templates were used for the $B-$ and $V-$bands and filter-vague templates were used for the $g'r'i'$ bands. The first two principal components account for more that 90\% of the variability of the light curves \citep[see Fig. 1 in ][]{He_2018}. So we only use the first two principal components to fit the light curves.
The light curve is then fully parameterized by $(m_0, t_0, \beta_1, \beta_2)$ in our case. The best coefficients are found by minimizing $\chi^2$ using the least-squares fitting package \texttt{pycmpfit}\footnote{https://github.com/cosmonaut/pycmpfit}, with additional constraints that the fitted light curve must monotonically increase before maximum, and monotonically decrease 35 days after maximum. $\Delta m_{15}(B)$ is then measured from the fitted light curves (For error calculation on $\Delta m_{15}(B)$, see Appendix in Aldoroty et al. 2022 (in prep)). The fitted parameters and measured $\Delta m_{15}(B)$ are listed in \autoref{tab:parameters_91T_head} and \autoref{tab:parameters_normal_head}.

\subsection{K-corrections}
\label{k_corrections}
\begin{figure}
\centering
\resizebox{.7\textwidth}{!}{%
\includegraphics[height=3cm]{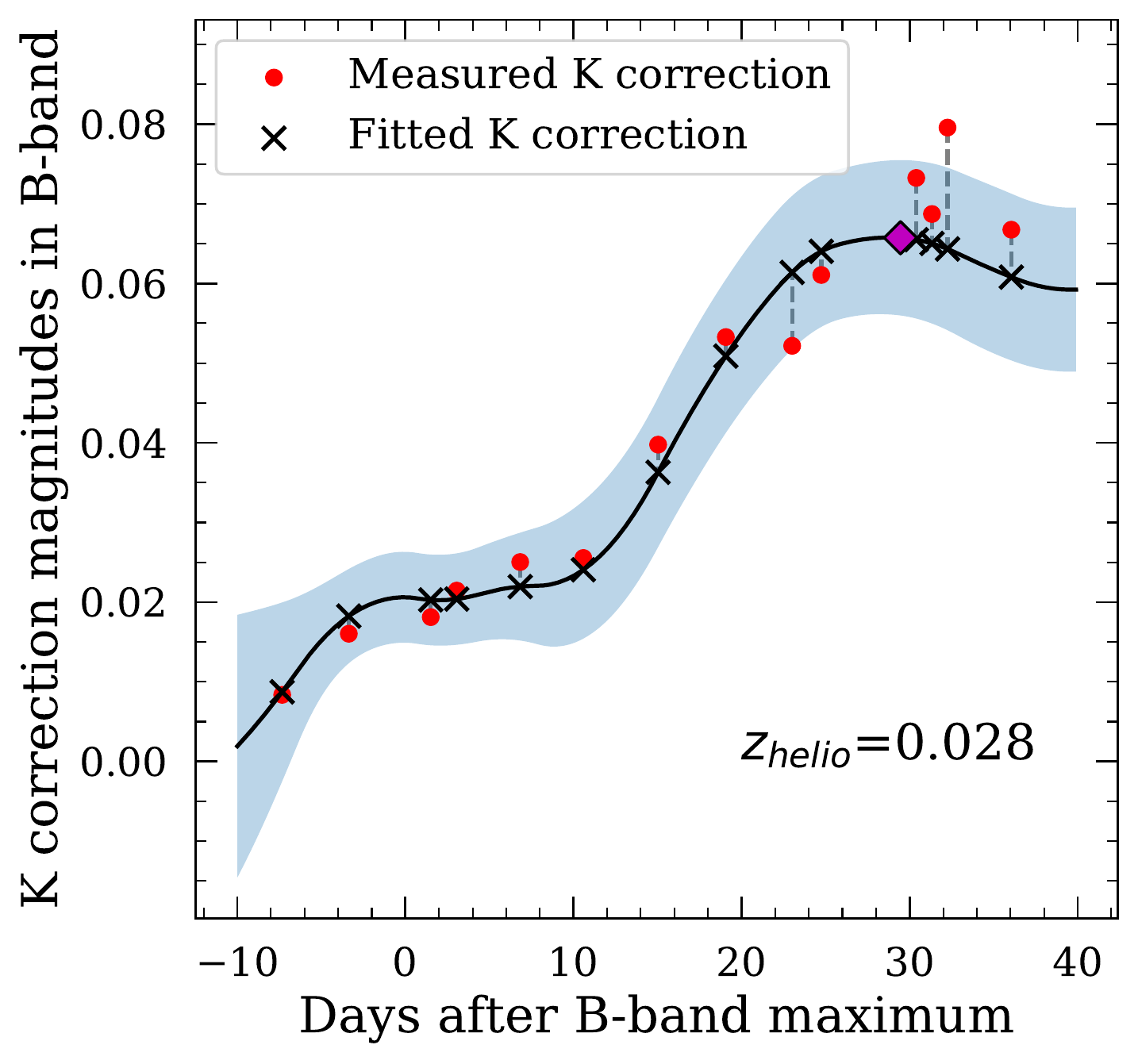}%
\quad
\includegraphics[height=3cm]{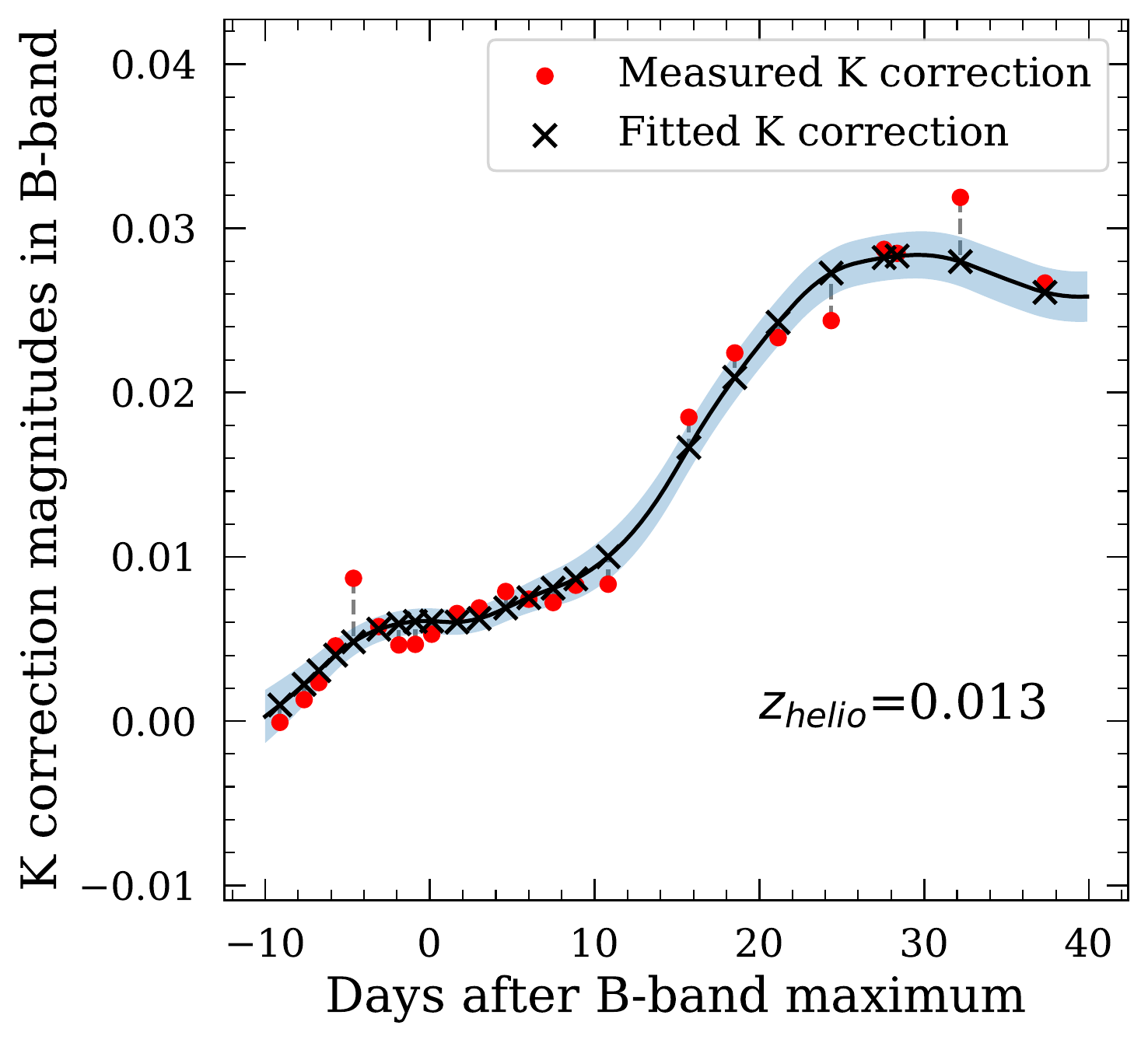}%
}
\caption{Examples of K-correction calculation for SN2016gcl (left) and SN2019vrq (right). The red data points are the K-correction values measured from mangled Hsiao's template. The black line is the FPCA fit to the red data points. The shaded region is the error of the FPCA fit. And the black crosses are the adopted K-correction values derived from the fitted FPCA curve. By fitting a smooth curve to the measured K-correction, we can not only get the error estimates, but also the K-correction values of the epochs when colors are not available to which the spectra need to be mangled to match (one example is the purple diamond on the left panel).}
\label{fig:kcorr}
\end{figure}

To compare photometry of SNe at different redshifts, K-corrections are needed to account for the effect of the redshift on the observed magnitudes \citep{Hamuy_etal_1993,Kim_etal_1996}.

Before we calculate and apply the K-correction to the observed photometric light curves, the light curves are first corrected for extinction along the line-of-sight in the Milky Way (MW) using the MW dust map given by \citet{Schlegel_etal_1998}, assuming the MW $R_{V}=3.1$ and a \citet{Fitzpatrick_1999} reddening law. Then the light curves are further corrected by removing the time dilation due to Hubble expansion. Corrected light curves are then fitted using the FPCA light curve fitting technique described above. The time of $B$-band maximum is then used to determine the rest-frame phase of the SN photometry to select the corresponding \citet{Hsiao_etal_2007} template spectrum to be used for K-corrections. In practice, the reddening is unknown before K-correction. The template is mangled using a smooth function to match the observed $B-V$ color when determining the $B$-band K-corrections, and $V-i'/I/R$ when determining the $V$-band K-corrections to minimize extrapolation errors. The smooth function used to mangle the template spectra is the \citet{Fitzpatrick_Massa_1999} reddening law assuming $R_V=3.1$. The K-corrections are then measured by the difference in the synthetic magnitudes of the mangled spectrum at redshift of zero and the given redshift of the SN (shown as red data points in \autoref{fig:kcorr}). The filter-vague FPCA decomposition of \cite{He_2018} shows that the light curves of SNe~Ia can be decomposed into a single set of FPCA vectors independent of the filter bands employed. This implies that the K-correction itself can also be decomposed into the same FPCA basis. The measured K-correction magnitudes at different epochs are further fitted using the \citet{He_2018} FPCA templates with the mean component taken out and the maximum epoch fixed by the previous FPCA fit using \texttt{pycmpfit} $\chi^2$ minimization. The final adopted values of the K-corrections (black cross in \autoref{fig:kcorr}) and their corresponding errors (shaded region in \autoref{fig:kcorr}) at desired epochs are then calculated from the fitted curve. 

\subsection{Host Reddening}
\label{subsec:host_extinction}
\begin{figure}
\gridline{
          \fig{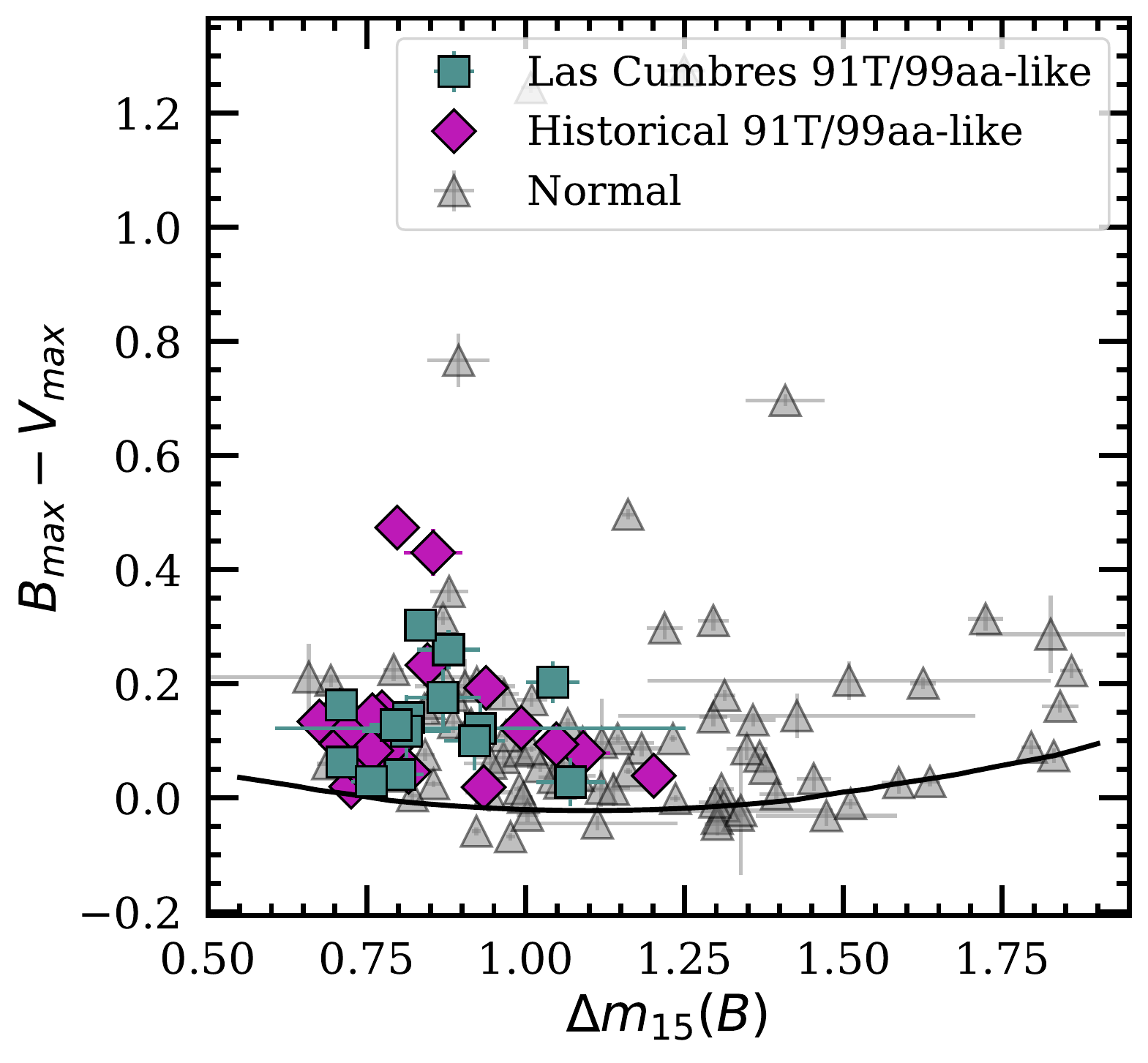}{0.5\textwidth}{}
          }
\caption{Observed  $B_{max}-V_{max}$ (MW extinction and K-corrected) versus  $\Delta m_{15}(B)$. Extinction can only cause $B_{max}-V_{max}$ to increase in value, thus the lower envelope (plotted as black solid line) can be used as an extinction free curve to correct for host extinction (\autoref{subsec:host_extinction}). 91T/99aa-like data points are plotted in green squares, and historical 91T/99aa-like data points are plotted in purple diamonds. Normal SNe Ia are plotted in grey triangles.}
\label{fig: host_extinction}
\end{figure}

Extinction can only make objects redder. Assuming that the intrinsic colors of SNe~Ia are uniquely determined by the light curve width as measured by the $\Delta m_{15}(B)$ values,  one can expect there to be a ``blue edge'' in the distribution of the observed colors when the observed colors are plotted together with the $\Delta m_{15}(B)$ (Fig.~\ref{fig: host_extinction}). With a large sample of SNe~Ia, we may assume this blue edge to represent the intrinsic colors of SNe~Ia for a given $\Delta m_{15}(B)$, and the reddening due to dust in the host galaxy $E(B-V)_{host}$ can be deduced by the difference between the observed colors and the this ridge line. Fig. \ref{fig: host_extinction} shows the MW- and K-corrected $B_{max}-V_{max}$ vs. $\Delta m_{15}(B)$. We estimate the position of the ``blue edge'' by fitting a second order polynomial ($ B_{max}-V_{max}=A\times(\Delta m_{15}(B)-1.1)^2+C$ where A and C are parameters.) to the ridge line that divides the SN sample into two groups with the bluer group having 10 \% of the total number of the SNe in the sample. To do so robustly, 500 bootstrapped samples are generated with replacement from the data set, leading to 500 estimates of the blue ridge lines. The median and standard deviation of these 500 ridge lines are used as the line of the intrinsic color and its corresponding uncertainties, respectively. Uncertainties of the host galaxy extinction are then estimated by the quadrature sum of uncertainties in $B_{max}-V_{max}$ and the uncertainties of the lower boundaries of the intrinsic colors. The equation for the final ridge line is
$$(B_{max}-V_{max})(mag)=0.188\times(\Delta m_{15}(B) (mag)-1.1)^2-0.021.$$

\section{Spectra}
\label{sec:spectra}

Spectra of our 91T/99aa-like sample are plotted in \autoref{fig:spectra_sequence} at different epochs relative to the $B$ band maximum.  For SN2018apo we do not have $B$-band photometry, so the maximum date was determined using $g$ band photometry. These spectra were collected from the Open Supernova Catalogue \citep{Guillochon_etal_2017}. 
The spectra are grouped into 5 groups each showing the data within 2 days of (a)  $-$13 days, (b) $-$8 days, (c) $-$4 days, (d) 0 days, and (e) +6 days from $B$-band maximum for our 91T/99aa-like sample. The strength of the Si II $6355 \pm 5 \AA$ around maximum increases from top to bottom for each group. Spectra of SN1991T, SN1999aa, and the spectroscopically normal SN 2004eo at the relevant phases are plotted in bold lines together with each group for comparisons.

Not all 91T/99aa-like events classified by their spectra are slow decliners. The 91T/99aa-like objects with $\Delta m_{15}(B)>1.0 $ mag are plotted in red in \autoref{fig:spectra_sequence}. It is readily seen that some of the SNe with the shallowest \Si lines are actually not slow decliners. The slowest decliners among 91T/99aa-like objects with $\Delta m_{15}(B)<0.8$ mag are plotted in dashed grey line. Although still showing shallower Si II lines, the spectra of 91T/99aa-like objects show all the major spectral features of normal SNe~Ia at about 6 days after maximum. This suggests strongly that they share similar ejecta structures with spectroscopically normal SNe~Ia and are likely originate from similar physical systems and processes.

\subsection{Pseudo-equivalent widths}
\label{method: pEW}
The true continuum in a SN~Ia spectrum can not be formally determined, but the strength of an absorption line can be measured with respect to the flux at wavelengths shorter and longer than that of a given line.  This is called the pseudo-equivalent width (pEW).  It is defined as follows:

\begin{equation}
pEW \; = \; \sum_{i=1}^{N}\left (1-\frac{f(\lambda_i)}{f_c(\lambda_i)} \right ) \; \; ,
\end{equation}

\parindent = 0 mm

where $\lambda_1$ and $\lambda_N$ define the boundaries between which the pEW is calculated, $f(\lambda_i)$ is the spectral flux at wavelength $\lambda_i$, and $f_c(\lambda_i)$ is the pseudo continuum flux level defined by the line connecting the endpoints at $\lambda_1$ and $\lambda_N$. 

\parindent = 9 mm

To measure the pEW of the \Si line, the endpoints are found by selecting the point with maximum flux of binned spectra (with a bin size of 40$\AA$) in the wavelength ranges [5820 $\AA$, 6000 $\AA$] and [6200 $\AA$, 6540 $\AA$] for the red side and blue side, respectively. The endpoints are further inspected by eye. We use a Monte Carlo method to calculate the errors of the pEW measurements. First we find the signal-to-noise ratio at each wavelength by dividing the original spectrum with a heavily smoothed spectrum (smoothed by the wavelet transformation method described by \citealp{Wagers_2010}). Then the standard deviation of the signal-to-noise ratio at each pixel is found by the standard deviation of signal-to-noise ratios within 50 $\AA$. Next, we create a large number of simulated spectra with random noise generated using a Gaussian distribution with this calculated standard deviation. Finally, the pEW measurement is obtained for each of the simulated spectra, and the mean and the standard deviation of these pEW measurements give our final pEW value and its corresponding 1-$\sigma$ error.

\newsavebox{
\myimage
}

\newcommand{\width}{0.27}
\begin{figure}[ht]

  \centering
  \savebox{\myimage}{
        \fig{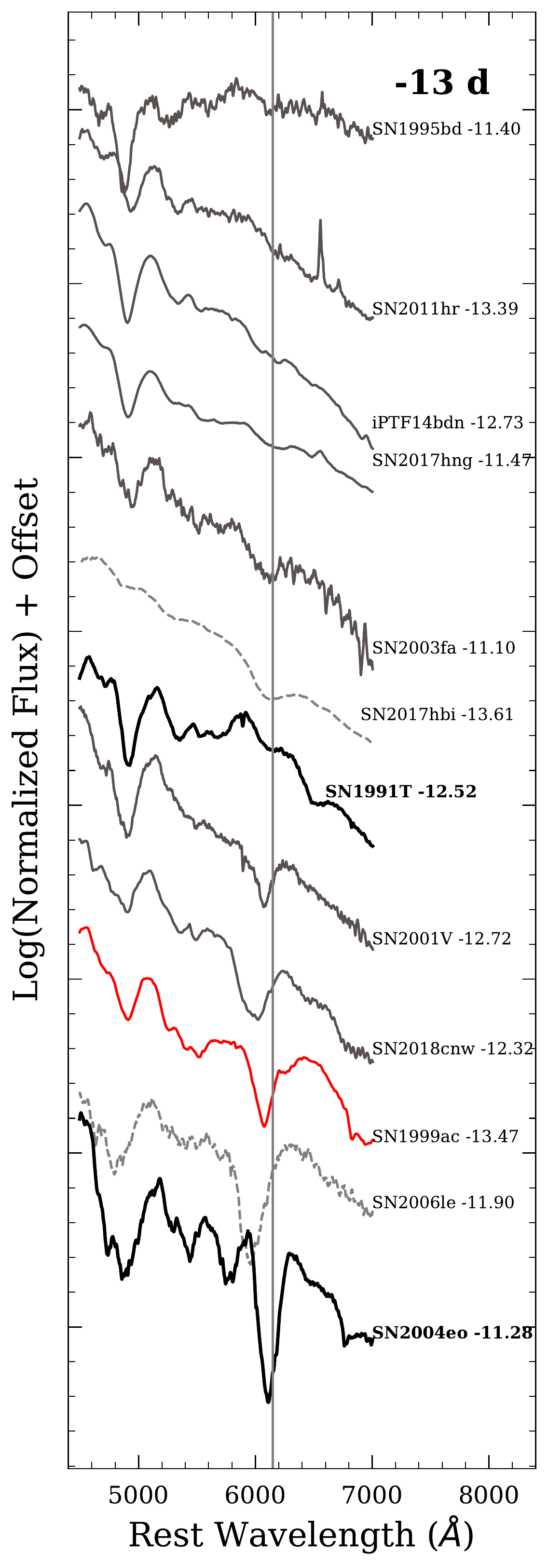}{\width\textwidth}{(a)}
        \fig{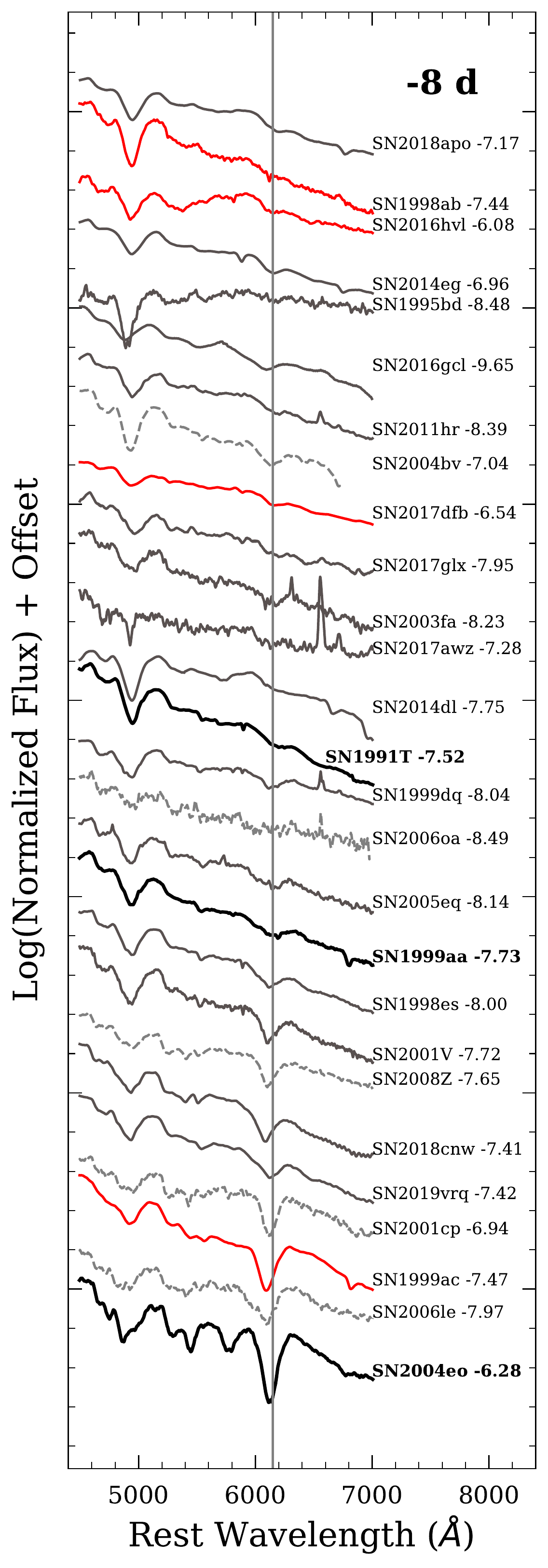}{\width\textwidth}{(b)}
        \fig{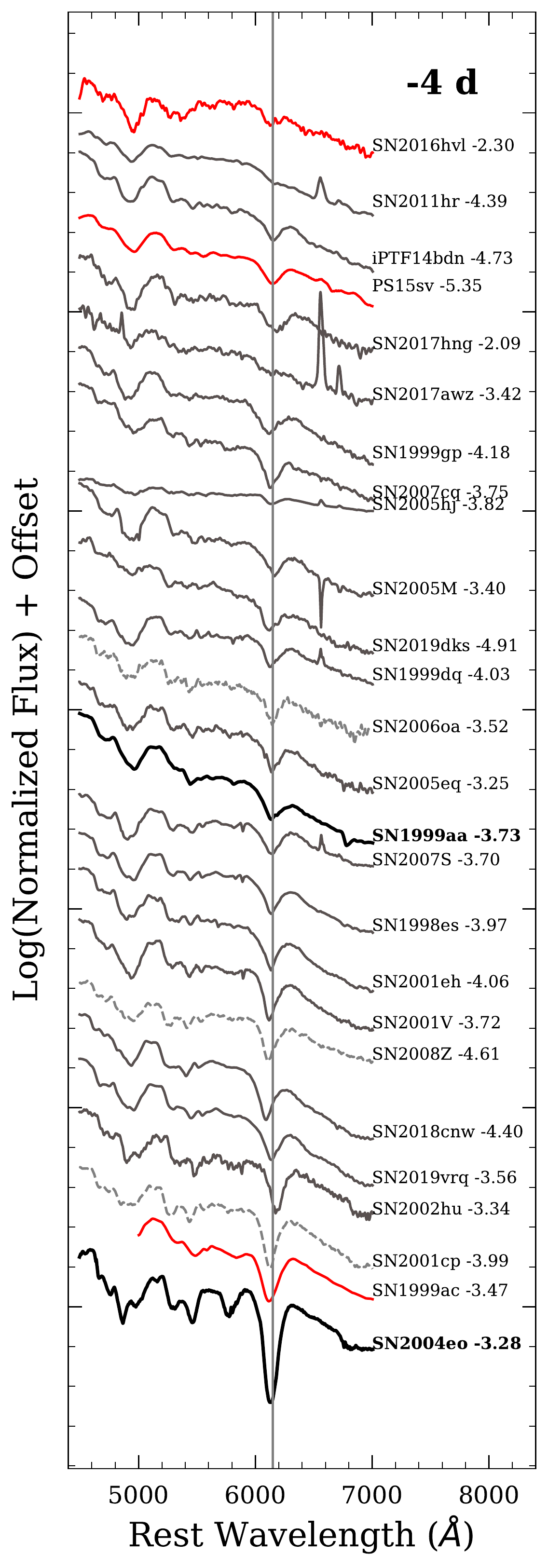}{\width\textwidth}{(c)}
        \fig{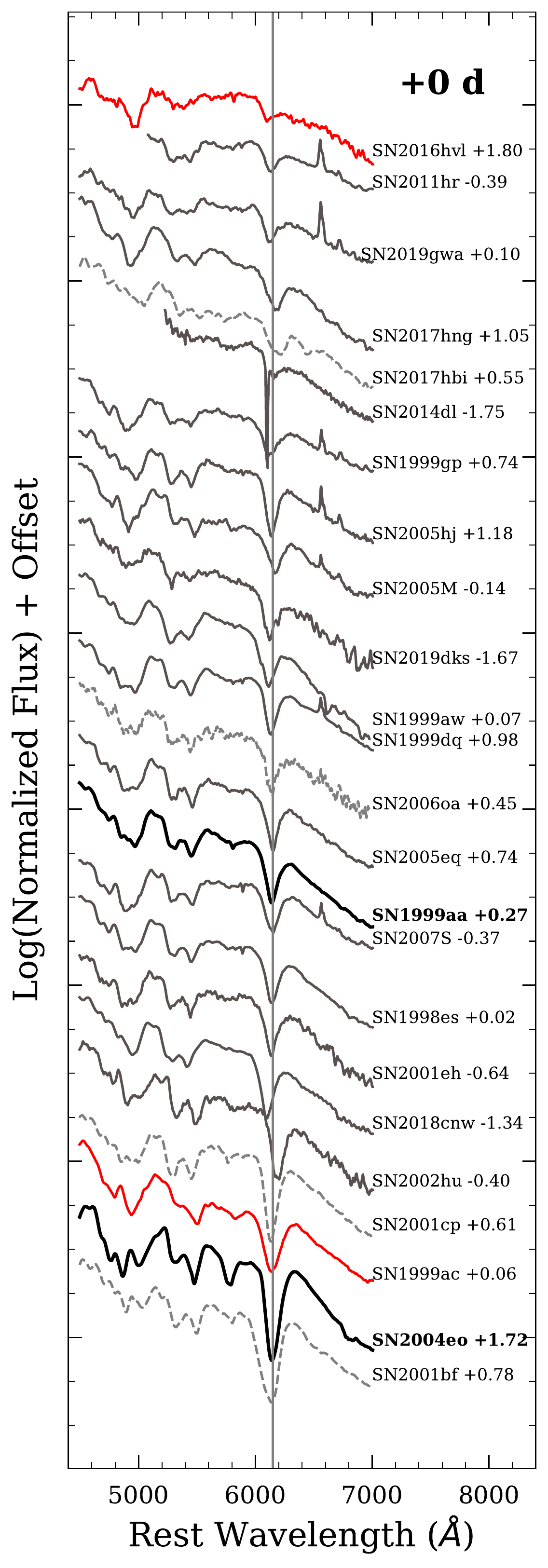}{\width\textwidth}{(d)}
        \fig{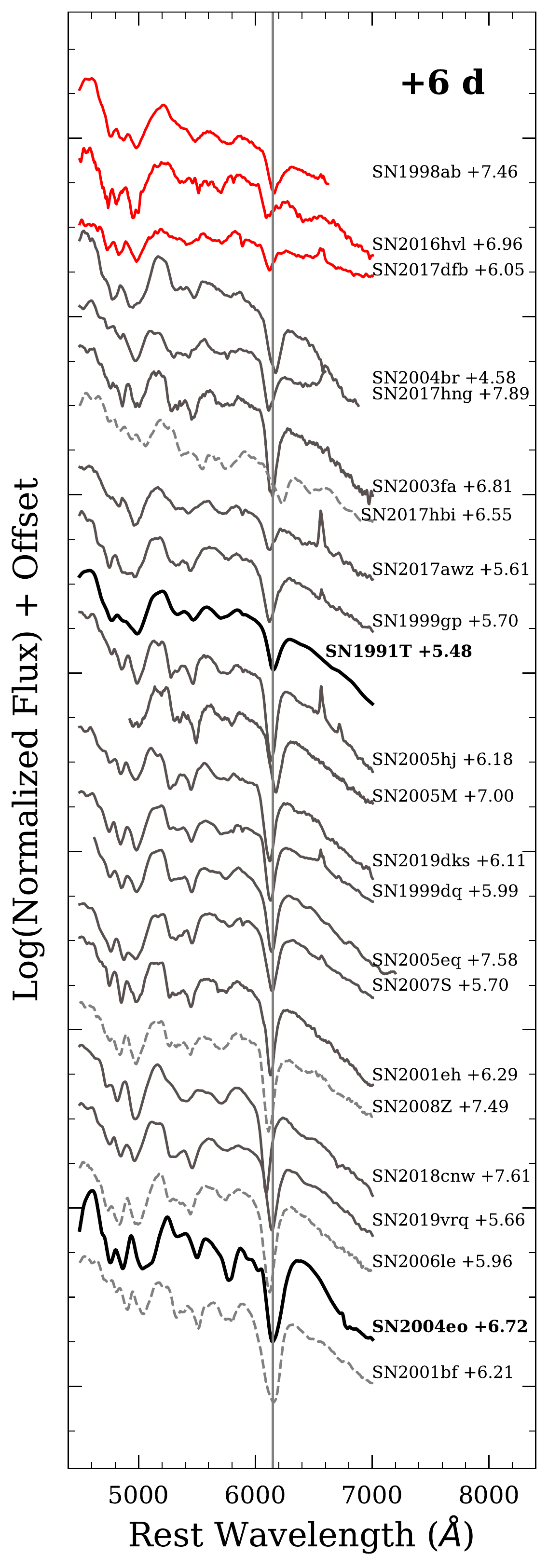}{\width\textwidth}{(e)}
        
  }
  \rotatebox{90}{
    \begin{minipage}{\wd\myimage}
      \usebox{\myimage}
      \caption{Spectra within 2 days of (a) $-$13 days; (b) $-$8 days; (c) $-$4 days; (d) 0 days; and (e) +6 days of our 91T/99aa-like sample. The Si II $6355 \pm 5 \AA$ around maximum increases from top to bottom for each panel. SN 1991T, SN 1999aa, and the typical normal Type Ia SN 2004eo are plotted in bold. 91T/99aa-like objects with $\Delta m_{15}(B)>1.0$ mag are plotted in red. And normal objects in our sample with $\Delta m_{15}(B)<0.8$ mag are also plotted in dashed grey lines. A vertical grey line at 6150 $\AA$ is also plotted to guide the eye.}
      \label{fig:spectra_sequence}
    \end{minipage}}

\end{figure}

\begin{figure}
\centering
\resizebox{.9\textwidth}{!}{%
\includegraphics[height=3cm]{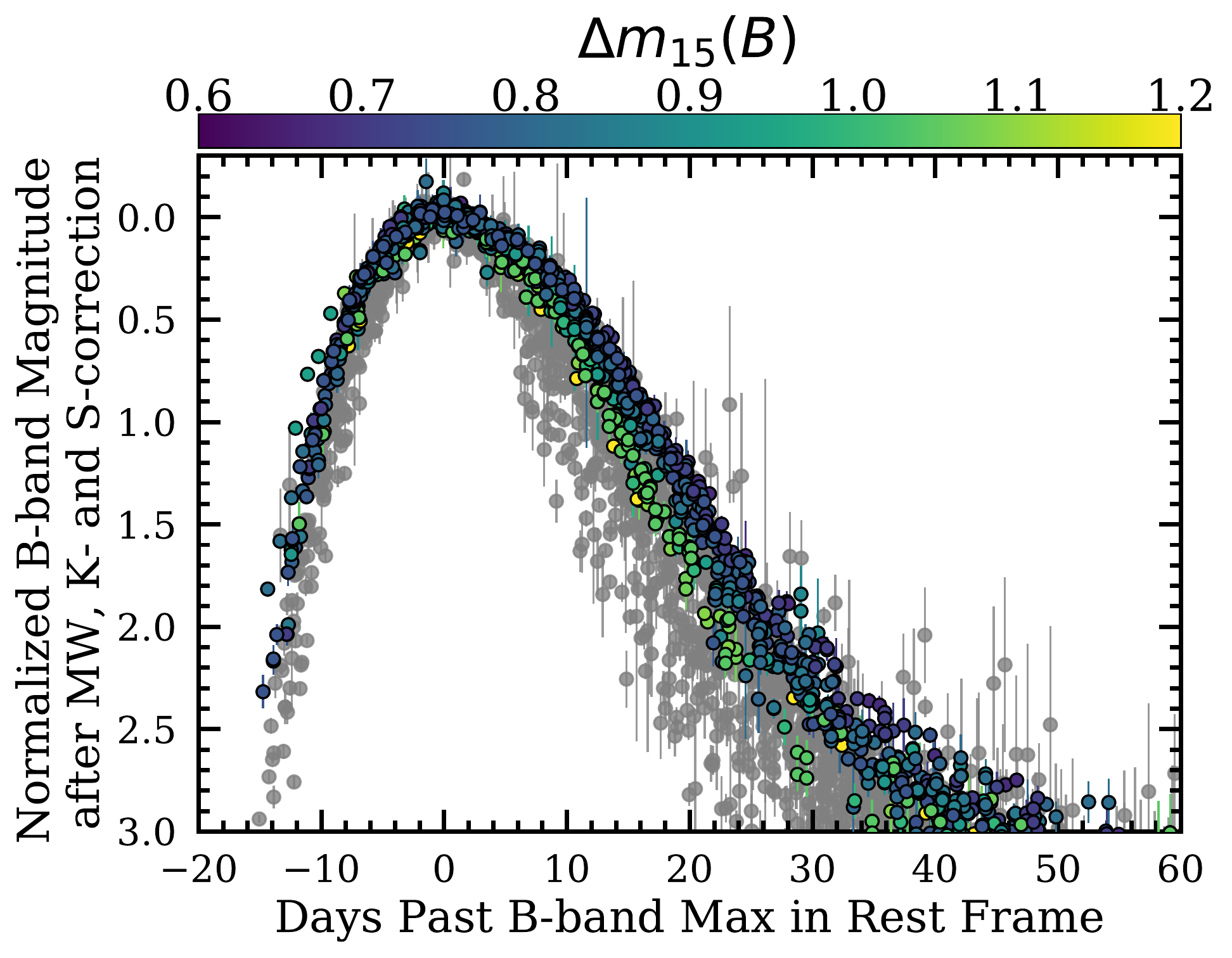}%
\quad
\includegraphics[height=3cm]{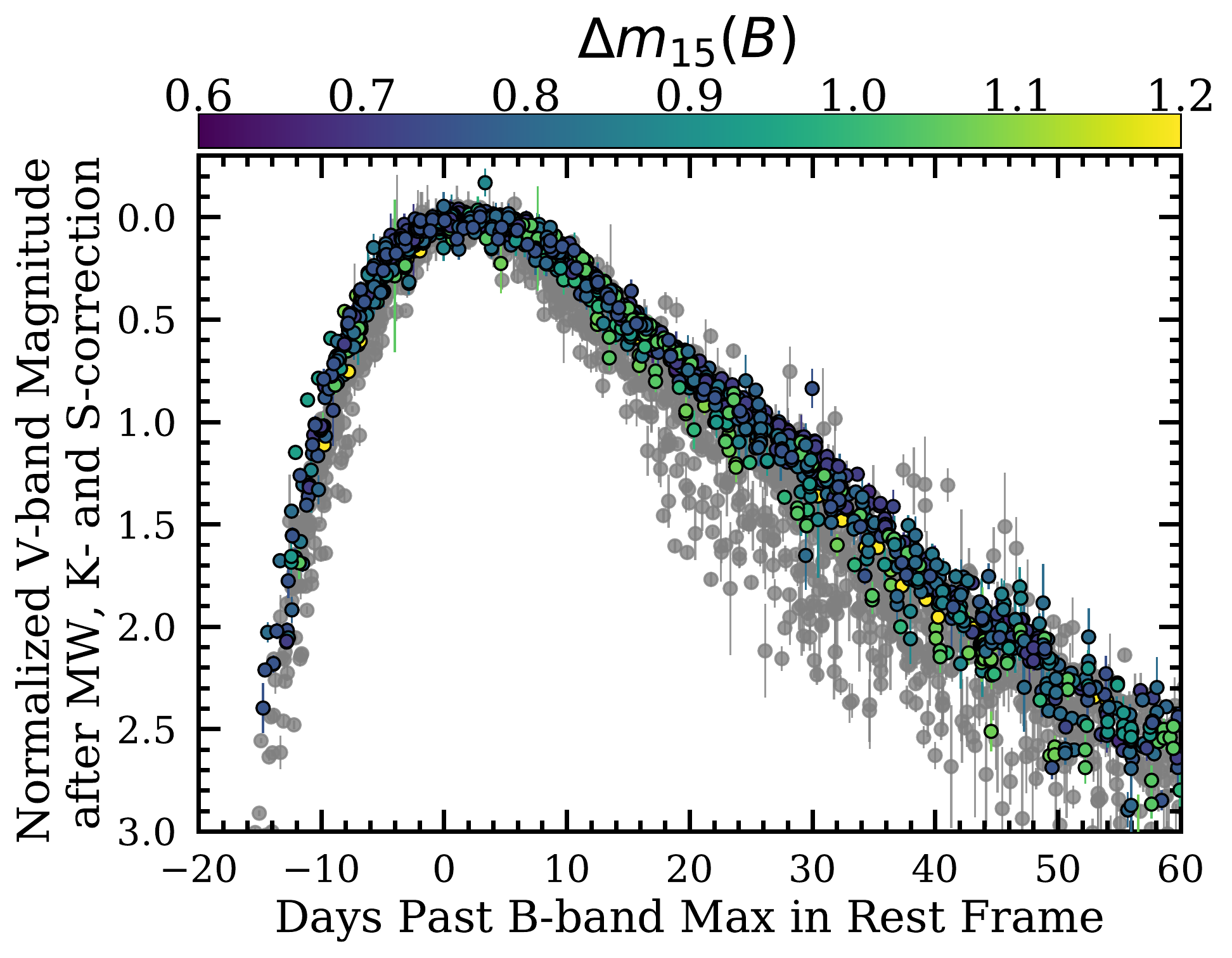}%
}
\caption{$B$- and $V$-band light curves for the SNe used in this paper. The light curves have been shifted vertically for better visualization. The grey data points are normal SNe~Ia, while the data points of 91T/99aa-like SNe~Ia are color coded by $\Delta m_{15}(B)$.}
\label{fig: lc_all}
\end{figure}

\section{Standardization}
\label{sec:standardization}
The discovery of the accelerating expansion of the universe using SNe~Ia was based on the discovery of the relationship between the SNe~Ia peak luminosity magnitude and the light curve decline rate \citep{Phillips_1993} and the peak luminosity magnitude and color relationship found later \citep{Hamuy_etal_1996H0, Tripp_1998}, which make SNe~Ia standardizable candles. 91T/99aa-like SNe~Ia are inevitably over-represented in flux-limited surveys. Understanding the intrinsic properties of these SNe is therefore crucial to controlling the systematic errors when using SNe~Ia to measure distances. 

A Hubble diagram using SNe~Ia can be used to constrain the cosmological acceleration parameters parameters, but it requires a high-redshift SNe~Ia sample to be able to differentiate between different cosmology models \citep{1995ApJ...450...14G}. However, a nearby SNe~Ia sample is also necessary to evaluate the precision of SNe~Ia as standardizable candles, specifically, the standardizability of 91T/99aa-like objects in our case. 

In this section, we evaluate the precision to which 91T/99aa-like SNe~Ia may be used as standardizable candles. Recently, more sophisticated modeling and parameterization of SNe~Ia are used for SNe~Ia standardization, for example SALT2 \citep{Guy_etal_2007,Guy_etal_2010,Betoule_etal_2014}, SNEMO \citep{Saunders_2018_SNEMO}, and SUGAR \citep{Leget_2020_SNEMO}. Here we adopt the simple two-parameter parameterization using the decline rate $\Delta m_{15}(B)$ and host color excess $E(B-V)_{host}$ for standardization:
\begin{equation}
\label{eq:calibrate_eqn}
    m_B^* \; = \; m_B(0)^* \; + \; b[\Delta m_{15}(B)-1.1]+R_B E(B-V)_{host} \; \; ,
\end{equation}

\parindent = 0 mm

where $m_B^* = m_B - R_B E(B-V)_{gal} - K_B$ is the MW and K-corrected apparent peak magnitudes of SNe~Ia in the $B$-band, $m_B(0)^*$ is the expected apparent peak magnitude of our fiducial SN~Ia at redshift $z$, with $\Delta m_{15}(B)$ = 1.1 and $E(B-V)=0$. Here we adopt the approximation of $m_B(0)^*=5log_{10}(cz)+k$ as all SNe in our sample have $z\lesssim 0.1$. Thus, we immediately have $H_0 = 10^{0.2(M_B(0)+k+25)}$, where $M_B(0)$ is the absolute peak magnitude of a fiducial SN~Ia. 
The global parameters are then ($b$, $R_B$, $k$) where $k$ is the intercept of the Hubble diagram. The Phillips luminosity-decline relationship \citep{Phillips_1993} is represented by the parameter $b$, the luminosity-color relationship is determined by $R_B$, and $k$ contains the information of Hubble constant at present, or equivalently, the absolute magnitude of the fiducial SN~Ia.

In the analyses presented here, heliocentric redshifts were converted into the CMB frame using the velocity vector determined by \citet{Lineweaver_1997_cmb_helio} to calculate luminosity distances. The uncertainty in the redshifts due to peculiar velocities is assumed to be $\sigma_z=0.001$ (300 km/s in velocity).

Parameters in Equation \ref{eq:calibrate_eqn} are determined by fitting SNe in our sample with $z_{cmb}>0.01$ using $\chi^2$ minimization of

\begin{equation}
\chi^2 = \sum_{i=1}^N \frac{(m_{Bcorr}^* - m_B(0)^*)^2}{\sigma_{m_{Bcorr,i}^*}^2 + \sigma_{intrinsic}^2 + \sigma_{pec}^2} \; ,
\end{equation}
\parindent = 0 mm

where $m_{Bcorr}^*=m_B^*-b\Delta (m_{15}(B)-1.1)-R_B E(B-V)$,  $\sigma_{pec}$ accounts for peculiar velocity ($v_{pec}$ = 300 
km s$^{-1}$),

\begin{equation}
\sigma_{m_{Bcorr}^*}^2=\sigma_{m_B^*}^2+b^2 \sigma_{\Delta m_{15}(B)}^2 + R_B^2 \sigma_{E(B-V)}^2 \; , 
\end{equation}

and $\sigma_{intrinsic}^2$ accounts for possible intrinsic scatter in SNe~Ia, which is determined so that the final $\chi^2_{\nu}=1$.

\parindent = 9 mm

\begin{splitdeluxetable*}{ccccccccBcccccccc}
\tabletypesize{\scriptsize}
\tablecaption{Parameters of 91T/99aa-like SNe Ia in our sample.\label{tab:parameters_91T_head}}
\tablehead{\colhead{SN (subtype\edithere{\tablenotemark{a}})}\vspace{-0.3cm} & \colhead{$z_{CMB}$} & \colhead{$pEW_{nearmax}(\text{Si II }\lambda 6355)$} & \colhead{$E(B-V)_{host}$} & \colhead{Hubble Residual} & \colhead{$t_{max}(B)$} & \colhead{$\Delta m_{15}(B)$} & \colhead{$m_{max}(B)$} & \colhead{SN (subtype\edithere{\tablenotemark{a}})} & \colhead{PC1(B)} & \colhead{PC2(B)} & \colhead{$t_{max}(V)$} & \colhead{$\Delta m_{15}(V)$} & \colhead{$m_{max}(V)$} & \colhead{PC1(V)} & \colhead{PC2(V)}\\
 &  &\colhead{($\AA$)} & & \colhead{(mag)} & \colhead{(MJD)} & \colhead{(mag)} & \colhead{(mag)} &  & &  & \colhead{(MJD)} & \colhead{(mag)} & \colhead{(mag)} &  & 
}
\startdata
LSQ15aae & 0.0518 & ... & 0.27 (0.04) & -0.35 (0.31) & 57116.45 (0.35) & 0.88 (0.05) & 17.77 (0.03) & LSQ15aae & -0.28 (0.32) & -0.03 (0.20) & 57118.78 (0.35) & 0.58 (0.04) & 17.51 (0.02) & -0.39 (0.24) & 0.00 (0.12) \\
PS15sv (3) & 0.0333 & 32.0 & 0.05 (0.04) & -0.26 (0.33) & 57114.35 (0.85) & 1.07 (0.05) & 16.31 (0.04) & PS15sv (3) & 1.25 (0.51) & -0.35 (0.21) & 57115.85 (0.32) & 0.69 (0.03) & 16.28 (0.02) & 0.52 (0.19) & 0.04 (0.08) \\
SN1991T (4) & 0.0069 & 42.2 & 0.21 (0.02) & -2.00 (0.47) & 48375.52 (0.18) & 0.94 (0.01) & 11.60 (0.01) & SN1991T (4) & -0.13 (0.09) & -0.06 (0.04) & 48378.19 (0.12) & 0.62 (0.01) & 11.41 (0.01) & -0.34 (0.07) & 0.19 (0.04) \\
SN1995bd (2) & 0.0144 & 20.1 & 0.44 (0.04) & -0.38 (0.35) & 50085.74 (0.21) & 0.85 (0.05) & 15.51 (0.03) & SN1995bd (2) & -0.98 (0.31) & 0.24 (0.13) & 50087.50 (0.22) & 0.69 (0.04) & 15.08 (0.02) & -0.29 (0.23) & 0.43 (0.11) \\
SN1998ab (0) & 0.0279 & 4.5 & 0.10 (0.05) & -0.37 (0.34) & 50913.82 (0.16) & 1.09 (0.04) & 16.00 (0.03) & SN1998ab (0) & 1.48 (0.28) & -0.46 (0.09) & 50915.94 (0.25) & 0.62 (0.04) & 15.92 (0.03) & 0.12 (0.25) & -0.10 (0.11) \\
SN1998es (5) & 0.0096 & 53.0 & 0.14 (0.01) & -0.01 (0.35) & 51142.20 (0.04) & 0.71 (0.01) & 13.87 (0.01) & SN1998es (5) & -0.78 (0.08) & -0.67 (0.04) & 51144.30 (0.05) & 0.52 (0.01) & 13.72 (0.01) & -0.76 (0.05) & -0.10 (0.03) \\
SN1999aa (5) & 0.0152 & 52.4 & 0.01 (0.01) & 0.29 (0.32) & 51231.73 (0.04) & 0.73 (0.01) & 14.76 (0.01) & SN1999aa (5) & -0.55 (0.05) & -0.74 (0.03) & 51233.92 (0.05) & 0.54 (0.01) & 14.74 (0.00) & -0.51 (0.04) & -0.12 (0.02) \\
SN1999ac (8) & 0.0098 & 80.4 & 0.06 (0.02) & 0.01 (0.37) & 51249.47 (0.03) & 1.20 (0.01) & 14.07 (0.01) & SN1999ac (8) & 1.09 (0.05) & 0.20 (0.03) & 51252.63 (0.07) & 0.65 (0.01) & 14.03 (0.00) & 0.41 (0.04) & -0.12 (0.03) \\
SN1999aw (4) & 0.0392 & 48.7 & 0.12 (0.02) & -0.11 (0.29) & 51253.93 (0.23) & 0.68 (0.01) & 16.72 (0.01) & SN1999aw (4) & -1.90 (0.11) & 0.15 (0.04) & 51255.63 (0.24) & 0.57 (0.02) & 16.59 (0.01) & -1.27 (0.10) & 0.45 (0.07)\\
\enddata
\edithere{\tablenotetext{a}{Sub-type classification based on the strength of \Simax. The classification is the floor hexadecimal integer of \Simax/10 (see \autoref{subsec: Results_pEW} for more details).}}
\end{splitdeluxetable*}

\begin{splitdeluxetable*}{ccccccccBcccccccc}
\tabletypesize{\scriptsize}
\tablecaption{Parameters of normal SNe Ia in our sample.\label{tab:parameters_normal_head}}
\tablehead{\colhead{SN (subtype\edithere{\tablenotemark{a}})}\vspace{-0.3cm} & \colhead{$z_{CMB}$} & \colhead{$pEW_{nearmax}(\text{Si II }\lambda 6355)$} & \colhead{$E(B-V)_{host}$} & \colhead{Hubble Residual} & \colhead{$t_{max}(B)$} & \colhead{$\Delta m_{15}(B)$} & \colhead{$m_{max}(B)$} & \colhead{SN (subtype\edithere{\tablenotemark{a}})} & \colhead{PC1(B)} & \colhead{PC2(B)} & \colhead{$t_{max}(V)$} & \colhead{$\Delta m_{15}(V)$} & \colhead{$m_{max}(V)$} & \colhead{PC1(V)} & \colhead{PC2(V)}\\
 &  &\colhead{($\AA$)} & & \colhead{(mag)} & \colhead{(MJD)} & \colhead{(mag)} & \colhead{(mag)} &  & &  & \colhead{(MJD)} & \colhead{(mag)} & \colhead{(mag)} &  & 
}
\startdata
SN1998dh (C) & 0.0078 & 121.6 & 0.12 (0.02) & 0.09 (0.25) & 51029.19 (0.03) & 1.23 (0.01) & 13.88 (0.00) & SN1998dh (C) & 2.01 (0.05) & -0.40 (0.03) & 51031.57 (0.04) & 0.69 (0.01) & 13.77 (0.01) & 0.90 (0.04) & -0.25 (0.02) \\
SN1998dm (6) & 0.0060 & 68.4 & 0.33 (0.02) & 1.07 (0.32) & 51060.38 (0.07) & 0.87 (0.01) & 14.68 (0.01) & SN1998dm (6) & 0.14 (0.10) & -0.61 (0.04) & 51062.06 (0.05) & 0.60 (0.01) & 14.37 (0.01) & -0.08 (0.06) & -0.07 (0.02) \\
SN1999cp (9) & 0.0099 & 99.6 & 0.04 (0.02) & 0.10 (0.25) & 51363.46 (0.04) & 0.99 (0.01) & 13.93 (0.00) & SN1999cp (9) & 0.90 (0.06) & -0.60 (0.04) & 51364.71 (0.05) & 0.66 (0.01) & 13.91 (0.00) & 0.15 (0.05) & 0.09 (0.03) \\
SN1999dk (C) & 0.0140 & 126.4 & 0.12 (0.02) & -0.15 (0.18) & 51414.78 (0.13) & 1.09 (0.02) & 14.79 (0.01) & SN1999dk (C) & 1.48 (0.16) & -0.54 (0.09) & 51416.88 (0.15) & 0.68 (0.02) & 14.69 (0.01) & 0.63 (0.13) & -0.13 (0.06) \\
SN2000cn (B) & 0.0227 & 118.6 & 0.17 (0.02) & -0.05 (0.12) & 51707.17 (0.04) & 1.63 (0.02) & 16.53 (0.01) & SN2000cn (B) & 4.78 (0.23) & -0.62 (0.15) & 51709.58 (0.08) & 0.96 (0.01) & 16.33 (0.01) & 3.79 (0.22) & -0.61 (0.11) \\
SN2000cw (B) & 0.0288 & 111.4 & 0.11 (0.02) & 0.11 (0.12) & 51748.18 (0.19) & 1.18 (0.03) & 16.66 (0.01) & SN2000cw (B) & 2.31 (0.46) & -0.64 (0.34) & 51751.05 (0.22) & 0.69 (0.02) & 16.57 (0.01) & 1.23 (0.34) & -0.45 (0.19) \\
SN2000dk (C) & 0.0160 & 121.6 & -0.01 (0.02) & 0.05 (0.15) & 51812.27 (0.06) & 1.64 (0.02) & 15.30 (0.01) & SN2000dk (C) & 4.29 (0.15) & -0.22 (0.09) & 51814.42 (0.06) & 0.95 (0.01) & 15.28 (0.01) & 3.33 (0.09) & -0.44 (0.06) \\
SN2000dn (9) & 0.0307 & 98.7 & 0.08 (0.04) & 0.06 (0.15) & 51824.74 (0.14) & 1.06 (0.04) & 16.55 (0.02) & SN2000dn (9) & 1.32 (0.36) & -0.47 (0.25) & 51827.22 (0.17) & 0.69 (0.03) & 16.49 (0.02) & 0.94 (0.24) & -0.23 (0.14) \\
SN2000dr (C) & 0.0180 & 124.9 & 0.08 (0.03) & -0.04 (0.16) & 51834.31 (0.09) & 1.84 (0.03) & 15.94 (0.01) & SN2000dr (C) & 5.00 (0.00) & 0.27 (0.16) & 51836.42 (0.14) & 1.00 (0.02) & 15.77 (0.01) & 3.52 (0.19) & -0.32 (0.12) \\
SN2000fa (9) & 0.0215 & 90.9 & 0.19 (0.02) & -0.05 (0.13) & 51891.96 (0.14) & 0.89 (0.02) & 15.91 (0.01) & SN2000fa (9) & 0.30 (0.13) & -0.59 (0.07) & 51893.77 (0.22) & 0.60 (0.01) & 15.73 (0.01) & -0.03 (0.09) & -0.13 (0.05) \\
\enddata
\edithere{\tablenotetext{a}{Sub-type classification based on the strength of \Simax. The classification is the floor hexadecimal integer of \Simax/10 (see \autoref{subsec: Results_pEW} for more details).}}
\end{splitdeluxetable*}

\section{Results}
\label{sec:results}
\subsection{Photometric properties of 91T/99aa-like SNe~Ia}
The $B$- and $V$-band light curves of the SNe used in this analysis are plotted together in \autoref{fig: lc_all}. Gray points are for normal SNe~Ia while 91T/99aa-like objects are color coded by the value of $\Delta m_{15}(B)$. As expected, light curves of 91T/99aa-like SNe are typically broader compared to normal SNe~Ia. This trend is more pronounced in the $V$-band.

\begin{figure}
\centering
\resizebox{.9\textwidth}{!}{%
\includegraphics[height=3cm]{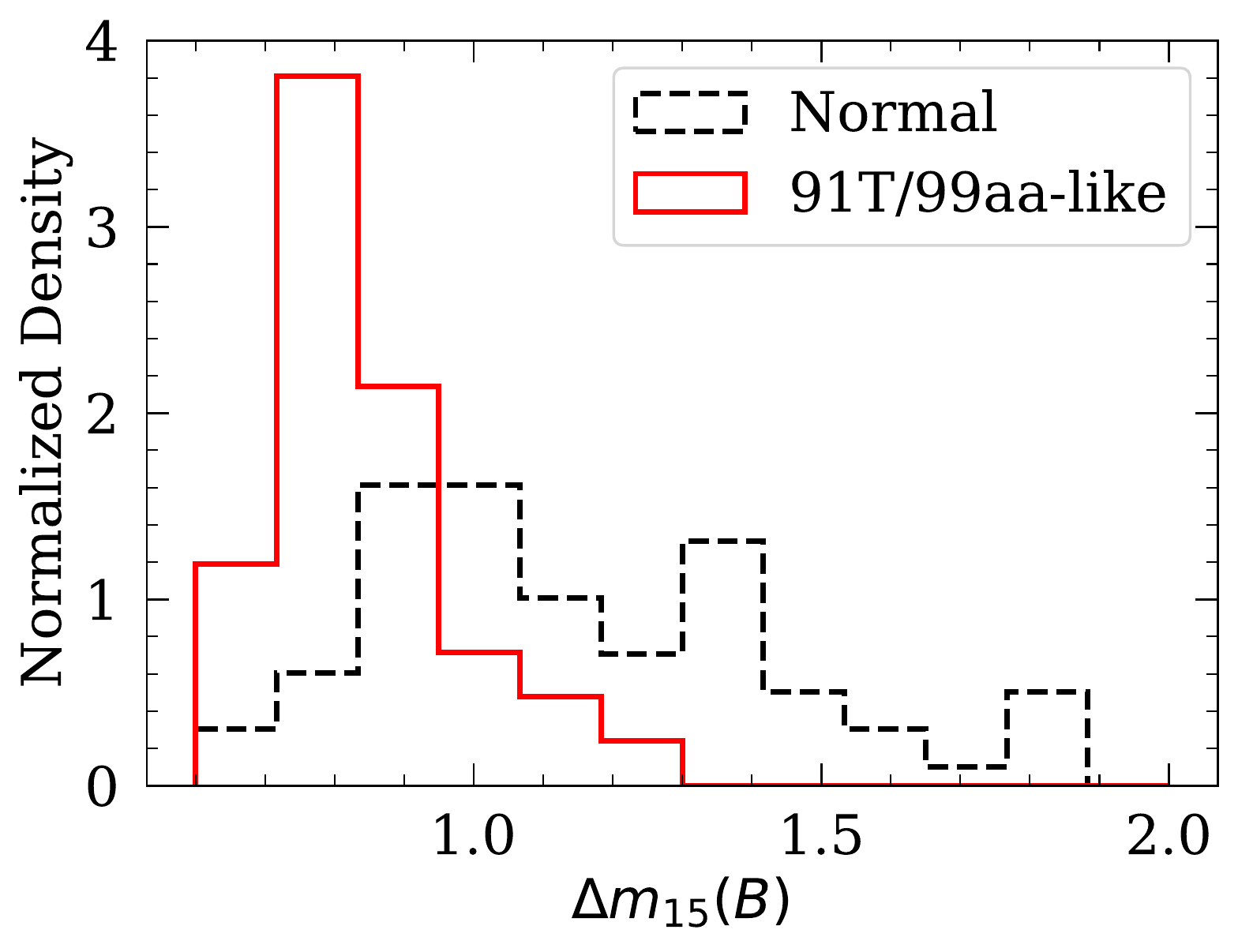}%
\quad
\includegraphics[height=3cm]{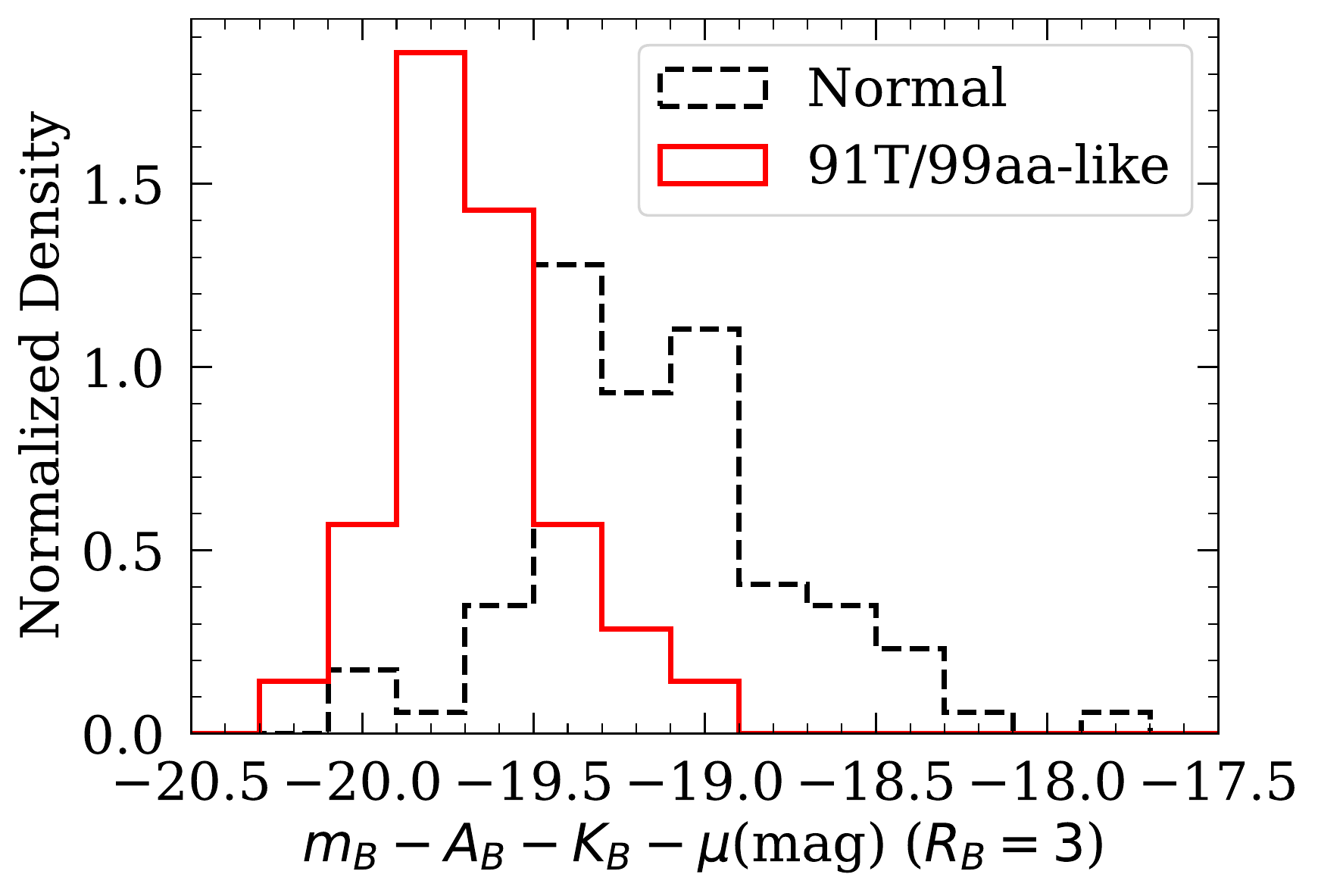}%
}
\caption{\emph{Left}: Histogram of $\Delta m_{15}$ in $B$ band. \emph{Right}: Histogram of absolute magnitudes at maximum in $B$ band assuming a constant $R_B=3$. }
\label{fig: histogram_dm15_Babsmag}
\end{figure}

Host galaxy extinction needs to be taken out if one wants to compare the intrinsic brightness of 91T/99aa-like SNe to normal SNe Ia. The extinction is commonly characterized by the total-to-selective extinction ratio $R_V=A_V/E(B-V)$, where $E(B-V) = A_B-A_V$. Previous studies of the host extinction from SNe Ia yielded diverse values of $R_V$ ranging from $R_V=1$ to $R_V=3.5$ \citep[e.g.][]{ Tripp_1998,Phillips_etal_1999,Conley_etal_2007,Hicken_2009,Wang_etal_2009_HV}. For a rough comparison of the intrinsic luminosity distributions between 91T/99aa-like SNe and normal SNe Ia without any standardization, we take a moderate value of $R_{V(host)}=2$ with $E(B-V)_{host}$ calculated from \autoref{subsec:host_extinction}.
In \autoref{fig: histogram_dm15_Babsmag} we plot the distribution of the decline rate parameter $\Delta m_{15}(B)$, and the extinction- and K-corrected absolute $B-$band magnitudes when assuming host $R_V=2$ for all 123 SNe~Ia included in this work (SN2018apo is not included in this figure as we do not have SN2018apo B-band data.). The distance moduli are calculated using aforementioned assumed cosmology\footnote{Since our supernovae are nearby ($z < 0.1$), luminosity distances are effectively independent of any sensible combination of $\Omega _M$ and $\Omega _{\Lambda}$.}. The weighted mean and standard deviation of the full sample's absolute peak $B$-band magnitude and $\Delta m_{15}(B)$ are listed in \autoref{tab:hist_stats}. 

The spectroscopically peculiar 91T/99aa-like objects are $\sim$ 0.4 mag brighter than normal SNe~Ia after applying K-corrections and correcting for Milky Way extinction and host galaxy extinction. Moreover, these overluminous events seem to have a more uniform peak luminosity and a more uniform distribution of values of the decline rate parameter $\Delta m_{15}(B)$, with a standard deviation of about 0.13 mag compared to 0.32 mag for normal SN~Ia.

\begin{deluxetable}{cccccc}
\tablecaption{Weighted means and standard deviations of absolute peak magnitudes and $\Delta m_{15}(B)$ in \autoref{fig: histogram_dm15_Babsmag}.}
\tablehead{SN subclass & $N_{SN}$ & $\overline{M_B}$ & $\sigma_{M_B}$ & $\overline{\Delta m_{15}(B)}$ & $\sigma_{\Delta m_{15}(B)}$}
\startdata
Normal & 87 & -19.30 & 0.47 & 1.09 & 0.32 \\
91T/99aa-like & 36 & -19.70 & 0.32 & 0.82 & 0.13
\label{tab:hist_stats}
\enddata
\end{deluxetable}

\subsection{The Hubble diagram}

We standardize our sample using the method described in \autoref{sec:standardization} in two cases: for one we assume 91T/99aa-like and normal SNe~Ia share the same intrinsic scatter $\sigma_{int}$, and for the other we assume each group has a different intrinsic scatters $\sigma_{int}(\text{91T/99aa-like})$ and $\sigma_{int}(\text{normal})$ (so that for each group, $\chi^2_\nu\approx1$ after fitting). Final fitting parameters for both cases are summarized in \autoref{tab:hubble_stats}. And the Hubble residuals for the case with different intrinsic scatters are listed in \autoref{tab:parameters_91T_head} and \autoref{tab:parameters_normal_head}.
Below we focus on the case with different intrinsic scatters.

\begin{figure}[!htp]
\centering
\includegraphics[width=0.8\textwidth]{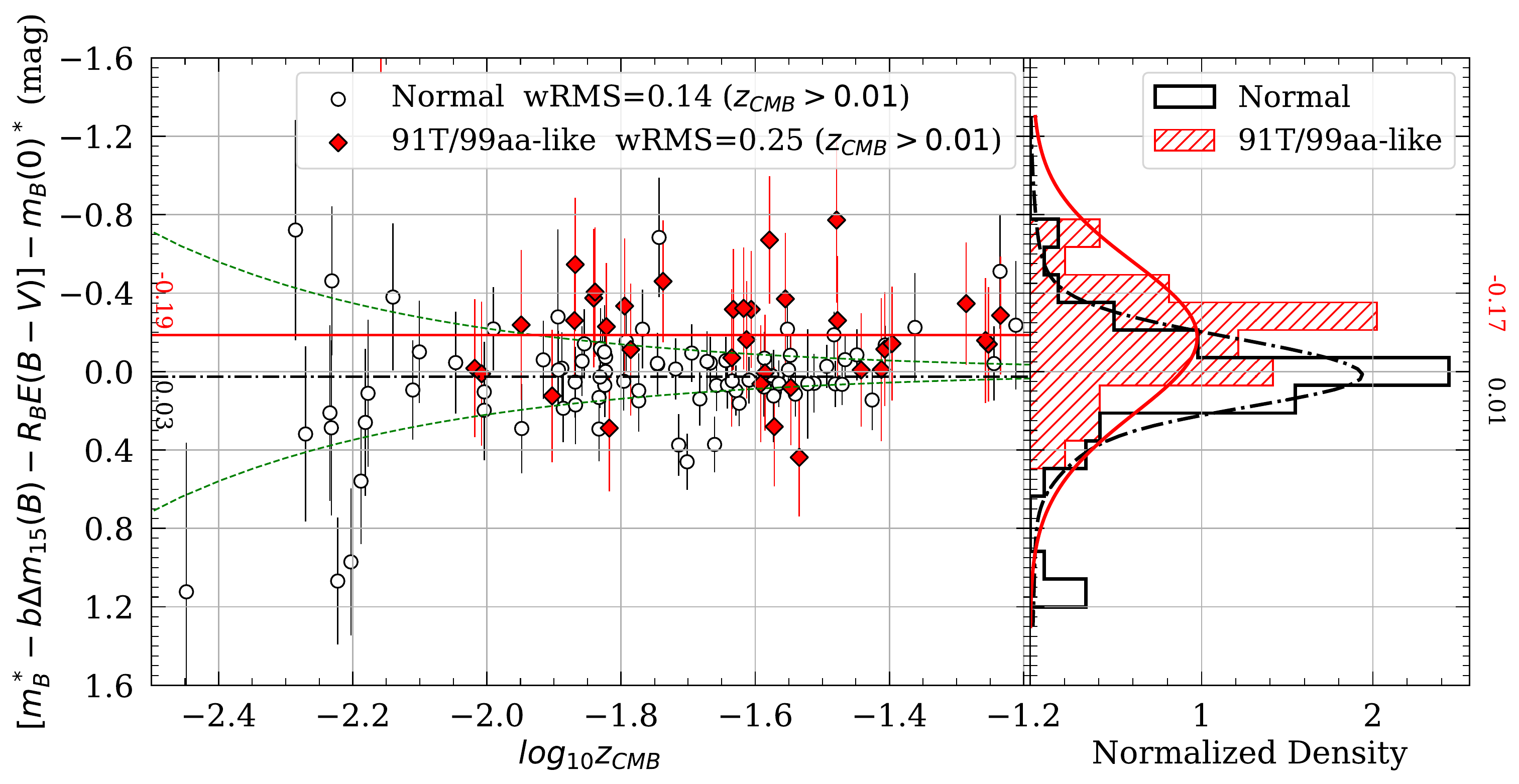}
\caption{Hubble Diagram Residual (left) and residual histogram (right) in B band for 36 91T/99aa-like SNe~Ia (in filled circles) and 87 normal SNe~Ia (in open circles) when assuming these two groups have different intrinsic scatters (0.28 mag for 91T/99aa-like, and 0.05 mag for normal Iae). \emph{Left:} The weighted mean Hubble residuals for 91T-like SNe~Ia and normal SNe~Ia are 0.03 mag and -0.18 mag, which are also plotted as horizontal lines on the plot (solid line for 91T-like SNe, dash dotted line for normal SNe~Ia). Dashed line denotes the peculiar velocity effect with $v_{pec}=300 km/s$. \emph{Right:} Distribution of Hubble residuals. Gaussian ideograms of each group are also plotted (solid line for 91T/99aa-like, dash dotted line for normal SNe~Ia). The peak of resulting Gaussian ideograms are 0.01 mag and -0.17 mag for normal SNe~Ia and 91T/99aa-like SNe~Ia respectively. }
\label{fig:hubble_diag}
\end{figure}

In \autoref{fig:hubble_diag}, we present the $B$ band residual Hubble diagram, and the histogram of Hubble residuals along with the fitted Gaussian ideogram \footnote{See \href{https://pdg.lbl.gov/2015/reviews/rpp2015-rev-rpp-intro.pdf}{https://pdg.lbl.gov/2015/reviews/rpp2015-rev-rpp-intro.pdf} for more discussion on ideogram.}. The weighted mean Hubble residual for normal SNe~Ia and 91T/99aa-like SNe~Ia with $z_{cmb} > 0.01$ are 0.03 mag and -0.19 mag respectively, and the peak in Gaussian ideogram for normal SNe~Ia and all 91T/99aa-like SNe~Ia in our sample are 0.01 mag and -0.17 mag respectively, suggesting that even fully corrected, 91T/99aa-like SNe~Ia still are $\sim 0.2$ mag brighter than normal SNe~Ia \footnote{A similar conclusion about the over-luminosity of 91T/99aa-like SNe has been drawn by M. M. Phillips (Phillips, et al 2023 submitted) based on data from the Carnegie Supernova Project.} \citep{Reindl_etal_2005,2021ApJ...912...71B}. Therefore careful classification of 91T/99aa-like SNe is needed to avoid possible pollution to the cosmological sample of SNe~Ia, or to introduce an additional parameter to account for the Hubble residual offset in 91T/99aa-like and normal SNe.

The intrinsic scatter for 91T/99aa-like is $\sigma_{int}\text{(91T/99aa-like)}=0.28$ mag, more than that of normal SNe~Ia ($\sigma_{int}\text{(normal)}=0.05$ mag). The weighted root mean square (wRMS) are $0.25\pm0.03$ mag and $0.14\pm0.02$ mag for 91T/99aa-like and normal SNe~Ia respectively. It suggests that 91T/99aa-like SNe are also excellent relative distance indicators, though not as precise as normal SNe~Ia, and can be used to delineate the Hubble flow to a precision of 12\%. However, an additional parameter needs to be introduced to bring 91T/99aa-like SNe to the same standard value as normal SNe~Ia. Note that for absolute distance determination, it requires independent estimates of nearby SN hosts \citep[e.g.][]{Riess_etal_2019}, which is beyond the scope of this paper.

\begin{deluxetable}{cccccccccc}
\tabletypesize{\footnotesize}

\tablecolumns{9}

\tablecaption{Fitting parameters of Hubble diagram.}
\tablehead{
\colhead{Condition on $\sigma_{int}$}\vspace{-0.2cm} &\colhead{b}& \colhead{ $R_B$} & k & \multicolumn{2}{c}{$\sigma_{int}$}  & \multicolumn{2}{c}{RMS ($z_{cmb}>0.01$)} & \multicolumn{2}{c}{wRMS ($z_{cmb}>0.01$)}\\
& & & & \multicolumn{2}{c}{(mag)} & \multicolumn{2}{c}{(mag)} & \multicolumn{2}{c}{(mag)} \\
& & & & normal & 91T/99aa-like & normal & 91T/99aa-like & normal & 91T/99aa-like
}
\startdata
$\sigma_{int(normal)}=\sigma_{int(91T/99aa)}$&$0.88 \pm 0.06$ & $3.13 \pm 0.17$ &  $-3.58 \pm 0.03$ & 0.14 & 0.14 & $0.18\pm0.03$ &$0.25\pm0.03$ & $0.16\pm0.02$ & $0.24\pm0.03$\\
$\sigma_{int(normal)}\neq\sigma_{int(91T/99aa)}$&$0.81 \pm 0.05$ & $3.35 \pm 0.15$ &  $-3.57 \pm 0.02$ & 0.05 & 0.28 & $0.17\pm0.02$ & $0.26\pm0.03$ & $0.14\pm0.02$ & $0.25\pm0.03$
\label{tab:hubble_stats}
\enddata
\vspace{-0.8cm}
\tablecomments{comments}

\end{deluxetable}

\subsection{\Si pEW}
\label{subsec: Results_pEW}
We used pEW measurements of the \Si line from the spectra in our database that are closest to---and within 10 days from---$B$ band maximum as our \Si pEW measurement around maximum. In total, we have such measurements of 79 normal SNe~Ia and 35 91T/99aa-like SNe. Final pEW measurements are listed in \autoref{tab:parameters_91T_head} and \autoref{tab:parameters_normal_head}. Hubble residuals versus \Simax are plotted in \autoref{fig: residual_vs_pEW}.(See also M. M. Phillips 2023 for further discussion).
As \Simax\ increases from 0~\AA~to 150~\AA, Hubble residuals increase first and then stay nearly constant. Thus we fit a continuous broken line so that Hubble residual linearly increases with \Simax\ first, and stays constant beyond a \Simax\ value.
The line is fitted using $\chi^2$ minimization \citep[See section 15.3 in ][]{Numerical_Recipes} with \texttt{pycmpfit}. The best fitted result is

\begin{equation}
  \text{Hubble Residual (mag)} =
    \begin{cases}
      (0.009\pm0.003)\times\text{ pEW}(\text{Si II } \lambda \lambda 6355) (\AA)-(0.482\pm0.154), & \text{ pEW}(\text{Si II } \lambda \lambda 6355)<55.62\AA\\
      0.036, & \text{Otherwise.}

    \end{cases}       
\end{equation}

There does not seem to be a correlation between the Hubble residual and $\Delta m_{15}(B)$ as shown in the left panel in \autoref{fig: residual_vs_dm15_and_pEW}. Thus the broken linear correlation between the Hubble residual and \Si is unlikely due to the artifacts of standardizations.
As shown in the right panel of \autoref{fig: residual_vs_dm15_and_pEW}, the relation between $\Delta m_{15}(B)$ and \Simax\ is non-linear. Thus a linear correlation of $\Delta m_{15}(B)$ to Hubble residual cannot fully take out the dependency of Hubble residual on \Simax.

For our sample of 113 objects with $\text{pEW}(\text{Si II } \lambda \lambda 6355)$ near maximum and B-band light curve data, after taking out the fitted linear relationship between \Simax\ and light curve shape- and color- corrected Hubble residual, the wRMS of the Hubble residuals of the 113 objects decreases from 0.17 mag to 0.14 mag (for 34 91T/99aa-like objects, wRMS decreases from 0.26 mag to 0.22 mag, and for 79 normal objects, wRMS decreases from 0.129 mag to 0.125 mag).

The Hubble residual offset in 91T/99aa-like and normal SNe~Ia can then be explained by the positive linear correlation in between the Hubble residual and the pEW, as most SNe~Ia have \Simax $< 55.6\AA$, and 91T/99aa-like SNe are characterized by the shallow Si II lines.

We suggest a new sub-type classification scheme amongst SNe~Ia as follows. The SNe~Ia is subdivided using a hexadecimal number with 0 being the extreme SN1991T-like showing undetectable \Si line, and F being the extreme case with the \Simax\ beyond 160 \AA. In between these extreme cases, the sub-types are given by a hexadecimal number that is the integer part of \Simax/10. The sub-types of SNe in our sample are given in \autoref{tab:parameters_91T_head} and \autoref{tab:parameters_normal_head}.

\label{subsec:figures}
\begin{figure}
    \centering
    \includegraphics[height=10cm]{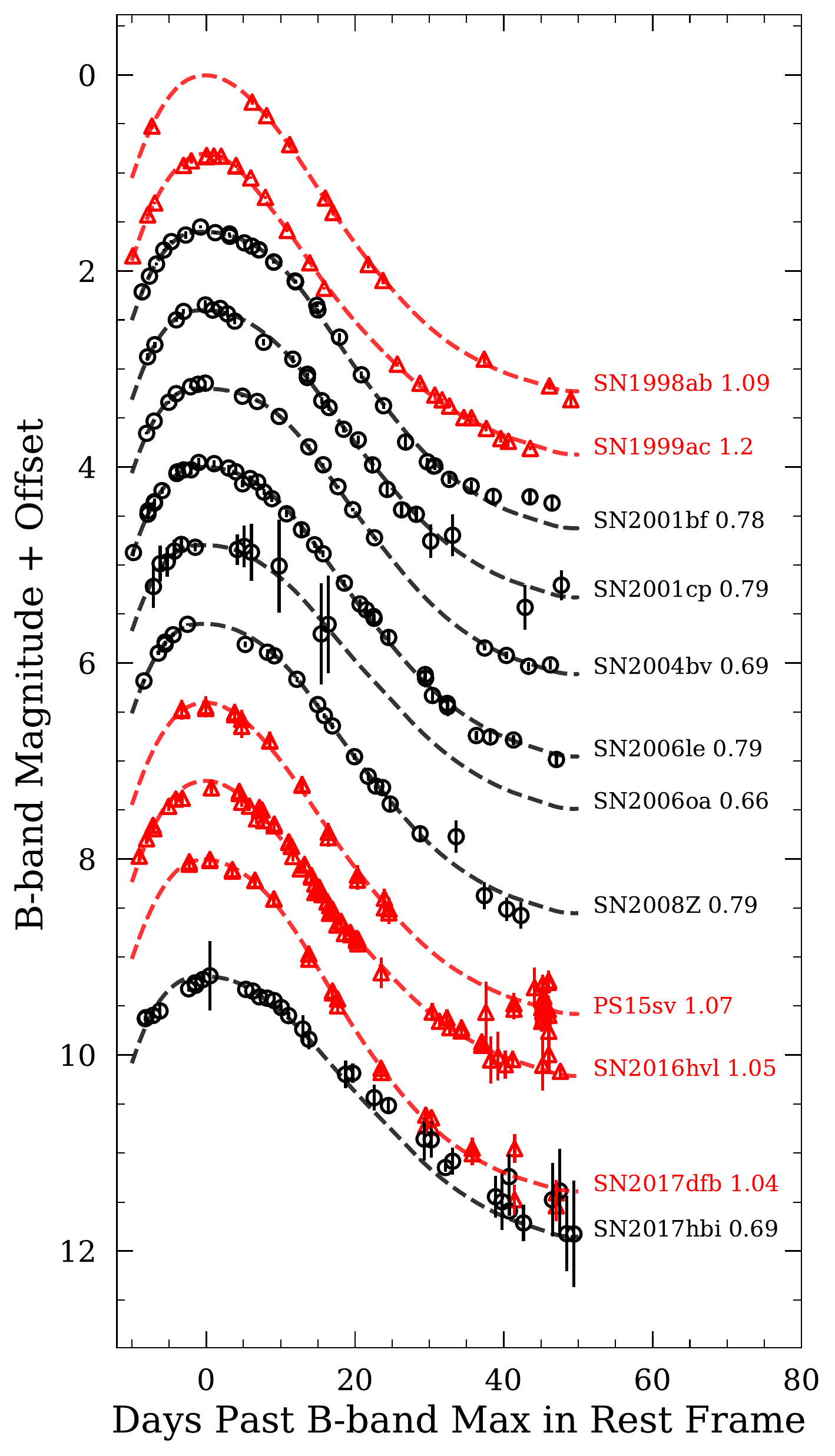}%
    \caption{B-band light curves and fits of normal SNe Ia (in black circles) with $\Delta m_{15}(B)<0.8$ mag and 91T/99aa-like SNe (in red triangles) with $\Delta m_{15}(B)>1.0$ mag. $\Delta m_{15}(B)$ values are labeled next to each light curve.}
    \label{fig:pec_lc}
\end{figure}

In the right panel of \autoref{fig: residual_vs_dm15_and_pEW} we plot the \Simax\ vs $\Delta m_{15}(B)$. As can be seen from the figure, $\Delta m_{15}(B)$ does not always increase as \Simax\ increases. There are a few objects with normal decline rate yet have very shallow Si II lines, and objects with slow decline rate but moderate strength of Si II lines. To show that the effect is not some artifacts from fitting, the light curves of normal objects with $\Delta m_{15}(B)<0.8$ mag and 91T/99aa-like objects with $\Delta m_{15}(B)>1.0$ mag are shown in \autoref{fig:pec_lc}, and their spectra are highlighted in \autoref{fig:spectra_sequence}.

\begin{figure}[!htp]
\centering
{%
\includegraphics[height=7cm]{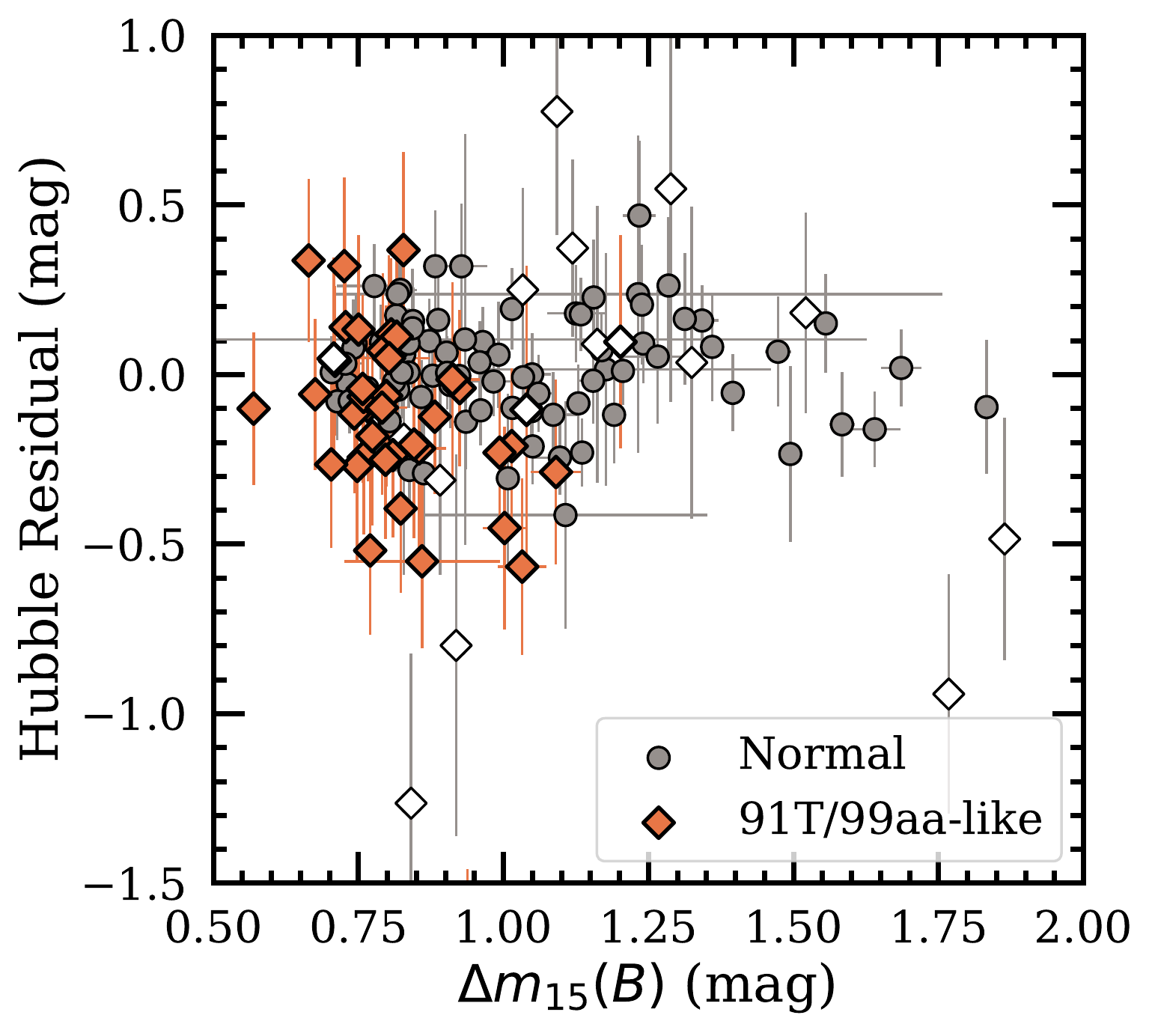}%
\includegraphics[height=7cm]{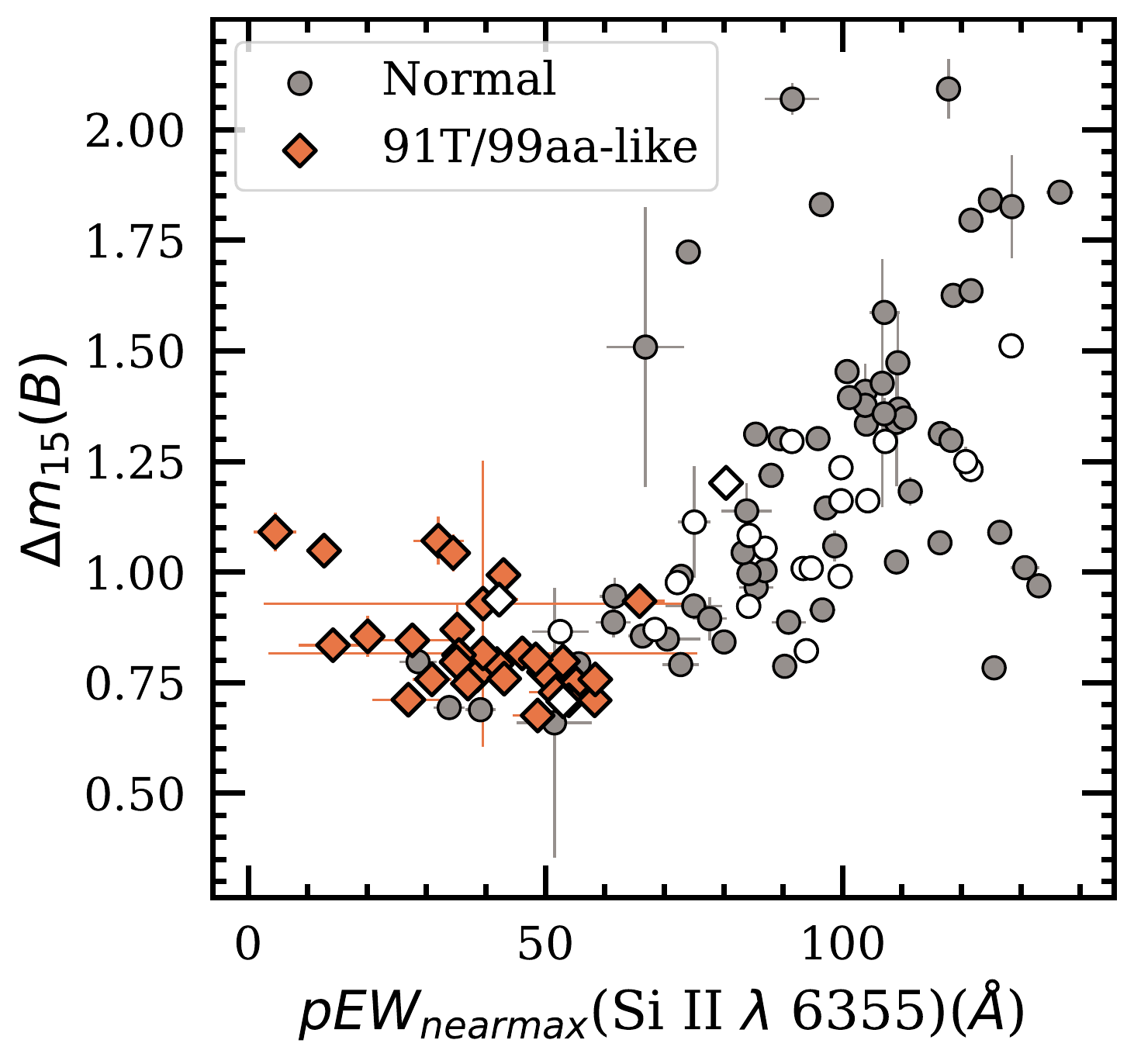}%
}
\caption{\emph{Left:} Hubble residual vs. $\Delta m_{15}(B)$. \emph{Right:} $\Delta m_{15}(B)$ vs. \Simax.}
\label{fig: residual_vs_dm15_and_pEW}
\end{figure}

\begin{figure}[!htp]
\centering
{%
\includegraphics[height=7cm]{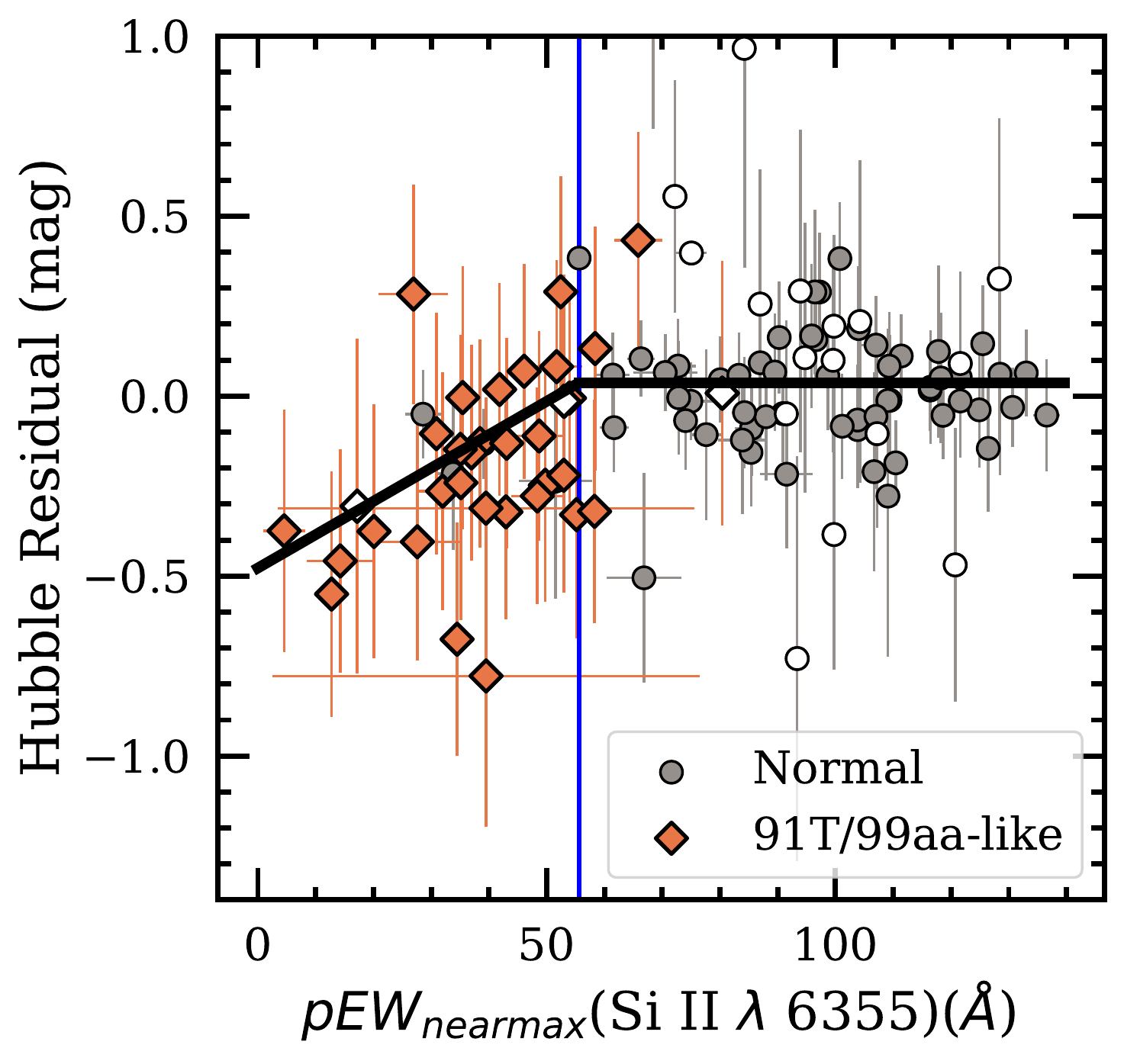}%
}
\caption{Hubble residual vs. \Simax. Open circles indicate objects with $z_{cmb}<0.01$. Blue vertical line is where pEW=55.62$\AA$. Black line shows the best linear fit for objects with pEW $<$ 55.62 $\AA$ (Hubble residual (mag) = $0.009\times pEW_{max}\text{Si II} \lambda 6355 (\AA)-0.482$), and the weighted mean (=0.036 mag) for objects with pEW $>$ 55.62 $\AA$. }
\label{fig: residual_vs_pEW}
\end{figure}

\section{Discussion and Conclusions}
\label{sec:discussion}
Spectroscopically confirmed peculiar SNe~Ia such as overluminous 91T/99aa-like SNe~Ia have generally been excluded from cosmological use of SNe~Ia. However, those overluminous events will be over-represented in future flux-limited surveys. Furthermore, the spectral evolution of 91T/99aa-like objects resembles that of normal SNe~Ia after maximum light, which makes keeping the homogeneity of the SNe~Ia cosmological sample more difficult. Hence, it is crucial to understand the intrinsic properties of these over-luminous SNe~Ia to avoid inevitable systematic errors in luminosity distances.

To address this problem, We obtained multi-epoch optical $BVg'r'i'$ photometric observations of 16 91T/99aa-like SNe~Ia taken by Las Cumbres Observatory in the redshift range $z=0.011 - 0.057$. This sample was analyzed together with 21 well sampled photometric observations of 91T/99aa-like SNe~Ia as well as 87 photometrically well sampled normal SNe~Ia from the literature. Our main conclusions are as follows:

1. When assuming host $R_V= 2$, after correcting for K-correction, MW extinction and host extinction, the spectroscopically peculiar 91T/99aa-like objects are $\sim$0.4 mag brighter than the normal SNe~Ia.

2. After fully corrected (see \autoref{sec:standardization}), 91T/99aa-like objects are still $\sim$0.2 mag brighter than normal SNe~Ia. 

3. 91T/99aa-like objects themselves are excellent relative distance indicators and can be used to delineate galaxies in the Hubble flow to 12\% in distance. 

4. $\Delta m_{15}(B)$ does not always increases as \Simax\ increases. For \Simax$ < 55.6\AA$, light curves tend to decline faster with shallower \Si.

5. After fully corrected, the Hubble residual still has a fairly strong linear correlation with \Simax\ for \Simax$ < 55.6\AA$ while the Hubble residual does not correlate with $\Delta m_{15}(B)$. This is due to the non-linear relation between $\Delta m_{15}(B)$ and \Simax. So a linear correction of $\Delta m_{15}(B)$ cannot fully take out the dependence of the Hubble residual on \Simax\ if any.

The current study is based on the optical data. In the NIR, the systematic offset of the Hubble residuals between 91T/99aa-like and normal SNe may be different from the optical. A careful analyses for the NIR properties of 91T/99aa-like SNe is needed.

The spectral evolution of 91T/99aa-like SNe~Ia argues that early spectral data are needed for their identification. This may cause the SN~Ia samples to be contaminated by mis-identifications of 91T/99aa-like SNe if spectral data before optical maximum are missing. 
This would lead to large systematic biases in cosmological analyses if the fraction of 91T/99aa-like SNe~Ia changes with redshift.
High-mass but passive hosts are observed to produce SNe~Ia over the full range of decline rates, whereas low-mass, star-forming hosts (which increase with redshift \citealp{Noeske_et_al_2007a,Daddi_et_al_2007,Elbaz_et_al_2007}) preferentially give rise to slower declining events like 91T/99aa-like SNe \citep{Sullivan_etal_2010,Uddin_et_al_2020}. 
The fractional evolution of 91T/99aa-like SNe~Ia will cause systematic errors of the SNe Ia Hubble residual to be as large as 0.2 mag in the extreme case at high redshifts where SNe Ia are dominated by 91T/99aa-like events. This will be very important on upcoming surveys like LSST/Rubin and WFIRST/Roman for which the projected uncertainties are smaller than the effects reported here.

Based on the post-maximum spectral similarities between 91T/99aa-like SNe~Ia and the normal SNe~Ia, it may be concluded that the 91T/99aa-like SNe~Ia and normal SNe~Ia could easily be confused with just photometric data. It is critical to identify them in future supernova surveys which aim at much higher precision than today in cosmological applications. 
The 0.2 mag correction between 91T/99aa-like SNe and normal SNe Ia also depends on the relative abundances of 91T/99aa-like objects at different \Simax\ since 91T/99aa-like objects with lower \Simax\ have brighter Hubble residuals. 
Therefore it will be necessary to construct a classification scheme of the 91T/99aa-like SNe. 
The correlation between the Hubble residual and pEW of \Si line seen in Figure~\ref{fig: residual_vs_pEW} suggests that the strength of the \Si line can be employed to sub-classify SNe~Ia. Instead of classifying SNe~Ia with descriptive monikers such as 91T/99aa-like or shallow silicon, we suggest 
dividing the SNe~Ia Types into sub-classes based on the values of \Simax. The measured pEW values are typically in the range from 0 to 160 (see Figure~\ref{fig: residual_vs_pEW}); A single hexadecimal digit can be used to group them in intervals of 10, with the last class being inclusive of all SNe with \Simax\ larger than 160 \AA. The sub-classification is natural as the presence of Si II is the basis for SN~Ia classification. As the Si II line changes rapidly around maximum light, in the case where it is not easy to obtain spectra around maximum light, we recommend using the data-driven spectral modelling code introduced in \citet{Hu_lstm_2022}, which is able to model the spectrum at maximum well given observed spectra at other phases.

Another systematic effect that is routinely addressed recently in cosmological sample is a $\sim$ 0.06 mag correction found from the empirical correlation in between SN~Ia corrected Hubble residuals with host galaxy mass \citep[the mass step;][]{Kelly_etal_2010,Sullivan_etal_2010,Lampeitl_etal_2010,Gupta_et_al_2011,Childress_etal_2013,Johansson_2013}. SNe~Ia in high-mass, passive hosts appear to be brighter than those in low mass, star-forming hosts. It can be easily seen that 91T/99aa-like SNe~Ia have a host mass step that is in the opposite direction since they are associated with active star formation \citep{Hakobyan_2020}. Thus the fraction of 91T/99aa-like SNe~Ia will affect the size of the mass step. These effects need to be carefully taken into account for future surveys to measure cosmological parameters with better control of systematics for future surveys. Our study suggests further that the shallow Si II line SNe Ia cannot be the source of the mass step effect, and that the mass step is caused by the SNe Ia with normal Si II line strength. However, for SNe at high redshifts, the fractional contribution of 91T/99aa-like events may become larger which will lead to a systematic effect that must be taken into account for precision cosmology measurements. 

The correlation between luminosity and the pEW of the Si II 6355 line may also suggest that the group of 91T/99aa-like events consists of SNe with related physical origins. One such possibility is that they are similar events that differ in such effects as arising from geometric orientations and subtle physical processes at the explosion. The SNe with more complete nuclear burning which converts Si to Fe group elements show shallower Si II lines (see \autoref{fig: residual_vs_dm15_and_pEW}, right panel). At the luminous end, with $\Delta m_{15} \lesssim 1$, the light curve shape is insensitive to these effects but the luminosity is correlated with the strength of the Si II 6355 line (see \autoref{fig: residual_vs_pEW}). In the context of delayed detonation, this may be further related to the central density at ignition and the mass of the progenitors \citep{Khokhlov_1991,Khokhlov_etal_1993,Hoeflich_etal_1996,Hoeflich_etal_2017}.  Quantitative modeling of these physical processes is beyond the scope of this study. If geometric effect dominates the observed luminosity and Si II 6355 line strength correlation, \autoref{fig: residual_vs_pEW} suggests that the side with more advanced nuclear burning shows weaker Si II lines when faced with the observer. The geometry of SNe Ia has been probed by spectropolarimetry, and the degree of continuum polarization is consistently low ($\lesssim 0.2\%$). The level of Si II 6355 line polarization is around 0.5\% in general \citep{Wang_Wheeler:annurev.astro.46.060407.145139,Cikota_etal_2019,Yang_2020_18gv}. There have been no published spectropolarimetry data that can place tight constraints on the geometric shapes of the 91T/99aa-like events, although recent data seem to indicate that they are very lowly polarized similar to other SNe Ia (Y. Yang, private communications). If geometric effect does play a role, spectroscopy during the late time nebular phase can also be of great diagnostic value. Interestingly, these observations are qualitatively consistent with the discovery that the SNe with larger light curve stretch values are both more luminous and less polarized \citep[][]{Wang_etal_2007,Cikota_etal_2019}. The more complete burning effectively removes the asymmetries arising from the instabilities during the deflagration phase preceding the detonation.

\begin{acknowledgments}
This work is enabled by observations made from the Las Cumbres Observatory Global Telescope network. The Las Cumbres Observatory team is supported by NSF grants AST-1911225 and AST-1911151. This work also used the computing resources provided by Texas A\&M High Performance Research Computing. 
J.Y. acknowledges generous support from the Texas A\&M University and the George P. and Cynthia Woods Institute for Fundamental Physics and Astronomy. We thank the Mitchell Foundation for their support of the Cook's Branch meetings on Supernovae. We thank Mark Phillips for discussions early on in this project pertaining to the over luminosity of 91T-like SNe Ia in the Hubble diagram and the relationship between the Hubble residual and the \Simax. J.Y. also thanks Yi Yang, Xiaofeng Wang, and Chris Burns for  helpful discussion. J.Y. acknowledges the generosity of Nikolaus Volgenau in helping us understand the Las Cumbres telescopes. L.W. is supported by the NSF grant AST-1817099.
P.J.B. is supported by NASA grant 80NSSC20K0456, ``SOUSA's Sequel: Improving Standard Candles by Improving UV Calibration.''
L.G. acknowledges financial support from the Spanish Ministerio de Ciencia e Innovaci\'on (MCIN), the Agencia Estatal de Investigaci\'on (AEI) 10.13039/501100011033, and the European Social Fund (ESF) "Investing in your future" under the 2019 Ram\'on y Cajal program RYC2019-027683-I and the PID2020-115253GA-I00 HOSTFLOWS project, from Centro Superior de Investigaciones Cient\'ificas (CSIC) under the PIE project 20215AT016, and the program Unidad de Excelencia Mar\'ia de Maeztu CEX2020-001058-M.
This work is supported by NSF through the grant AST-1613455 to N.S..
\end{acknowledgments}

\bibliography{references}{}

\begin{thebibliography}{}
\expandafter\ifx\csname natexlab\endcsname\relax\def\natexlab#1{#1}\fi
\providecommand{\url}[1]{\href{#1}{#1}}
\providecommand{\dodoi}[1]{doi:~\href{http://doi.org/#1}{\nolinkurl{#1}}}
\providecommand{\doeprint}[1]{\href{http://ascl.net/#1}{\nolinkurl{http://ascl.net/#1}}}
\providecommand{\doarXiv}[1]{\href{https://arxiv.org/abs/#1}{\nolinkurl{https://arxiv.org/abs/#1}}}

\bibitem[{Altavilla {et~al.}(2004)Altavilla, Fiorentino, Marconi, Musella,
  Cappellaro, Barbon, Benetti, Pastorello, Riello, Turatto, \&
  Zampieri}]{Altavilla_etal_2004}
Altavilla, G., Fiorentino, G., Marconi, M., {et~al.} 2004, MNRAS, 349, 1344,
  \dodoi{10.1111/j.1365-2966.2004.07616.x}

\bibitem[{Baltay {et~al.}(2021)Baltay, Grossman, Howard, Rabinowitz, Arcavi,
  Barbour, Burke, Contreras, Dilday, Graham, Hiramatsu, Hossenzadeh, Howell,
  McCully, McKinnon, Ment, Montesi, Pellegrino, \& Valenti}]{Baltay_2021}
Baltay, C., Grossman, L., Howard, R., {et~al.} 2021, Publications of the
  Astronomical Society of the Pacific, 133, 44002,
  \dodoi{10.1088/1538-3873/abd417}

\bibitem[{Benetti {et~al.}(2005)Benetti, Cappellaro, Mazzali, Turatto,
  Altavilla, Bufano, Elias-Rosa, Kotak, Pignata, Salvo, \&
  Stanishev}]{Benetti_etal_2005}
Benetti, S., Cappellaro, E., Mazzali, P.~A., {et~al.} 2005, \apj, 623, 1011,
  \dodoi{10.1086/428608}

\bibitem[{Bertin {et~al.}(2002)Bertin, Mellier, Radovich, Missonnier, Didelon,
  \& Morin}]{Bertin_2002}
Bertin, E., Mellier, Y., Radovich, M., {et~al.} 2002, in Astronomical Society
  of the Pacific Conference Series, Vol. 281, Astronomical Data Analysis
  Software and Systems XI, ed. D.~A. Bohlender, D.~Durand, \& T.~H. Handley,
  228

\bibitem[{Bessell(1990)}]{Bessell_1990}
Bessell, M.~S. 1990, \pasp, 102, 1181, \dodoi{10.1086/132749}

\bibitem[{{Betoule} {et~al.}(2014){Betoule}, {Kessler}, {Guy}, {Mosher},
  {Hardin}, {Biswas}, {Astier}, {El-Hage}, {Konig}, {Kuhlmann}, {Marriner},
  {Pain}, {Regnault}, {Balland}, {Bassett}, {Brown}, {Campbell}, {Carlberg},
  {Cellier-Holzem}, {Cinabro}, {Conley}, {D'Andrea}, {DePoy}, {Doi}, {Ellis},
  {Fabbro}, {Filippenko}, {Foley}, {Frieman}, {Fouchez}, {Galbany}, {Goobar},
  {Gupta}, {Hill}, {Hlozek}, {Hogan}, {Hook}, {Howell}, {Jha}, {Le Guillou},
  {Leloudas}, {Lidman}, {Marshall}, {M{\"o}ller}, {Mour{\~a}o}, {Neveu},
  {Nichol}, {Olmstead}, {Palanque-Delabrouille}, {Perlmutter}, {Prieto},
  {Pritchet}, {Richmond}, {Riess}, {Ruhlmann-Kleider}, {Sako}, {Schahmaneche},
  {Schneider}, {Smith}, {Sollerman}, {Sullivan}, {Walton}, \&
  {Wheeler}}]{Betoule_etal_2014}
{Betoule}, M., {Kessler}, R., {Guy}, J., {et~al.} 2014, \aap, 568, A22,
  \dodoi{10.1051/0004-6361/201423413}

\bibitem[{Blondin {et~al.}(2012)Blondin, Matheson, Kirshner, Mandel, Berlind,
  Calkins, Challis, Garnavich, Jha, Modjaz, Riess, \&
  Schmidt}]{Blondin_etal_2012}
Blondin, S., Matheson, T., Kirshner, R.~P., {et~al.} 2012, \aj, 143, 126,
  \dodoi{10.1088/0004-6256/143/5/126}

\bibitem[{Boone {et~al.}(2021)Boone, Aldering, Antilogus, Aragon, Bailey,
  Baltay, Bongard, Buton, Copin, Dixon, Fouchez, Gangler, Gupta, Hayden,
  Hillebrandt, Kim, Kowalski, K{\"{u}}sters, L{\'{e}}get, Mondon, Nordin, Pain,
  Pecontal, Pereira, Perlmutter, Ponder, Rabinowitz, Rigault, Rubin, Runge,
  Saunders, Smadja, Suzuki, Tao, Taubenberger, Thomas, \&
  Vincenzi}]{2021ApJ...912...71B}
Boone, K., Aldering, G., Antilogus, P., {et~al.} 2021, \apj, 912, 71,
  \dodoi{10.3847/1538-4357/abec3b}

\bibitem[{Branch {et~al.}(2006)Branch, Dang, Hall, Ketchum, Melakayil, Parrent,
  Troxel, Casebeer, Jeffery, \& Baron}]{Branch_etal_2006}
Branch, D., Dang, L.~C., Hall, N., {et~al.} 2006, \pasp, 118, 560,
  \dodoi{10.1086/502778}

\bibitem[{{Brown} {et~al.}(2014){Brown}, {Breeveld}, {Holland}, {Kuin}, \&
  {Pritchard}}]{Brown_etal_2014_SOUSA}
{Brown}, P.~J., {Breeveld}, A.~A., {Holland}, S., {Kuin}, P., \& {Pritchard},
  T. 2014, \apss, 354, 89, \dodoi{10.1007/s10509-014-2059-8}

\bibitem[{Brown {et~al.}(2013)Brown, Baliber, Bianco, Bowman, Burleson, Conway,
  Crellin, Depagne, De~Vera, Dilday, Dragomir, Dubberley, Eastman, Elphick,
  Falarski, Foale, Ford, Fulton, Garza, Gomez, Graham, Greene, Haldeman,
  Hawkins, Haworth, Haynes, Hidas, Hjelstrom, Howell, Hygelund, Lister,
  Lobdill, Martinez, Mullins, Norbury, Parrent, Paulson, Petry, Pickles,
  Posner, Rosing, Ross, Sand, Saunders, Shobbrook, Shporer, Street, Thomas,
  Tsapras, Tufts, Valenti, Vander~Horst, Walker, White, \&
  Willis}]{Brown_etal_2013_lcogt}
Brown, T.~M., Baliber, N., Bianco, F.~B., {et~al.} 2013, \pasp, 125, 1031,
  \dodoi{10.1086/673168}

\bibitem[{Burns {et~al.}(2011)Burns, Stritzinger, Phillips, Kattner, Persson,
  Madore, Freedman, Boldt, Campillay, Contreras, Folatelli, Gonzalez,
  Krzeminski, Morrell, Salgado, \& Suntzeff}]{Burns_etal_2011}
Burns, C.~R., Stritzinger, M., Phillips, M.~M., {et~al.} 2011, \aj, 141, 19,
  \dodoi{10.1088/0004-6256/141/1/19}

\bibitem[{Burrow {et~al.}(2020)Burrow, Baron, Ashall, Burns, Morrell,
  Stritzinger, Brown, Folatelli, Freedman, Galbany, Hoeflich, Hsiao,
  Krisciunas, Phillips, Piro, Suntzeff, \& Uddin}]{Burrow_2020_csp}
Burrow, A., Baron, E., Ashall, C., {et~al.} 2020, \apj, 901, 154,
  \dodoi{10.3847/1538-4357/abafa2}

\bibitem[{Childress {et~al.}(2013)Childress, Aldering, Antilogus, Aragon,
  Bailey, Baltay, Bongard, Buton, Canto, Cellier-Holzem, Chotard, Copin,
  Fakhouri, Gangler, Guy, Hsiao, Kerschhaggl, Kim, Kowalski, Loken, Nugent,
  Paech, Pain, Pecontal, Pereira, Perlmutter, Rabinowitz, Rigault, Runge,
  Scalzo, Smadja, Tao, Thomas, Weaver, \& Wu}]{Childress_etal_2013}
Childress, M., Aldering, G., Antilogus, P., {et~al.} 2013, ApJ, 770, 108,
  \dodoi{10.1088/0004-637X/770/2/108}

\bibitem[{{Cikota} {et~al.}(2019){Cikota}, {Patat}, {Wang}, {Wheeler}, {Bulla},
  {Baade}, {H{\"o}flich}, {Cikota}, {Clocchiatti}, {Maund}, {Stevance}, \&
  {Yang}}]{Cikota_etal_2019}
{Cikota}, A., {Patat}, F., {Wang}, L., {et~al.} 2019, \mnras, 490, 578,
  \dodoi{10.1093/mnras/stz2322}

\bibitem[{Conley {et~al.}(2007)Conley, Carlberg, Guy, Howell, Jha, Riess, \&
  Sullivan}]{Conley_etal_2007}
Conley, A., Carlberg, R.~G., Guy, J., {et~al.} 2007, ApJL, 664, L13,
  \dodoi{10.1086/520625}

\bibitem[{Daddi {et~al.}(2007)Daddi, Dickinson, Morrison, Chary, Cimatti,
  Elbaz, Frayer, Renzini, Pope, Alexander, Bauer, Giavalisco, Huynh, Kurk, \&
  Mignoli}]{Daddi_et_al_2007}
Daddi, E., Dickinson, M., Morrison, G., {et~al.} 2007, \apj, 670, 156,
  \dodoi{10.1086/521818}

\bibitem[{DePoy {et~al.}(2003)DePoy, Atwood, Belville, Brewer, Byard, Gould,
  Mason, O'Brien, Pappalardo, Pogge, Steinbrecher, \&
  Teiga}]{2003SPIE.4841..827D}
DePoy, D.~L., Atwood, B., Belville, S.~R., {et~al.} 2003, in Society of
  Photo-Optical Instrumentation Engineers (SPIE) Conference Series, Vol. 4841,
  Instrument Design and Performance for Optical/Infrared Ground-based
  Telescopes, ed. M.~Iye \& A.~F.~M. Moorwood, 827--838,
  \dodoi{10.1117/12.459907}

\bibitem[{Elbaz {et~al.}(2007)Elbaz, Daddi, Le~Borgne, Dickinson, Alexander,
  Chary, Starck, Brandt, Kitzbichler, MacDonald, Nonino, Popesso, Stern, \&
  Vanzella}]{Elbaz_et_al_2007}
Elbaz, D., Daddi, E., Le~Borgne, D., {et~al.} 2007, \aap, 468, 33,
  \dodoi{10.1051/0004-6361:20077525}

\bibitem[{Filippenko {et~al.}(1992{\natexlab{a}})Filippenko, Richmond,
  Matheson, Shields, Burbidge, Cohen, Dickinson, Malkan, Nelson, Pietz,
  Schlegel, Schmeer, Spinrad, Steidel, Tran, \&
  Wren}]{Filippenko_etal_1992_91T}
Filippenko, A.~V., Richmond, M.~W., Matheson, T., {et~al.} 1992{\natexlab{a}},
  ApJL, 384, L15, \dodoi{10.1086/186252}

\bibitem[{Filippenko {et~al.}(1992{\natexlab{b}})Filippenko, Richmond, Branch,
  Gaskell, Herbst, Ford, Treffers, Matheson, Ho, Dey, Sargent, Small, \& van
  Breugel}]{Filippenko_etal_1992_91bg}
Filippenko, A.~V., Richmond, M.~W., Branch, D., {et~al.} 1992{\natexlab{b}},
  AJ, 104, 1543, \dodoi{10.1086/116339}

\bibitem[{Fitzpatrick(1999)}]{Fitzpatrick_1999}
Fitzpatrick, E.~L. 1999, \pasp, 111, 63, \dodoi{10.1086/316293}

\bibitem[{Fitzpatrick \& Massa(1999)}]{Fitzpatrick_Massa_1999}
Fitzpatrick, E.~L., \& Massa, D. 1999, \apj, 525, 1011, \dodoi{10.1086/307944}

\bibitem[{Foley {et~al.}(2018)Foley, Scolnic, Rest, Jha, Pan, Riess, Challis,
  Chambers, Coulter, Dettman, Foley, Fox, Huber, Jones, Kilpatrick, Kirshner,
  Schultz, Siebert, Flewelling, Gibson, Magnier, Miller, Primak, Smartt, Smith,
  Wainscoat, Waters, \& Willman}]{Foley_etal_2018_foundation}
Foley, R.~J., Scolnic, D., Rest, A., {et~al.} 2018, \mnras, 475, 193,
  \dodoi{10.1093/mnras/stx3136}

\bibitem[{Freedman {et~al.}(2001)Freedman, Madore, Gibson, Ferrarese, Kelson,
  Sakai, Mould, Kennicutt~Jr., Ford, Graham, Huchra, Hughes, Illingworth,
  Macri, \& Stetson}]{Freedman_etal_2001}
Freedman, W.~L., Madore, B.~F., Gibson, B.~K., {et~al.} 2001, ApJ, 553, 47,
  \dodoi{10.1086/320638}

\bibitem[{Fukugita {et~al.}(1996)Fukugita, Ichikawa, Gunn, Doi, Shimasaku, \&
  Schneider}]{Fukugita_etal_1996}
Fukugita, M., Ichikawa, T., Gunn, J.~E., {et~al.} 1996, AJ, 111, 1748,
  \dodoi{10.1086/117915}

\bibitem[{Ganeshalingam {et~al.}(2010)Ganeshalingam, Li, Filippenko, Anderson,
  Foster, Gates, Griffith, Grigsby, Joubert, Leja, Lowe, Macomber, Pritchard,
  Thrasher, \& Winslow}]{Ganeshalingam_2010}
Ganeshalingam, M., Li, W., Filippenko, A.~V., {et~al.} 2010, The Astrophysical
  Journal Supplement Series, 190, 418, \dodoi{10.1088/0067-0049/190/2/418}

\bibitem[{Garavini {et~al.}(2004)Garavini, Folatelli, Goobar, Nobili, Aldering,
  Amadon, Amanullah, Astier, Balland, Blanc, Burns, Conley, Dahl{\'{e}}n,
  Deustua, Ellis, Fabbro, Fan, Frye, Gates, Gibbons, Goldhaber, Goldman, Groom,
  Haissinski, Hardin, Hook, Howell, Kasen, Kent, Kim, Knop, Lee, Lidman,
  Mendez, Miller, Moniez, Mour{\~{a}}o, Newberg, Nugent, Pain, Perdereau,
  Perlmutter, Prasad, Quimby, Raux, Regnault, Rich, Richards, Ruiz-Lapuente,
  Sainton, Schaefer, Schahmaneche, Smith, Spadafora, Stanishev, Walton, Wang,
  Wood-Vasey, \& {Supernova Cosmology Project}}]{Garavini_etal_2004}
Garavini, G., Folatelli, G., Goobar, A., {et~al.} 2004, \aj, 128, 387,
  \dodoi{10.1086/421747}

\bibitem[{Goldhaber {et~al.}(2001)Goldhaber, Groom, Kim, Aldering, Astier,
  Conley, Deustua, Ellis, Fabbro, Fruchter, Goobar, Hook, Irwin, Kim, Knop,
  Lidman, McMahon, Nugent, Pain, Panagia, Pennypacker, Perlmutter,
  Ruiz-Lapuente, Schaefer, Walton, \& York}]{Goldhaber_etal_2001}
Goldhaber, G., Groom, D.~E., Kim, A., {et~al.} 2001, ApJ, 558, 359,
  \dodoi{10.1086/322460}

\bibitem[{Goobar \& Perlmutter(1995)}]{1995ApJ...450...14G}
Goobar, A., \& Perlmutter, S. 1995, \apj, 450, 14, \dodoi{10.1086/176113}

\bibitem[{Guillochon {et~al.}(2017)Guillochon, Parrent, Kelley, \&
  Margutti}]{Guillochon_etal_2017}
Guillochon, J., Parrent, J., Kelley, L.~Z., \& Margutti, R. 2017, \apj, 835,
  64, \dodoi{10.3847/1538-4357/835/1/64}

\bibitem[{Gupta {et~al.}(2011)Gupta, D'Andrea, Sako, Conroy, Smith, Bassett,
  Frieman, Garnavich, Jha, Kessler, Lampeitl, Marriner, Nichol, \&
  Schneider}]{Gupta_et_al_2011}
Gupta, R.~R., D'Andrea, C.~B., Sako, M., {et~al.} 2011, \apj, 740, 92,
  \dodoi{10.1088/0004-637X/740/2/92}

\bibitem[{Guy {et~al.}(2007)Guy, Astier, Baumont, Hardin, Pain, Regnault, Basa,
  Carlberg, Conley, Fabbro, Fouchez, Hook, Howell, Perrett, Pritchet, Rich,
  Sullivan, Antilogus, Aubourg, Bazin, Bronder, Filiol, Palanque-Delabrouille,
  Ripoche, \& Ruhlmann-Kleider}]{Guy_etal_2007}
Guy, J., Astier, P., Baumont, S., {et~al.} 2007, \aap, 466, 11,
  \dodoi{10.1051/0004-6361:20066930}

\bibitem[{Guy {et~al.}(2010)Guy, Sullivan, Conley, Regnault, Astier, Balland,
  Basa, Carlberg, Fouchez, Hardin, Hook, Howell, Pain, Palanque-Delabrouille,
  Perrett, Pritchet, Rich, Ruhlmann-Kleider, Balam, Baumont, Ellis, Fabbro,
  Fakhouri, Fourmanoit, Gonz{\'{a}}lez-Gait{\'{a}}n, Graham, Hsiao, Kronborg,
  Lidman, Mourao, Perlmutter, Ripoche, Suzuki, \& Walker}]{Guy_etal_2010}
Guy, J., Sullivan, M., Conley, A., {et~al.} 2010, \aap, 523, A7,
  \dodoi{10.1051/0004-6361/201014468}

\bibitem[{Hakobyan {et~al.}(2020)Hakobyan, Barkhudaryan, Karapetyan, Gevorgyan,
  Mamon, Kunth, Adibekyan, \& Turatto}]{Hakobyan_2020}
Hakobyan, A.~A., Barkhudaryan, L.~V., Karapetyan, A.~G., {et~al.} 2020, \mnras,
  499, 1424, \dodoi{10.1093/mnras/staa2940}

\bibitem[{Hamuy(2003)}]{Hamuy_2003}
Hamuy, M. 2003, ApJ, 582, 905, \dodoi{10.1086/344689}

\bibitem[{Hamuy {et~al.}(1996{\natexlab{a}})Hamuy, Phillips, Suntzeff,
  Schommer, Maza, \& Aviles}]{Hamuy_1996_hubble}
Hamuy, M., Phillips, M.~M., Suntzeff, N.~B., {et~al.} 1996{\natexlab{a}}, \aj,
  112, 2398, \dodoi{10.1086/118191}

\bibitem[{Hamuy {et~al.}(1996{\natexlab{b}})Hamuy, Phillips, Suntzeff,
  Schommer, Maza, \& Aviles}]{Hamuy_etal_1996abs}
---. 1996{\natexlab{b}}, AJ, 112, 2391, \dodoi{10.1086/118190}

\bibitem[{Hamuy {et~al.}(1996{\natexlab{c}})Hamuy, Phillips, Suntzeff,
  Schommer, Maza, \& Aviles}]{Hamuy_etal_1996H0}
---. 1996{\natexlab{c}}, AJ, 112, 2398, \dodoi{10.1086/118191}

\bibitem[{Hamuy {et~al.}(1993)Hamuy, Phillips, Wells, \&
  Maza}]{Hamuy_etal_1993}
Hamuy, M., Phillips, M.~M., Wells, L.~A., \& Maza, J. 1993, PASP, 105, 787

\bibitem[{He {et~al.}(2018)He, Wang, \& Huang}]{He_2018}
He, S., Wang, L., \& Huang, J.~Z. 2018, \apj, 857, 110,
  \dodoi{10.3847/1538-4357/aab0a8}

\bibitem[{Henden {et~al.}(2016)Henden, Templeton, Terrell, Smith, Levine, \&
  Welch}]{Henden_2016}
Henden, A.~A., Templeton, M., Terrell, D., {et~al.} 2016, VizieR Online Data
  Catalog, 2336

\bibitem[{Hicken {et~al.}(2009)Hicken, Challis, Jha, Kirshner, Matheson,
  Modjaz, Rest, Wood-Vasey, Bakos, Barton, Berlind, Bragg, Brice{\~{n}}o,
  Brown, Caldwell, Calkins, Cho, Ciupik, Contreras, Dendy, Dosaj, Durham,
  Eriksen, Esquerdo, Everett, Falco, Fernandez, Gaba, Garnavich, Graves, Green,
  Groner, Hergenrother, Holman, Hradecky, Huchra, Hutchison, Jerius, Jordan,
  Kilgard, Krauss, Luhman, Macri, Marrone, McDowell, McIntosh, McNamara,
  Megeath, Mochejska, Munoz, Muzerolle, Naranjo, Narayan, Pahre, Peters,
  Peterson, Rines, Ripman, Roussanova, Schild, Sicilia-Aguilar, Sokoloski,
  Smalley, Smith, Spahr, Stanek, Barmby, Blondin, Stubbs, Szentgyorgyi, Torres,
  Vaz, Vikhlinin, Wang, Westover, Woods, \& Zhao}]{Hicken_2009}
Hicken, M., Challis, P., Jha, S., {et~al.} 2009, The Astrophysical Journal,
  700, 331, \dodoi{10.1088/0004-637x/700/1/331}

\bibitem[{Hicken {et~al.}(2012)Hicken, Challis, Kirshner, Rest, Cramer,
  Wood-Vasey, Bakos, Berlind, Brown, Caldwell, Calkins, Currie, de~Kleer,
  Esquerdo, Everett, Falco, Fernandez, Friedman, Groner, Hartman, Holman,
  Hutchins, Keys, Kipping, Latham, Marion, Narayan, Pahre, Pal, Peters,
  Perumpilly, Ripman, Sipocz, Szentgyorgyi, Tang, Torres, Vaz, Wolk, \&
  Zezas}]{Hicken_2012}
Hicken, M., Challis, P., Kirshner, R.~P., {et~al.} 2012, The Astrophysical
  Journal Supplement Series, 200, 12, \dodoi{10.1088/0067-0049/200/2/12}

\bibitem[{Hoeflich {et~al.}(2017)Hoeflich, Hsiao, Ashall, Burns, Diamond,
  Phillips, Sand, Stritzinger, Suntzeff, Contreras, Krisciunas, Morrell, \&
  Wang}]{Hoeflich_etal_2017}
Hoeflich, P., Hsiao, E.~Y., Ashall, C., {et~al.} 2017, \apj, 846, 58,
  \dodoi{10.3847/1538-4357/aa84b2}

\bibitem[{H{\"{o}}flich {et~al.}(1996)H{\"{o}}flich, Khokhlov, Wheeler,
  Phillips, Suntzeff, \& Hamuy}]{Hoeflich_etal_1996}
H{\"{o}}flich, P., Khokhlov, A., Wheeler, J.~C., {et~al.} 1996, ApJL, 472,
  L81+, \dodoi{10.1086/310363}

\bibitem[{Howell {et~al.}(2006)Howell, Sullivan, Nugent, Ellis, Conley,
  Le~Borgne, Carlberg, Guy, Balam, Basa, Fouchez, Hook, Hsiao, Neill, Pain,
  Perrett, \& Pritchet}]{Howell_etal_2006}
Howell, D.~A., Sullivan, M., Nugent, P.~E., {et~al.} 2006, Nature, 443, 308,
  \dodoi{10.1038/nature05103}

\bibitem[{Hsiao {et~al.}(2007)Hsiao, Conley, Howell, Sullivan, Pritchet,
  Carlberg, Nugent, \& Phillips}]{Hsiao_etal_2007}
Hsiao, E.~Y., Conley, A., Howell, D.~A., {et~al.} 2007, ApJ, 663, 1187,
  \dodoi{10.1086/518232}

\bibitem[{Hu {et~al.}(2022{\natexlab{a}})Hu, Chen, \& Wang}]{Hu_lstm_2022}
Hu, L., Chen, X., \& Wang, L. 2022{\natexlab{a}}, The Astrophysical Journal,
  930, 70, \dodoi{10.3847/1538-4357/ac5c48}

\bibitem[{Hu {et~al.}(2022{\natexlab{b}})Hu, Wang, Chen, \& Yang}]{Hu_2022}
Hu, L., Wang, L., Chen, X., \& Yang, J. 2022{\natexlab{b}}, The Astrophysical
  Journal, 936, 157, \dodoi{10.3847/1538-4357/ac7394}

\bibitem[{Iben~Jr. \& Tutukov(1984)}]{Iben_Tutukov_1984}
Iben~Jr., I., \& Tutukov, A.~V. 1984, ApJs, 54, 335, \dodoi{10.1086/190932}

\bibitem[{Jha {et~al.}(2006)Jha, Kirshner, Challis, Garnavich, Matheson,
  Soderberg, Graves, Hicken, Alves, Arce, Balog, Barmby, Barton, Berlind,
  Bragg, Brice{\~{n}}o, Brown, Buckley, Caldwell, Calkins, Carter, Concannon,
  Donnelly, Eriksen, Fabricant, Falco, Fiore, Garcia, G{\'{o}}mez, Grogin,
  Groner, Groot, Karl E.~Haisch, Hartmann, Hergenrother, Holman, Huchra,
  Jayawardhana, Jerius, Kannappan, Kim, Kleyna, Kochanek, Koranyi,
  Krockenberger, Lada, Luhman, Luu, Macri, Mader, Mahdavi, Marengo, Marsden,
  McLeod, McNamara, Megeath, Moraru, Mossman, Muench, Mu{\~{n}}oz, Muzerolle,
  Naranjo, Nelson-Patel, Pahre, Patten, Peters, Peters, Raymond, Rines, Schild,
  Sobczak, Spahr, Stauffer, Stefanik, Szentgyorgyi, Tollestrup,
  V{\"{a}}is{\"{a}}nen, Vikhlinin, Wang, Willner, Wolk, Zajac, Zhao, \&
  Stanek}]{Jha_2006}
Jha, S., Kirshner, R.~P., Challis, P., {et~al.} 2006, The Astronomical Journal,
  131, 527, \dodoi{10.1086/497989}

\bibitem[{Johansson {et~al.}(2013)Johansson, Thomas, Pforr, Maraston, Nichol,
  Smith, Lampeitl, Beifiori, Gupta, \& Schneider}]{Johansson_2013}
Johansson, J., Thomas, D., Pforr, J., {et~al.} 2013, \mnras, 435, 1680,
  \dodoi{10.1093/mnras/stt1408}

\bibitem[{Jordi {et~al.}(2006)Jordi, Grebel, \& Ammon}]{Jordi_2006_transform}
Jordi, K., Grebel, E.~K., \& Ammon, K. 2006, \aap, 460, 339,
  \dodoi{10.1051/0004-6361:20066082}

\bibitem[{Kelly {et~al.}(2010)Kelly, Hicken, Burke, Mandel, \&
  Kirshner}]{Kelly_etal_2010}
Kelly, P.~L., Hicken, M., Burke, D.~L., Mandel, K.~S., \& Kirshner, R.~P. 2010,
  ApJ, 715, 743, \dodoi{10.1088/0004-637X/715/2/743}

\bibitem[{{Khokhlov} {et~al.}(1993){Khokhlov}, {Mueller}, \&
  {Hoeflich}}]{Khokhlov_etal_1993}
{Khokhlov}, A., {Mueller}, E., \& {Hoeflich}, P. 1993, \aap, 270, 223

\bibitem[{Khokhlov(1991)}]{Khokhlov_1991}
Khokhlov, A.~M. 1991, \aap, 245, 114

\bibitem[{Kim {et~al.}(1996)Kim, Goobar, \& Perlmutter}]{Kim_etal_1996}
Kim, A., Goobar, A., \& Perlmutter, S. 1996, PASP, 108, 190,
  \dodoi{10.1086/133709}

\bibitem[{Krisciunas {et~al.}(2000)Krisciunas, Hastings, Loomis, McMillan,
  Rest, Riess, \& Stubbs}]{Krisciunas_etal_2000}
Krisciunas, K., Hastings, N.~C., Loomis, K., {et~al.} 2000, ApJ, 539, 658,
  \dodoi{10.1086/309263}

\bibitem[{Krisciunas {et~al.}(2003)Krisciunas, Suntzeff, Candia, Arenas,
  Espinoza, Gonzalez, Gonzalez, H{\"{o}}flich, Landolt, Phillips, \&
  Pizarro}]{Krisciunas_2003_Scorr}
Krisciunas, K., Suntzeff, N.~B., Candia, P., {et~al.} 2003, \aj, 125, 166,
  \dodoi{10.1086/345571}

\bibitem[{Krisciunas {et~al.}(2017)Krisciunas, Contreras, Burns, Phillips,
  Stritzinger, Morrell, Hamuy, Anais, Boldt, Busta, Campillay, Castell{\'{o}}n,
  Folatelli, Freedman, Gonz{\'{a}}lez, Hsiao, Krzeminski, Persson, Roth,
  Salgado, Ser{\'{o}}n, Suntzeff, Torres, Filippenko, Li, Madore, DePoy,
  Marshall, Rheault, \& Villanueva}]{Krisciunas_2017}
Krisciunas, K., Contreras, C., Burns, C.~R., {et~al.} 2017, The Astronomical
  Journal, 154, 211, \dodoi{10.3847/1538-3881/aa8df0}

\bibitem[{Lampeitl {et~al.}(2010)Lampeitl, Smith, Nichol, Bassett, Cinabro,
  Dilday, Foley, Frieman, Garnavich, Goobar, Im, Jha, Marriner, Miquel, Nordin,
  {\"{O}}stman, Riess, Sako, Schneider, Sollerman, \&
  Stritzinger}]{Lampeitl_etal_2010}
Lampeitl, H., Smith, M., Nichol, R.~C., {et~al.} 2010, ApJ, 722, 566,
  \dodoi{10.1088/0004-637X/722/1/566}

\bibitem[{L{\'{e}}get {et~al.}(2020)L{\'{e}}get, Gangler, Mondon, Aldering,
  Antilogus, Aragon, Bailey, Baltay, Barbary, Bongard, Boone, Buton, Chotard,
  Copin, Dixon, Fagrelius, Feindt, Fouchez, Hayden, Hillebrandt, Kim, Kowalski,
  Kuesters, Lombardo, Lin, Nordin, Pain, Pecontal, Pereira, Perlmutter, Ponder,
  Pruzhinskaya, Rabinowitz, Rigault, Runge, Rubin, Saunders, Says, Smadja,
  Sofiatti, Suzuki, Taubenberger, Tao, \& Thomas}]{Leget_2020_SNEMO}
L{\'{e}}get, P.~F., Gangler, E., Mondon, F., {et~al.} 2020, \aap, 636, A46,
  \dodoi{10.1051/0004-6361/201834954}

\bibitem[{Leibundgut {et~al.}(1993)Leibundgut, Kirshner, Phillips, Wells,
  Suntzeff, Hamuy, Schommer, Walker, Gonzalez, Ugarte, Williams, Williger,
  Gomez, Marzke, Schmidt, Whitney, Coldwell, Peters, Chaffee, Foltz, Rehner,
  Siciliano, Barnes, Cheng, Hintzen, Kim, Maza, Parker, Porter, Schmidtke, \&
  Sonneborn}]{Leibundgut_etal_1993}
Leibundgut, B., Kirshner, R.~P., Phillips, M.~M., {et~al.} 1993, AJ, 105, 301,
  \dodoi{10.1086/116427}

\bibitem[{Li {et~al.}(2001)Li, Filippenko, Treffers, Riess, Hu, \&
  Qiu}]{Li_etal_2001}
Li, W., Filippenko, A.~V., Treffers, R.~R., {et~al.} 2001, ApJ, 546, 734,
  \dodoi{10.1086/318299}

\bibitem[{Li {et~al.}(2003)Li, Filippenko, Chornock, Berger, Berlind, Calkins,
  Challis, Fassnacht, Jha, Kirshner, Matheson, Sargent, Simcoe, Smith, \&
  Squires}]{Li_etal_2003}
Li, W., Filippenko, A.~V., Chornock, R., {et~al.} 2003, PASP, 115, 453

\bibitem[{Lineweaver(1997)}]{Lineweaver_1997_cmb_helio}
Lineweaver, C.~H. 1997, in Microwave Background Anisotropies, Vol.~16, 69--75

\bibitem[{Lira {et~al.}(1998)Lira, Suntzeff, Phillips, Hamuy, Maza, Schommer,
  Smith, Wells, Avil{\'{e}}s, Baldwin, Elias, Gonz{\'{a}}lez, Layden,
  Navarrete, Ugarte, Walker, Williger, Baganoff, Crotts, Rich, Tyson, Dey,
  Guhathakurta, Hibbard, Kim, Rehner, Siciliano, Roth, Seitzer, \&
  Williams}]{Lira_1998}
Lira, P., Suntzeff, N.~B., Phillips, M.~M., {et~al.} 1998, The Astronomical
  Journal, 115, 234, \dodoi{10.1086/300175}

\bibitem[{McCully {et~al.}(2018)McCully, Volgenau, Harbeck, Lister, Saunders,
  Turner, Siiverd, \& Bowman}]{McCully_2018}
McCully, C., Volgenau, N.~H., Harbeck, D.-R., {et~al.} 2018, in Software and
  Cyberinfrastructure for Astronomy V, ed. J.~C. Guzman \& J.~Ibsen, Vol.
  10707, International Society for Optics and Photonics (SPIE), 141--149,
  \dodoi{10.1117/12.2314340}

\bibitem[{Noeske {et~al.}(2007)Noeske, Weiner, Faber, Papovich, Koo,
  Somerville, Bundy, Conselice, Newman, Schiminovich, Le~Floc'h, Coil, Rieke,
  Lotz, Primack, Barmby, Cooper, Davis, Ellis, Fazio, Guhathakurta, Huang,
  Kassin, Martin, Phillips, Rich, Small, Willmer, \&
  Wilson}]{Noeske_et_al_2007a}
Noeske, K.~G., Weiner, B.~J., Faber, S.~M., {et~al.} 2007, \apjl, 660, L43,
  \dodoi{10.1086/517926}

\bibitem[{Perlmutter {et~al.}(1999)Perlmutter, Aldering, Goldhaber, Knop,
  Nugent, Castro, Deustua, Fabbro, Goobar, Groom, Hook, Kim, Kim, Lee, Nunes,
  Pain, Pennypacker, Quimby, Lidman, Ellis, Irwin, McMahon, Ruiz-Lapuente,
  Walton, Schaefer, Boyle, Filippenko, Matheson, Fruchter, Panagia, Newberg,
  Couch, \& {The Supernova Cosmology Project}}]{Perlmutter_etal_1999}
Perlmutter, S., Aldering, G., Goldhaber, G., {et~al.} 1999, ApJ, 517, 565,
  \dodoi{10.1086/307221}

\bibitem[{Phillips(1993)}]{Phillips_1993}
Phillips, M.~M. 1993, ApJL, 413, L105, \dodoi{10.1086/186970}

\bibitem[{Phillips {et~al.}(1999)Phillips, Lira, Suntzeff, Schommer, Hamuy, \&
  Maza}]{Phillips_etal_1999}
Phillips, M.~M., Lira, P., Suntzeff, N.~B., {et~al.} 1999, AJ, 118, 1766,
  \dodoi{10.1086/301032}

\bibitem[{Phillips {et~al.}(1992)Phillips, Wells, Suntzeff, Hamuy, Leibundgut,
  Kirshner, \& Foltz}]{Phillips_etal_1992}
Phillips, M.~M., Wells, L.~A., Suntzeff, N.~B., {et~al.} 1992, AJ, 103, 1632,
  \dodoi{10.1086/116177}

\bibitem[{Phillips {et~al.}(2019)Phillips, Contreras, Hsiao, Morrell, Burns,
  Stritzinger, Ashall, Freedman, Hoeflich, Persson, Piro, Suntzeff, Uddin,
  Anais, Baron, Busta, Campillay, Castell{\'{o}}n, Corco, Diamond, Gall,
  Gonzalez, Holmbo, Krisciunas, Roth, Ser{\'{o}}n, Taddia, Torres, Anderson,
  Baltay, Folatelli, Galbany, Goobar, Hadjiyska, Hamuy, Kasliwal, Lidman,
  Nugent, Perlmutter, Rabinowitz, Ryder, Schmidt, Shappee, \&
  Walker}]{Phillips_2019}
Phillips, M.~M., Contreras, C., Hsiao, E.~Y., {et~al.} 2019, \pasp, 131, 14001,
  \dodoi{10.1088/1538-3873/aae8bd}

\bibitem[{Press {et~al.}(2007)Press, Teukolsky, Vetterling, \&
  Flannery}]{Numerical_Recipes}
Press, W.~H., Teukolsky, S.~A., Vetterling, W.~T., \& Flannery, B.~P. 2007,
  Numerical Recipes 3rd Edition: The Art of Scientific Computing, 3rd edn.
  (Cambridge University Press).
\newblock
  \url{http://www.amazon.com/Numerical-Recipes-3rd-Scientific-Computing/dp/0521880688/ref=sr_1_1?ie=UTF8&s=books&qid=1280322496&sr=8-1}

\bibitem[{Reindl {et~al.}(2005)Reindl, Tammann, Sandage, \&
  Saha}]{Reindl_etal_2005}
Reindl, B., Tammann, G.~A., Sandage, A., \& Saha, A. 2005, ApJ, 624, 532,
  \dodoi{10.1086/429218}

\bibitem[{{Riess} {et~al.}(2019){Riess}, {Casertano}, {Yuan}, {Macri}, \&
  {Scolnic}}]{Riess_etal_2019}
{Riess}, A.~G., {Casertano}, S., {Yuan}, W., {Macri}, L.~M., \& {Scolnic}, D.
  2019, \apj, 876, 85, \dodoi{10.3847/1538-4357/ab1422}

\bibitem[{Riess {et~al.}(1996)Riess, Press, \& Kirshner}]{Riess_etal_1996_mlcs}
Riess, A.~G., Press, W.~H., \& Kirshner, R.~P. 1996, ApJ, 473, 88,
  \dodoi{10.1086/178129}

\bibitem[{Riess {et~al.}(1998)Riess, Filippenko, Challis, Clocchiatti, Diercks,
  Garnavich, Gilliland, Hogan, Jha, Kirshner, Leibundgut, Phillips, Reiss,
  Schmidt, Schommer, Smith, Spyromilio, Stubbs, Suntzeff, \&
  Tonry}]{Riess_etal_1998}
Riess, A.~G., Filippenko, A.~V., Challis, P., {et~al.} 1998, AJ, 116, 1009,
  \dodoi{10.1086/300499}

\bibitem[{Riess {et~al.}(1999)Riess, Kirshner, Schmidt, Jha, Challis,
  Garnavich, Esin, Carpenter, Grashius, Schild, Berlind, Huchra, Prosser,
  Falco, Benson, Brice{\~{n}}o, Brown, Caldwell, Dell'Antonio, Filippenko,
  Goodman, Grogin, Groner, Hughes, Green, Jansen, Kleyna, Luu, Macri, McLeod,
  McLeod, McNamara, McLean, Milone, Mohr, Moraru, Peng, Peters, Prestwich,
  Stanek, Szentgyorgyi, \& Zhao}]{Riess_1999}
Riess, A.~G., Kirshner, R.~P., Schmidt, B.~P., {et~al.} 1999, \aj, 117, 707,
  \dodoi{10.1086/300738}

\bibitem[{Riess {et~al.}(2021)Riess, Yuan, Macri, Scolnic, Brout, Casertano,
  Jones, Murakami, Breuval, Brink, Filippenko, Hoffmann, Jha, Kenworthy,
  Mackenty, Stahl, \& Zheng}]{Riess_2021}
Riess, A.~G., Yuan, W., Macri, L.~M., {et~al.} 2021, arXiv e-prints,
  arXiv:2112.04510

\bibitem[{Saunders {et~al.}(2018)Saunders, Aldering, Antilogus, Bailey, Baltay,
  Barbary, Baugh, Boone, Bongard, Buton, Chen, Chotard, Copin, Dixon,
  Fagrelius, Fakhouri, Feindt, Fouchez, Gangler, Hayden, Hillebrandt, Kim,
  Kowalski, K{\"{u}}sters, Leget, Lombardo, Nordin, Pain, Pecontal, Pereira,
  Perlmutter, Rabinowitz, Rigault, Rubin, Runge, Smadja, Sofiatti, Suzuki, Tao,
  Taubenberger, Thomas, Vincenzi, \& Nearby
  Supernova~Factory}]{Saunders_2018_SNEMO}
Saunders, C., Aldering, G., Antilogus, P., {et~al.} 2018, \apj, 869, 167,
  \dodoi{10.3847/1538-4357/aaec7e}

\bibitem[{Scalzo {et~al.}(2012)Scalzo, Aldering, Antilogus, Aragon, Bailey,
  Baltay, Bongard, Buton, Canto, Cellier-Holzem, Childress, Chotard, Copin,
  Fakhouri, Gangler, Guy, Hsiao, Kerschhaggl, Kowalski, Nugent, Paech, Pain,
  Pecontal, Pereira, Perlmutter, Rabinowitz, Rigault, Runge, Smadja, Tao,
  Thomas, Weaver, \& Wu}]{Scalzo_2012}
Scalzo, R., Aldering, G., Antilogus, P., {et~al.} 2012, The Astrophysical
  Journal, 757, 12, \dodoi{10.1088/0004-637x/757/1/12}

\bibitem[{Schlegel {et~al.}(1998)Schlegel, Finkbeiner, \&
  Davis}]{Schlegel_etal_1998}
Schlegel, D.~J., Finkbeiner, D.~P., \& Davis, M. 1998, ApJ, 500, 525,
  \dodoi{10.1086/305772}

\bibitem[{Silverman {et~al.}(2012)Silverman, Foley, Filippenko, Ganeshalingam,
  Barth, Chornock, Griffith, Kong, Lee, Leonard, Matheson, Miller, Steele,
  Barris, Bloom, Cobb, Coil, Desroches, Gates, Ho, Jha, Kandrashoff, Li,
  Mandel, Modjaz, Moore, Mostardi, Papenkova, Park, Perley, Poznanski, Reuter,
  Scala, Serduke, Shields, Swift, Tonry, Van~Dyk, Wang, \&
  Wong}]{Silverman_etal_2012}
Silverman, J.~M., Foley, R.~J., Filippenko, A.~V., {et~al.} 2012, MNRAS, 425,
  1789, \dodoi{10.1111/j.1365-2966.2012.21270.x}

\bibitem[{Smitka {et~al.}(2015)Smitka, Brown, Suntzeff, Zhang, Zhai, Wang, Mo,
  \& Zhang}]{Smitka_etal_2015}
Smitka, M.~T., Brown, P.~J., Suntzeff, N.~B., {et~al.} 2015, \apj, 813, 30,
  \dodoi{10.1088/0004-637X/813/1/30}

\bibitem[{Stahl {et~al.}(2019)Stahl, Zheng, de~Jaeger, Filippenko, Bigley,
  Blanchard, Blanchard, Brink, Cargill, Casper, Channa, Choi, Choksi, Chu,
  Clubb, Cohen, Ellison, Falcon, Fazeli, Fuller, Ganeshalingam, Gates, Gould,
  Halevi, Hayakawa, Hestenes, Jeffers, Joubert, Kandrashoff, Kim, Kim, Kislak,
  Kleiser, Kong, de~Kouchkovsky, Krishnan, Kumar, Leja, Leonard, Li, Li, Lu,
  Mason, Molloy, Pina, Rex, Ross, Stegman, Tang, Thrasher, Wang, Wilkins, Yuk,
  Yunus, \& Zhang}]{Stahl_2019}
Stahl, B.~E., Zheng, W., de~Jaeger, T., {et~al.} 2019, \mnras, 490, 3882,
  \dodoi{10.1093/mnras/stz2742}

\bibitem[{Stetson(1987)}]{Stetson_1987}
Stetson, P.~B. 1987, \pasp, 99, 191, \dodoi{10.1086/131977}

\bibitem[{Stritzinger {et~al.}(2002)Stritzinger, Hamuy, Suntzeff, Smith,
  Phillips, Maza, Strolger, Antezana, Gonz{\'{a}}lez, Wischnjewsky, Candia,
  Espinoza, Gonz{\'{a}}lez, Stubbs, Becker, Rubenstein, \&
  Galaz}]{Stritzinger_etal_2002}
Stritzinger, M., Hamuy, M., Suntzeff, N.~B., {et~al.} 2002, AJ, 124, 2100,
  \dodoi{10.1086/342544}

\bibitem[{Strolger {et~al.}(2002)Strolger, Smith, Suntzeff, Phillips, Aldering,
  Nugent, Knop, Perlmutter, Schommer, Ho, Hamuy, Krisciunas, Germany,
  Covarrubias, Candia, Athey, Blanc, Bonacic, Bowers, Conley, Dahl{\'{e}}n,
  Freedman, Galaz, Gates, Goldhaber, Goobar, Groom, Hook, Marzke, Mateo,
  McCarthy, M{\'{e}}ndez, Muena, Persson, Quimby, Roth, Ruiz-Lapuente, Seguel,
  Szentgyorgyi, von Braun, Wood-Vasey, \& York}]{Strolger_2002}
Strolger, L.~G., Smith, R.~C., Suntzeff, N.~B., {et~al.} 2002, \aj, 124, 2905,
  \dodoi{10.1086/343058}

\bibitem[{Sullivan {et~al.}(2010)Sullivan, Conley, Howell, Neill, Astier,
  Balland, Basa, Carlberg, Fouchez, Guy, Hardin, Hook, Pain,
  Palanque-Delabrouille, Perrett, Pritchet, Regnault, Rich, Ruhlmann-Kleider,
  Baumont, Hsiao, Kronborg, Lidman, Perlmutter, \& Walker}]{Sullivan_etal_2010}
Sullivan, M., Conley, A., Howell, D.~A., {et~al.} 2010, MNRAS, 406, 782,
  \dodoi{10.1111/j.1365-2966.2010.16731.x}

\bibitem[{Suntzeff {et~al.}(1999)Suntzeff, Phillips, Covarrubias, Navarrete,
  P{\'{e}}rez, Guerra, Acevedo, Doyle, Harrison, Kane, Long, Maza, Miller,
  Piatti, Clari{\'{a}}, Ahumada, Pritzl, \& Winkler}]{Suntzeff_1999}
Suntzeff, N.~B., Phillips, M.~M., Covarrubias, R., {et~al.} 1999, \aj, 117,
  1175, \dodoi{10.1086/300771}

\bibitem[{Taam(1980)}]{Taam_1980}
Taam, R.~E. 1980, \apj, 237, 142, \dodoi{10.1086/157852}

\bibitem[{Tripp(1998)}]{Tripp_1998}
Tripp, R. 1998, \aap, 331, 815

\bibitem[{Uddin {et~al.}(2020)Uddin, Burns, Phillips, Suntzeff, Contreras,
  Hsiao, Morrell, Galbany, Stritzinger, Hoeflich, Ashall, Piro, Freedman,
  Persson, Krisciunas, \& Brown}]{Uddin_et_al_2020}
Uddin, S.~A., Burns, C.~R., Phillips, M.~M., {et~al.} 2020, \apj, 901, 143,
  \dodoi{10.3847/1538-4357/abafb7}

\bibitem[{Valenti {et~al.}(2016)Valenti, Howell, Stritzinger, Graham,
  Hosseinzadeh, Arcavi, Bildsten, Jerkstrand, McCully, Pastorello, Piro, Sand,
  Smartt, Terreran, Baltay, Benetti, Brown, Filippenko, Fraser, Rabinowitz,
  Sullivan, \& Yuan}]{Valenti_etal_2016}
Valenti, S., Howell, D.~A., Stritzinger, M.~D., {et~al.} 2016, \mnras, 459,
  3939, \dodoi{10.1093/mnras/stw870}

\bibitem[{Wagers {et~al.}(2010)Wagers, Wang, \& Asztalos}]{Wagers_2010}
Wagers, A., Wang, L., \& Asztalos, S. 2010, 711, 711,
  \dodoi{10.1088/0004-637x/711/2/711}

\bibitem[{Wang {et~al.}(2007)Wang, Baade, \& Patat}]{Wang_etal_2007}
Wang, L., Baade, D., \& Patat, F. 2007, Science, 315, 212,
  \dodoi{10.1126/science.1121656}

\bibitem[{Wang \& Wheeler(2008)}]{Wang_Wheeler:annurev.astro.46.060407.145139}
Wang, L., \& Wheeler, J.~C. 2008, Annual Review of Astronomy and Astrophysics,
  46, 433, \dodoi{10.1146/annurev.astro.46.060407.145139}

\bibitem[{Wang {et~al.}(2009)Wang, Filippenko, Ganeshalingam, Li, Silverman,
  Wang, Chornock, Foley, Gates, Macomber, Serduke, Steele, \&
  Wong}]{Wang_etal_2009_HV}
Wang, X., Filippenko, A.~V., Ganeshalingam, M., {et~al.} 2009, ApJL, 699, L139,
  \dodoi{10.1088/0004-637X/699/2/L139}

\bibitem[{Webbink(1984)}]{Webbink_1984}
Webbink, R.~F. 1984, ApJ, 277, 355, \dodoi{10.1086/161701}

\bibitem[{Whelan \& Iben~Jr.(1973)}]{Whelan_Iben_1973}
Whelan, J., \& Iben~Jr., I. 1973, ApJ, 186, 1007, \dodoi{10.1086/152565}

\bibitem[{Woosley \& Weaver(1994)}]{Woosley_Weaver_1994}
Woosley, S.~E., \& Weaver, T.~A. 1994, ApJ, 423, 371, \dodoi{10.1086/173813}

\bibitem[{Yang {et~al.}(2020)Yang, Hoeflich, Baade, Maund, Wang, Brown,
  Stevance, Arcavi, Burke, Cikota, Clocchiatti, Gal-Yam, Graham, Hiramatsu,
  Hosseinzadeh, Howell, Jha, McCully, Patat, Sand, Schulze, Spyromilio,
  Valenti, Vink{\'{o}}, Wang, Wheeler, Yaron, \& Zhang}]{Yang_2020_18gv}
Yang, Y., Hoeflich, P., Baade, D., {et~al.} 2020, \apj, 902, 46,
  \dodoi{10.3847/1538-4357/aba759}

\bibitem[{Zhang {et~al.}(2016)Zhang, Wang, Sasdelli, Zhang, Liu, Mazzali, Meng,
  Maeda, Chen, Huang, Zhao, Zhang, Zhai, Pian, Wang, Chang, Yi, Wang, Wang,
  Xin, Wang, Lun, Zheng, Zhang, Fan, \& Bai}]{Zhang2016}
Zhang, J.-J., Wang, X.-F., Sasdelli, M., {et~al.} 2016, The Astrophysical
  Journal, 817, 114, \dodoi{10.3847/0004-637X/817/2/114}

\end{thebibliography}
\bibliographystyle{aasjournal}

\end{document}